%% file: main.tex
\documentclass{aastex631}
\input{commands.tex}

\usepackage{amsmath}
\newcommand\numberthis{\addtocounter{equation}{1}\tag{\theequation}}
\usepackage{cleveref}
\usepackage{multirow}

\newcommand{\qub}{Astrophysics Research Centre, School of Mathematics and Physics, Queen's University Belfast, Belfast, BT7 1NN, UK}
\newcommand{\warwick}{Department of Physics, University of Warwick, Gibbet Hill Road, Coventry CV4 7AL, UK}
\newcommand{\cehwar}{Centre for Exoplanets and Habitability, University of Warwick, Gibbet Hill Road, Coventry CV4 7AL, UK}
\newcommand{\leic}{School of Physics and Astronomy, University of Leicester, University Road, Leicester, LE1 7RH, UK}
\newcommand{\norte}{Instituto de Astronom\'ia, Universidad Cat\'olica del Norte, Angamos 0610, 1270709, Antofagasta, Chile}
\newcommand{\diego}{Instituto de Estudios Astrof\'{i}sicos, Facultad de Ingenier\'{i}a y Ciencias, Universidad Diego Portales, Av. Ej\'{e}rcito Libertador 441, Santiago, Chile}
\newcommand{\cata}{Centro de Excelencia en Astrofísica y Tecnologías Afines (CATA), Camino El Observatorio 1515, Las Condes, Santiago, Chile}
\newcommand{\chile}{Departamento de Astronom\'{i}a, Universidad de Chile, Casilla 36-D, Santiago, Chile}
\newcommand{\oxford}{Department of Physics, University of Oxford, Keble Rd, Oxford OX1 3RH, UK}
\newcommand{\flatiron}{Center for Computational Astrophysics, Flatiron Institute, New York, NY 10010, USA}
\newcommand{\adler}{Science Engagement Division, The Adler Planetarium, Chicago, IL 60605, USA}
\newcommand{\esogar}{European Southern Observatory, Karl-Schwarzschild-Stra{\ss}e 2, 85748 Garching bei M{\"u}nchen, Germany}
\newcommand{\mssl}{Mullard Space Science Laboratory, University College London, Holmbury St Mary, Dorking, Surrey RH5 6NT, UK}
\newcommand{\geneva}{Observatoire Astronomique de l'Universit\'{e} de Gen\`{e}ve, Chemin Pegasi 51, 1290 Versoix, Switzerland}
\newcommand{\bern}{Space Research and Planetary Sciences, Physics Institute, University of Bern, Gesellschaftsstrasse 6, 3012 Bern, Switzerland}
\newcommand{\geminin}{Gemini Observatory/NSF's NOIRLab, Hilo, HI, United States}
\newcommand{\geminis}{Gemini Observatory/NSF's NOIRLab, Casilla 603, La Serena, Chile}
\newcommand{\qmul}{Astronomy Unit, Queen Mary University of London, Mile End Road, London E1 4NS, UK}
\newcommand{\citizen}{Planet Hunters NGTS Citizen Scientist}
\newcommand{\esa}{European Space Agency (ESA), European Space Research and Technology Centre (ESTEC), Keplerlaan 1, 2201 AZ Noordwijk, The Netherlands}
\newcommand{\saao}{South African Astronomical Observatory, P.O Box 9, Observatory 7935, Cape Town, South Africa}

\begin{document}

\title{Planet Hunters NGTS: New Planet Candidates from a Citizen Science Search of the \\ Next Generation Transit Survey Public Data}

\correspondingauthor{Sean~M.~O'Brien}
\email{sobrien27@qub.ac.uk}

%Author list. Paste below paper title
\author[0000-0001-7367-1188]{Sean M. O'Brien}
\affiliation{\qub}

\author[0000-0003-4365-1455]{Megan E. Schwamb}
\affiliation{\qub}

\author[0000-0002-4259-0155]{Samuel Gill}
\affiliation{\warwick}
\affiliation{\cehwar}

\author[0000-0002-9718-3266]{Christopher A. Watson}
\affiliation{\qub}

\author[0000-0003-0684-7803]{Matthew R. Burleigh}
\affiliation{\leic}

\author[0009-0006-0719-9229]{Alicia Kendall}
\affiliation{\leic}

\author[0000-0003-2478-0120]{Sarah L. Casewell}
\affiliation{\leic}

\author[0000-0001-7416-7522]{David R. Anderson}
\affiliation{\norte}

\author[0000-0002-1896-2377]{Jos\'{e} I. Vines}
\affiliation{\norte}

\author[0000-0003-2733-8725]{James S. Jenkins}
\affiliation{\diego}
\affiliation{\cata}

\author[0000-0002-5619-2502]{Douglas R. Alves}
\affiliation{\chile}

\author[0000-0002-1113-4122]{Laura Trouille}
\affiliation{\adler}

\author[0000-0003-2417-7006]{Sol\`{e}ne Ulmer-Moll}
\affiliation{\geneva}
\affiliation{\bern}

\author[0000-0001-7904-4441]{Edward M. Bryant}
\affiliation{\mssl}

\author[0009-0004-7473-4573]{Ioannis Apergis}
\affiliation{\warwick}
\affiliation{\cehwar}

\author[0000-0002-1357-9774]{Matthew P. Battley}
\affiliation{\geneva}

\author[0000-0001-6023-1335]{Daniel Bayliss}
\affiliation{\warwick}
\affiliation{\cehwar}

\author[0000-0002-9138-9028]{Nora L. Eisner}
\affiliation{\flatiron}

\author[0000-0003-2851-3070]{Edward Gillen}
\affiliation{\qmul}

\author[0000-0002-2908-7360]{Michael R. Goad}
\affiliation{\leic}

\author[0000-0002-3164-9086]{Maximilian N. G{\"u}nther}
\affiliation{\esa}

\author[0000-0002-2813-3994]{Beth A. Henderson}
\affiliation{\leic}

\author[0000-0003-2530-3000]{Jeong-Eun Heo}
\affiliation{\geminis}

\author[0000-0003-0343-7905]{David G. Jackson}
\affiliation{\qub}

\author[0000-0001-5578-359X]{Chris Lintott}
\affiliation{\oxford}

\author[0000-0003-1631-4170]{James McCormac}
\affiliation{\warwick}
\affiliation{\cehwar}

\author[0000-0002-7927-9555]{Maximiliano Moyano}
\affiliation{\norte}

\author[0000-0002-5254-2499]{Louise D. Nielsen}
\affiliation{\esogar}

\author[0000-0002-5899-7750]{Ares Osborn}
\affiliation{\warwick}
\affiliation{\cehwar}

\author[0000-0001-8018-0264]{Suman Saha}
\affiliation{\diego}
\affiliation{\cata}

\author[0000-0003-3904-6754]{Ramotholo R. Sefako}
\affiliation{\saao}

\author[0000-0002-4434-2307]{Andrew W. Stephens}
\affiliation{\geminin}

\author[0000-0002-2023-4029]{Rosanna H. Tilbrook}
\affiliation{\leic}

\author[0000-0001-7576-6236]{Stéphane Udry}
\affiliation{\geneva}

\author[0000-0001-6604-5533]{Richard G. West}
\affiliation{\warwick}
\affiliation{\cehwar}

\author[0000-0003-1452-2240]{Peter J. Wheatley}
\affiliation{\warwick}
\affiliation{\cehwar}

\author{Tafadzwa Zivave}
\affiliation{\warwick}
\affiliation{\cehwar}

\author[0009-0008-0576-556X]{See Min Lim}
\affiliation{\citizen}

\author[0000-0003-4864-5484]{Arttu Sainio}
\affiliation{\citizen}

\begin{abstract}
We present the results from the first two years of the Planet Hunters NGTS citizen science project, which searches for transiting planet candidates in data from the Next Generation Transit Survey (NGTS) by enlisting the help of members of the general public. Over 8,000 registered volunteers reviewed 138,198 light curves from the NGTS Public Data Releases 1 and 2. We utilize a user weighting scheme to combine the classifications of multiple users to identify the most promising planet candidates not initially discovered by the NGTS team. We highlight the five most interesting planet candidates detected through this search, which are all candidate short-period giant planets. This includes the TIC-165227846 system that, if confirmed, would be the lowest-mass star to host a close-in giant planet. We assess the detection efficiency of the project by determining the number of confirmed planets from the NASA Exoplanet Archive and TESS Objects of Interest (TOIs) successfully recovered by this search and find that 74\% of confirmed planets and 63\% of TOIs detected by NGTS are recovered by the Planet Hunters NGTS project. The identification of new planet candidates shows that the citizen science approach can provide a complementary method to the detection of exoplanets with ground-based surveys such as NGTS. 
\end{abstract}


\section{Introduction}\label{sec:intro}
\input sections/1_intro.tex

\section{Next Generation Transit Survey}\label{sec:ngts}
\input sections/2_ngts.tex

\section{Planet Hunters NGTS}\label{sec:phngts}
\input sections/3_phngts.tex

\section{Identifying Candidates}\label{sec:id_cands}
\input sections/4_idcands.tex

\section{Planet Candidates and Interesting Systems}\label{sec:planet_cands}
\input sections/5_planetcands.tex

\section{Detection Efficiency}\label{sec:det_eff}
\input sections/6_deteff.tex

\section{Conclusions}\label{sec:conclusions}
\input sections/7_conclusions.tex

\section*{Acknowledgments}\label{sec:acknowledgments}
\input sections/ack.tex

\appendix

\section{Distribution of Scores and Weights for the Additional Workflows}\label{sec:scoresweights}
\input sections/8_scoresweights.tex

\section{Tables of PHNGTS Classifications}\label{sec:classtables}
\input sections/9_classtables.tex

\bibliography{biblio}{}
\bibliographystyle{aasjournal}

\end{document}

%% file: commands.tex
\newcommand{\kepler}{{\it Kepler}}

\newcommand{\tess}{{\it TESS}}

\newcommand{\gaia}{{\it Gaia}}

\newcommand{\kms}{km\,s$^{-1}$}

\newcommand{\mjup}{\mbox{M$_{J}$}}
\newcommand{\rjup}{\mbox{R$_{J}$}}

\newcommand{\msun}{\mbox{M$_{\odot}$}}
\newcommand{\rsun}{\mbox{R$_{\odot}$}}
\newcommand{\lsun}{\mbox{L$_{\odot}$}}

\newcommand{\phngts}{PHNGTS}
\newcommand{\mysim}{\mathord{\sim}}

\newcommand{\orion}{\texttt{ORION}}
\newcommand{\triceratops}{\texttt{TRICERATOPS}}
\newcommand{\allesfitter}{\texttt{allesfitter}}

\newcommand{\ariadne}{\texttt{ARIADNE}}
\newcommand{\bruce}{\texttt{BRUCE}}

\newcommand{\fernandoid}{TIC-125925505}
\newcommand{\fernandoplrad}{$0.83^{+0.079}_{-0.076}$}
\newcommand{\fernandoplmass}{9.07}

\newcommand{\cosworthid}{TIC-135251751}

\newcommand{\cosworthplrad}{$1.32^{+0.046}_{-0.037}$}
\newcommand{\cosworthplmass}{2.74}

\newcommand{\nigelid}{TIC-165227846}
\newcommand{\nigelplrad}{$1.61^{+0.13}_{-0.13}$}

\newcommand{\felipeid}{TIC-125759305}
\newcommand{\felipeplrad}{$1.02^{+0.02}_{-0.01}$}

\newcommand{\sebastianid}{TIC-180997904}
\newcommand{\sebastianplrad}{$1.07^{+0.10}_{-0.02}$}

%% file: sections/1_intro.tex
Since the detection of the first exoplanet orbiting a Sun-like star \citep{MayorQueloz1995_51Pegb} over 5000 exoplanets have been discovered to date\footnote{\url{https://exoplanetarchive.ipac.caltech.edu/}} \citep{Akeson2013NASAExoArchive,Christiansen20225000Exoplanets}. The most prolific discovery method thus far has been the transit method, where one observes a decrease in the brightness of a star as a planet passes in front of its host \citep[][and references therein]{Borucki1984TransitMethod,Winn2010TransitBook}. This method relies on the detection of multiple, periodic transit signals, from which we can determine the planetary orbital period as well as the orbital inclination and the planetary radius (provided we have an accurate measurement of the stellar radius), among other parameters \citep[e.g.][]{Burke2014KeplerPCsQ1to8,Mayo2018K2Candidates,Guerrero2021TOIRelease}. This information can be combined with radial velocity \citep[RV;][and references therein]{Lovis2010RadialVel} observations, which provide a measurement of the planetary mass when combined with accurate measurement of the stellar mass, and hence allows the planetary bulk density to be determined -- enabling constraints to then be placed on the planetary composition \citep[e.g.][]{Seager2007MassRadiusRelation,Spiegel2014StructureExoplanets,Zeng2016MRRelationPREM}. Precisely characterizing exoplanets in this way provides insights into the formation mechanisms and evolutionary histories of the diverse range of planetary systems that have been discovered \citep[e.g.][and references therein]{Kley2012PlanetDiskInteractionReview,Zhu2021ExoStatsARA}.

The vast majority of transit searches utilize algorithms such as the Box-fitting Least Squares \citep[BLS;][]{Kovacs2002BLS} and Transiting Planet Search \citep[TPS;][]{Jenkins2002VariabilityTransitDetection,Jenkins2020KeplerTPS} algorithms, which detect repeated decreases in the brightness of stars to identify the strongest periodic signals \citep[e.g.][]{Bakos2004HATNet,Pollacco2006WASP,Vanderburg2016K2Yr1Candidates}. These candidates are visually vetted by small teams of professional astronomers to determine whether the dips in star light are characteristic of exoplanet transits or are due to false positives such as instrumental systematics or eclipsing binaries \citep{Morton2016KOIFPPs,Collins2018KELT_FPCat,Schanche2019SuperWASP_FPCat,Guerrero2021TOIRelease}. The most promising planet candidates are selected for follow-up observations such as RV measurements to attempt to validate or confirm the planetary nature of the candidate, provided they are amenable to such observations \citep[e.g.][]{Konacki2005OGLETR10bConf,Bakos2007HATP1b,Esposito2019HD219666bTESSSec1}.

However, the visual vetting stage is time-intensive and transit surveys are becoming increasingly limited by computing resources and the number of hours humans can contribute to the search, rather than being data-limited \citep{Kohler2019BigDataK2EE,Baron2019MLinAstro}. The process of visually vetting light curves for exoplanet transit-like features is an exercise in pattern recognition, which the human brain excels at \citep{Dashti2010PatternRecognition}. Indeed, with minimal training, non-expert volunteers can be trained to classify these dips, allowing us to discover exoplanets that were missed by traditional vetting techniques \citep[e.g.][]{Fischer2012PHPCs,Christiansen2018ExoExplorersK2138,Eisner2021PHT2Overview}. Citizen science has been employed in many astronomical \citep[e.g.][]{Lintott2008GalaxyZoo,Schwamb2012PHOverview,Kuchner2017BackyardWorlds,Aye2019Planet4} and non-astronomical fields \citep[e.g.][]{Jones2018PenguinWatch, Blickhan2019AntiSlavery, semenzin2020, Spiers2021} to efficiently search through these large datasets and also serendipitously spot peculiar phenomena that would be missed by automated algorithms \citep{Lintott2009HannyVoorwerp,Cardamone2009GalaxyZooGreenPeas}.

The successes of these citizen science projects, in particular the Planet Hunters \citep{Schwamb2012PHOverview}, Exoplanet Explorers \citep{Christiansen2018ExoExplorersK2138} and Planet Hunters TESS \citep{Eisner2021PHT2Overview} projects, which use data from the spaced-based \kepler\ \citep{Borucki2010Kepler}, K2 \citep{Howell2014K2Mission} and \textit{Transiting Exoplanet Survey Satellite} \citep[\tess;][]{Ricker2015TESS} missions, respectively, has led to the launch of the Planet Hunters NGTS (\phngts) citizen science project. \phngts\ uses data from the Next Generation Transit Survey (NGTS; \citet[][]{Wheatley2018NGTS}; see Section~\ref{sec:ngts}), that has been surveying large sections of the sky in search of exoplanet transits since 2015. To date, the facility has discovered tens of exoplanets \citep[e.g.][]{Bayliss2018NGTS1b,Tilbrook2021NGTS15.18b,Jackson2023NGTS22.25b} primarily through the visual vetting of candidates identified by \orion, a custom BLS algorithm adapted from \citet{CollierCameron2006TransitAlgorithm}. While the primary objective of the Planet Hunters and Planet Hunters TESS projects is to identify transit events that were not detected by the standard detection pipelines, \phngts\ uses phase-folded light curves from the \orion\ algorithm with the aim being to classify transit-like events in the NGTS data and identify planet candidates that were missed in the initial vetting of these data.

In this paper, we outline the \phngts\ project and describe the key results from the analysis of data from the first two NGTS Data Releases (DR1 \& DR2), including the discovery of 5 planet candidates not previously detected in these data. The layout of this paper is as follows. Section~\ref{sec:ngts} describes the photometric data from the NGTS facility that is used in the \phngts\ project. We give an overview of the \phngts\ project in Section~\ref{sec:phngts}. We describe the process of how we identify planet candidates in Section~\ref{sec:id_cands} and provide details of the most promising planet candidates discovered and describe the follow-up data obtained in Section~\ref{sec:planet_cands}. We evaluate the detection efficiency of the citizen science project in Section~\ref{sec:det_eff}. Finally, in Section~\ref{sec:conclusions}, we outline the conclusions.

%% file: sections/2_ngts.tex
The Next Generation Transit Survey \citep[NGTS;][]{Wheatley2018NGTS} is an array of twelve robotically-operated telescopes located at the Paranal Observatory in Chile. Each telescope has a 20\,cm aperture and a field of view of $8\deg^2$. The primary goal of NGTS is to survey large sections of the sky in search of exoplanet transits \citep[][]{Bayliss2018NGTS1b}. Since 2018, NGTS has also been used extensively for exoplanet follow-up observations, particularly for \tess\ mission candidates \citep[e.g.][]{Armstrong2020TOI849RemnantCore}. NGTS operates using a custom filter with a bandpass from 520 to 890\,nm, maximising sensitivity to late K and early M dwarf stars. The NGTS facility is designed to be sensitive to transit depths of 0.1\,per\,cent in order to detect Neptune-sized planets around Sun-like stars. The discovery of NGTS-4b with a transit depth of 0.13\,per\,cent represents the shallowest transiting system ever discovered from the ground \citep{West2019NGTS4b_SubNeptune}.

The details of the NGTS survey strategy and detection pipeline are given in \citet{Wheatley2018NGTS}, here we provide a brief summary. The twelve NGTS telescopes each survey large ($8\deg^2$) sections of the sky (`fields') for $\mysim250$\,nights during an observing season, which results in $\mysim500$\,hours of coverage per field. The NGTS telescopes operate with exposure times of 10\,s, which limits survey targets to $I$-band magnitudes brighter than approximately 16\,mag. The data for all survey fields observed up until 2018 are available via the ESO Archive Science Portal\footnote{\url{http://archive.eso.org/scienceportal/home}} \citep{NGTS2018NGTSDR1,NGTS2020NGTSDR2}, hereafter referred to as the `NGTS Public Data'. The results presented in this work are based on the analysis of the NGTS Public Data. The survey data are detrended using the \texttt{SysRem} algorithm \citep{Tamuz2005SysRemPaper,Mazeh2007SysRem} and searched for candidate transiting events using a custom Box-fitting Least Squares \citep[BLS;][]{Kovacs2002BLS} algorithm called \orion. The \orion\ algorithm computes a periodogram to detect the most significant periodic signals for each star, with up to 5 candidate periods identified for a given target. Each candidate period has an associated Signal Detection Efficiency (SDE), where a higher SDE value indicates a stronger periodic signal \citep[see][]{Alcock2000BLSSEDRef,Kovacs2002BLS}.

For context, we describe the internal vetting process that has been used in the discovery of 23 exoplanets to date by NGTS \citep[][Bouchy et al., in press]{Bayliss2018NGTS1b,Raynard2018NGTS2bInflatedHJ,Guenther2018NGTS3AbHJUnresolvedBinary,West2019NGTS4b_SubNeptune,Eigmueller2019NGTS5bInflatedHJ,Vines2019NGTS6b_USP_HJ,Costes2020NGTS8b9b_HJs,McCormac2020NGTS10b_USP_HJ,Bryant2020MNRASNGTS12b_subSaturn,Grieves2021NGTS13b_SupJup_Subgiant,SmithAMS2021NGTS14b_NeptuneDesert,Tilbrook2021NGTS15.18b,Alves2022MNRASNGTS21bInflatedSupJup,Jackson2023NGTS22.25b}. This process proceeds as follows: for each field, two professional astronomers (vetters) are assigned to independently review \orion\ candidates by scrolling through a webpage that displays all candidates for the given field. This interface displays, for each candidate: a light curve phase-folded on the most significant period (Peak 1); a plot of the BLS periodogram; and a thumbnail image of the region around the star to examine whether there are any nearby stars contaminating the photometric measurements. Further information such as the measured orbital period, SDE and estimated radius of the orbiting body are also given in this interface. In addition, each candidate has its own webpage that can be accessed from the main interface to view the light curves for individual nights and phase-folded data for any other significant periods identified by the algorithm. The candidate webpage can also be used to check whether secondary eclipse events are visible, or if there is a difference between the depths of the odd and even transits. These checks are designed to identify signals due to eclipsing binaries, a common false positive in transit surveys \citep{ODonovan2006TrES_BEBReject,Howell2011SpeckleKepler,LilloBox2014ComprehensiveFPRejection,Ciardi2015StellarMultiplicity,Lester2021SpeckleTESS}. Vetters can designate promising candidates for discussion that are then reviewed by the full vetting team to identify the best candidates for further follow-up observations. Typically, vetters are expected to review $\mysim1000-5000$ potential candidates that will contain a large number of false positives, due to the limitations of the search algorithm. The nature of this task leads to fatigue and therefore lower effectiveness in identifying potential planet candidates, as cognitive fatigue is common when undertaking repetitive tasks \citep{Robertson2010VigilantAttention,Langner2013SustainingAttention}.

%including ultra-short period hot Jupiters \citep{Vines2019NGTS6b_USP_HJ,McCormac2020NGTS10b_USP_HJ} and a sub-Neptune-sized planet orbiting in the Neptunian Desert \citep{West2019NGTS4b_SubNeptune}.

%, which is calculated by combining transit depth with the stellar radius as given by the TESS Input Catalog \citep{Stassun2019RevisedTIC},

%% file: sections/3_phngts.tex
\begin{figure*}
    \centering
    \includegraphics[width=\textwidth,height=0.84\textheight,keepaspectratio]{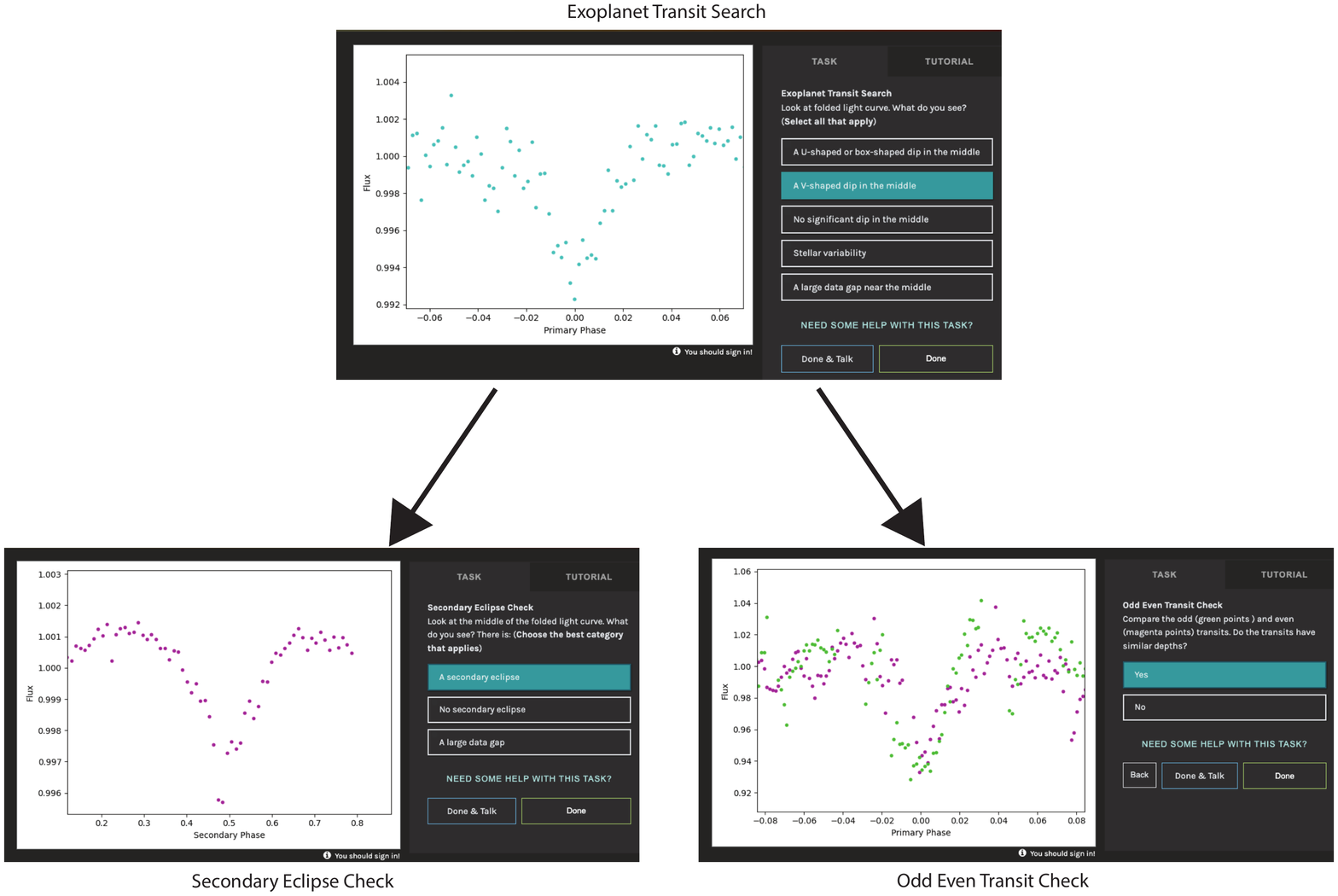}
    \caption{\phngts\ classification interfaces. \textbf{Top panel:} Exoplanet Transit Search interface showing a phase-folded light curve with the response `A V-shaped dip in the middle' selected. \textbf{Bottom left panel:} Secondary Eclipse Check interface showing the phase-folded light curve centered on phase=0.5 with the response `A secondary eclipse' selected. \textbf{Bottom right panel:} Second task of the Odd/Even Transit Check workflow asking whether the green points (odd transits) and magenta points (even transits) have similar depths, the selected response is `Yes'. Note the example images for each workflow do not show data for the same star.}
    \label{fig:combined_interface}
\end{figure*}

The \phngts\ project\footnote{\url{https://ngts.planethunters.org}} is the next iteration of the Planet Hunters citizen science project and is hosted on the Zooniverse platform \citep{Lintott2008GalaxyZoo,Lintott2011GalaxyZoo1,Fortson2012GalaxyZooZooniverse}. The aim of the project is to identify planet candidates that were missed in the initial vetting of the NGTS data. Volunteers who visit the website are shown phase-folded light curves (known as `subjects') and are asked questions to classify these subjects in a series of workflows. The project is divided into three workflows: the Exoplanet Transit Search (Figure~\ref{fig:combined_interface}: top panel); the Secondary Eclipse Check (Figure~\ref{fig:combined_interface}: bottom left panel); and the Odd/Even Transit Check (Figure~\ref{fig:combined_interface}: bottom right panel). Volunteers are free to select any of these three workflows to do. The Exoplanet Transit Search is the main classification interface where the primary transit in phase-folded light curves for \orion\ candidates are reviewed. If a subject is identified as a potential candidate via this workflow (see Section~\ref{sec:id_cands}) then the candidate is also classified in both the Secondary Eclipse Check and Odd/Even Transit Check.

First-time visitors to the \phngts\ site and users who are not logged in are prompted to read a short tutorial for each workflow that: outlines the aims of the \phngts\ project; explains the transit method and the phenomena each possible response is designed to identify; and shows examples of each of the possible responses for the given workflow. In addition to the tutorial, which can be accessed at any time, there is a help box and a `Field Guide' that show a wide variety of example light curves for each of the responses. After viewing this tutorial, volunteers can begin classifying real data from the NGTS facility. Once a volunteer chooses the response that best describes the features in the subject presented and selects the `Done' button, the classification is stored in the Zooniverse database and cannot be edited. The subject identifier, volunteer's anonymized IP (Internet Protocol) address, Zooniverse username and user ID number (if the user is logged in), timestamp, web browser and operating system information, and user response are recorded. Volunteers can classify either through a registered Zooniverse account or whilst not logged in. The classification process is identical for both user types although not-logged-in users are prompted to login/register for a Zooniverse account. Classifications made by registered users are linked by their Zooniverse username/ID while classifications made by users who are not logged in can be linked via the masked IP address. We note that non-logged-in classifications from a single IP address may not necessarily equate to a single individual, hence we refer to each unique IP address as `non-logged-in sessions' rather than users. Tables~\ref{tab:ets_annotations}, \ref{tab:sec_annotations} and \ref{tab:oec_annotations} in Appendix~\ref{sec:classtables} provide examples of the cleaned versions (duplicate classifications and auxiliary information columns removed) of the raw classifications submitted for each workflow. The full tables are available in the online supplementary material.

In the classification interface, each subject is presented without identifying information or stellar properties in order to reduce biases in the classifications \citep{Mayo1934Hawthorne,Adair1984Hawthorne,Levitt2009Hawthorne,Schwamb2012PHOverview}. Once a subject has been classified, users can choose to comment on the light curve, or discuss light curves classified by other volunteers, via the `Talk' discussion forum. Each subject has its own discussion page in the `Talk' forum where users can discuss the light curve features and see information about the subject including the stellar radius, if available, from the TESS Input Catalog (TIC) v8.2 \citep{Stassun2019RevisedTIC,Paegert2021TICv8.2}. The stellar radius, when combined with the transit depth (which can be estimated from the light curve presented), allows volunteers to estimate the radius of the transiting body and therefore highlight promising subjects to the attention of the science team via the `Talk' forum. The additional data for each subject does not include the TIC ID for the star. Each candidate receives a unique subject ID for each workflow but can be linked by the science team through their unique NGTS IDs or TIC ID where available.

The \phngts\ project utilizes phase-folded light curves for the most significant peak in the periodogram for each candidate (Peak 1) where the SDE is greater than 8. We elect to use this threshold for SDE as it has been found through the internal vetting process that below this limit the frequency of false positives due to systematics is excessive. The phase-folded light curves presented are binned to 10 minutes as it was found during testing that the subjects proved more straightforward to classify in this format, without possible transits being obscured. In addition, the light curves are displayed without error bars as in previous citizen science projects we have found that error bars do not influence the classification made by volunteers. The y-axis (normalized flux) spans a range from $1-1.5 \times \delta$ to $1 + \delta,$ where $\delta$ is the fractional transit depth measured by \orion. The x-axis (orbital phase) spans a range from $-\frac{3 \times T_{dur}}{P}$ to $\frac{3 \times T_{dur}}{P},$ where $T_{dur}$ and $P$ are the transit duration and orbital period measured by \orion, respectively. Here we describe each of the three workflows that constitute the \phngts\ project.

\subsection{Exoplanet Transit Search}\label{sec:exo_transit_search}
The Exoplanet Transit Search (Figure~\ref{fig:combined_interface}) is the main workflow of the \phngts\ project. There is no vetting stage similar to that described in Section~\ref{sec:ngts} that occurs prior to subjects being uploaded to the Exoplanet Transit Search. All \orion\ candidates with an SDE greater than 8 are classified in this workflow before potentially advancing to the additional workflows. Users are shown a phase-folded light curve centered on the location of the primary transit as identified by \orion\ and are asked to select any of the following responses that may apply: `A U-shaped or box-shaped dip in the middle' - typically indicative of an exoplanet transit; `A V-shaped dip in the middle' - typically indicative of an eclipsing binary transit; `No significant dip in the middle' - no clear transit visible; `Stellar variability' - identifies light curves that show out-of-transit variation; `A large data gap near the middle' - identifies light curves with large gaps in the data points around the location of the possible transit. Each subject in the Exoplanet Transit Search is seen by 20 volunteers before being retired from the workflow. We chose the number of classifications a subject must receive before retirement to balance the time it takes to fully classify the full sample while receiving a sufficient number of classifications to characterise the features present in a given subject. We elected to obtain 20 classifications per subject, an increase compared to previous citizen science projects such as Planet Hunters \citep{Schwamb2012PHOverview} and Planet Hunters TESS \citep{Eisner2021PHT2Overview} that use $\mysim8-15$ classifications per subject. We determined through beta testing of the project with a small number of Zooniverse volunteers that the increased complexity of the \phngts\ tasks required a greater number of classifiers. The Exoplanet Transit Search is designed to identify the most promising exoplanet candidates. This is achieved by combining the classifications of multiple users on each subject to select the candidates that are most likely to show a transit-like feature (see Section~\ref{sec:id_cands}). These candidates are then advanced to both the Secondary Eclipse Check and Odd/Even Transit Check for further vetting. These workflows are described below. We note that due to an error in uploading, 170 candidates were uploaded twice to the Exoplanet Transit Search. We merged the classifications for each of these candidates and removed duplicate classifications by the same users when applying the user weighting scheme (see Section~\ref{sec:id_cands}).

\subsection{Secondary Eclipse Check}\label{sec:secondary_eclipse}
The Secondary Eclipse Check (Figure~\ref{fig:combined_interface}) is designed to help check whether the candidate is an eclipsing binary system where we can observe both the primary transit and secondary eclipse. Users are presented with phase-folded light curves centered on orbital phase=0.5 with the x-axis spanning the same width as the Exoplanet Transit Search. We expect that given the short periods of these candidates the orbits are likely to be near-circularized \citep[][and references therein]{Winn2015OccurrenceArchitecture} and therefore any secondary eclipses will be at phase=0.5. Volunteers are asked to choose a single response from: `A secondary eclipse,' `No secondary eclipse' and `A large data gap.' This allows us to reject candidates where a clear secondary eclipse is observed. The NGTS telescopes are sensitive to transit depths of $\mysim0.1$\,per\,cent \citep{Wheatley2018NGTS,West2019NGTS4b_SubNeptune}, which is approximately the maximum depth that has been observed from the ground for optical secondary eclipses of hot Jupiters \citep{Wheatley2018NGTS}. Therefore, we do not expect the secondary eclipses of real exoplanets to be visible in the presented light curves. The middle row of Figure~\ref{fig:example_lcs} shows an example of a candidate (TIC-389932515) that was identified as displaying a U-shaped transit in the Exoplanet Transit Search but the Secondary Eclipse Check reveals that this system is an eclipsing binary with a clear secondary eclipse. Each subject in this workflow is seen by 15 volunteers before being retired. The number of classifications per subject before retirement is designed to balance speed of classifying the sample with obtaining enough classifications for accurate characterisation. We require less classifications for the Secondary Eclipse Check than the Exoplanet Transit Search as the task presented is more straightforward and, through the selection of the most promising candidates from the Exoplanet Transit Search (see Section~\ref{sec:id_cands}), the rate of false positives will be lower.
% \begin{figure}
% \gridline{\fig{images/ets_sec_eclipse_69499344.png}{0.7\textwidth}{(a) Exoplanet Transit Search light curve}}
% \gridline{\fig{images/sec_sec_eclipse_69757368.png}{0.7\textwidth}{(b) Secondary Eclipse Check light curve}}
% \caption{Light curves for the Exoplanet Transit Search and Secondary Eclipse Check for an eclipsing binary system that shows a regular U-shaped transit in the Exoplanet Transit Search but the Secondary Eclipse Check reveals a clear transit event at 0.5\,phase.}\label{fig:sec_eclipse}
% \end{figure}

\subsection{Odd/Even Transit Check}\label{sec:odd_even}
\begin{figure*}
    \centering
    \includegraphics[width=\textwidth]{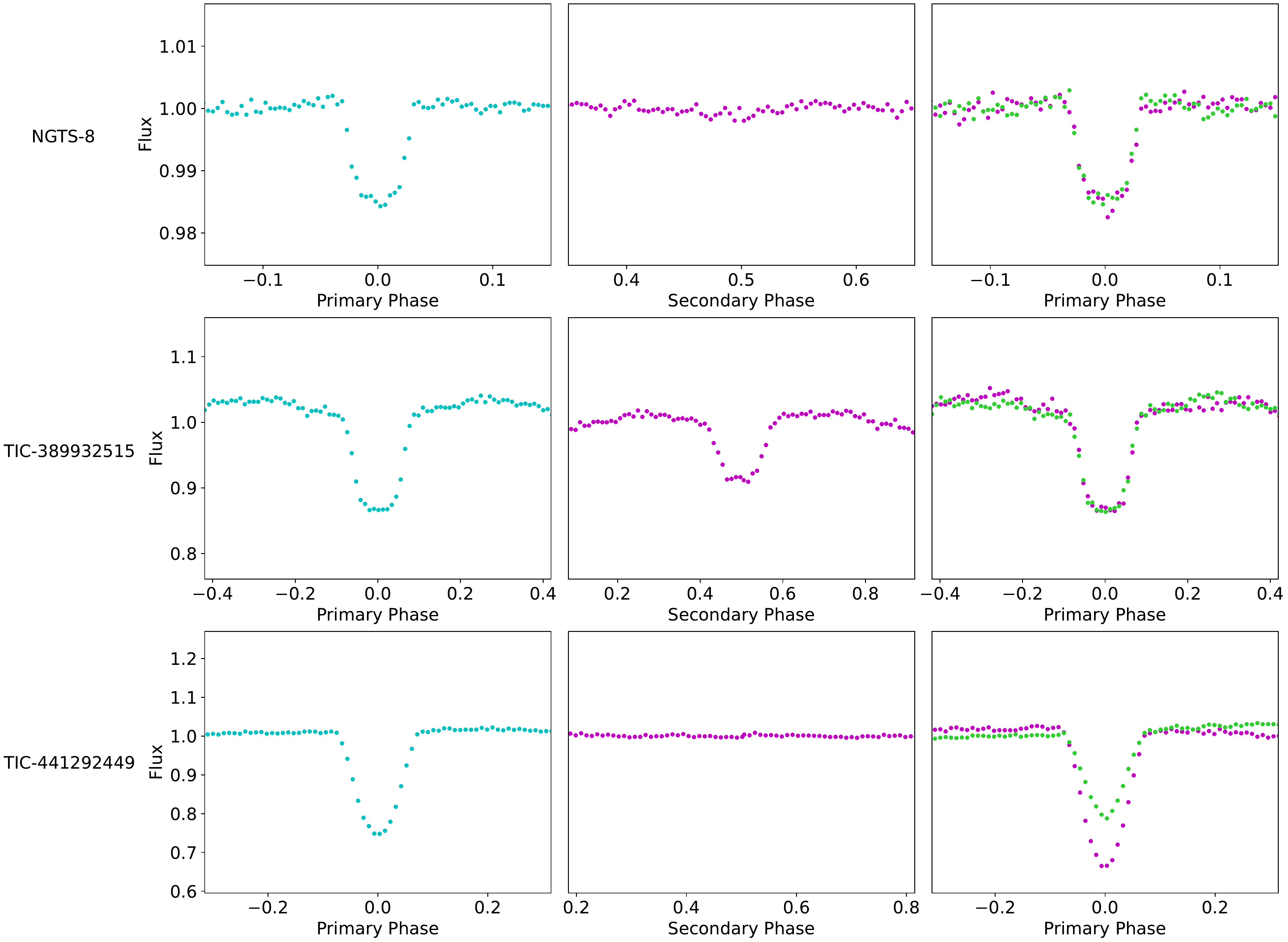}
    \caption{Light curves for three different stars as presented in each of the three workflows. \textbf{Top row:} Light curves for NGTS-8 which shows a clear U-shaped transit in the Exoplanet Transit Search. No secondary eclipse is visible in the Secondary Eclipse Check and the depths of the odd and even transits match in the Odd/Even Transit Check. \textbf{Middle row:} Light curves for TIC-389932515 which shows a U/V-shaped transit in the Exoplanet Transit Search. The depths of the odd and even transits match in the Odd/Even Transit Check, however a secondary eclipse is visible in the Secondary Eclipse Check indicating this is an eclipsing binary.  \textbf{Bottom row:}  Light curves for TIC-441292449 which shows a clear V-shaped transit in the Exoplanet Transit Search. No secondary eclipse is visible in the Secondary Eclipse Check however the depths of the odd and even transits do not match in the Odd/Even Transit Check indicating this is an eclipsing binary.}
    \label{fig:example_lcs}
\end{figure*}

The Odd/Even Transit Check (Figure~\ref{fig:combined_interface}) is designed to check whether the depths of the odd and even transits match. We can use this workflow to identify eclipsing binary systems where \orion\ has misidentified the orbital period as half the true period. The primary and secondary eclipse, which will correspond to the odd and even transits (or vice versa), will have different depths if the stars have differing luminosities. Volunteers are presented with the phase-folded light curve (centered on phase=0) with the odd-numbered transits (i.e. the 1st, 3rd, 5th, etc. consecutive transits) shown in green and the even transits shown in magenta. These colors were chosen to make the task accessible to volunteers with different types of color vision deficiency. The first task asks whether both sets of points cover the middle portion of the plot to check that there are no large gaps in the data. If large data gaps exist then it will not be possible to accurately determine whether the odd and even transit depths match. The second task (shown in Figure~\ref{fig:combined_interface}) asks whether the odd and even transits have similar depths. The bottom row of Figure~\ref{fig:example_lcs} shows an example of a candidate (TIC 441292449) where the Exoplanet Transit Search light curve appears as a regular V-shaped transit, while the Odd/Even Transit Check light curve shows the clear difference in depths.
The scenarios checked by both the Secondary Eclipse Check and Odd/Even Transit Check are common false positives in exoplanet transit searches \citep{ODonovan2006TrES_BEBReject,Howell2011SpeckleKepler,LilloBox2014ComprehensiveFPRejection,Ciardi2015StellarMultiplicity,Lester2021SpeckleTESS}. Each subject in this workflow is seen by 15 volunteers before being retired. This threshold for the number of classifications per subject is chosen for the same reasons described above in Section~\ref{sec:secondary_eclipse}.
%Clean example: NGTS-8b
%Secondary eclipse: TIC389932515
%Depth mistmatch: TIC441292449

\subsection{Site Statistics}\label{sec:site_stats}
\begin{figure}
    \centering
    \includegraphics[width=\columnwidth]{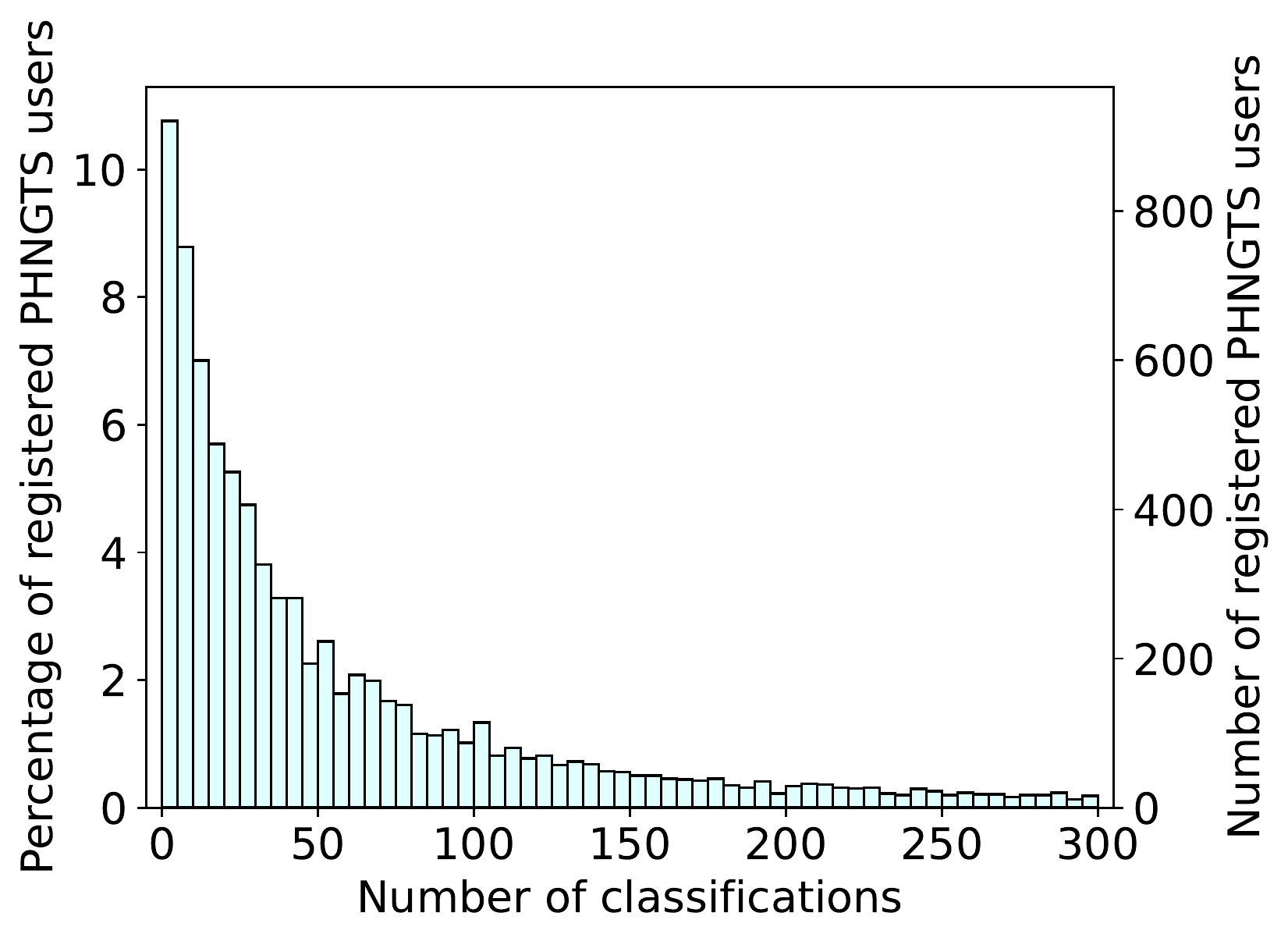}
    \caption{Histogram of the number of classifications by registered users, using a bin size of 5. The plotted distribution is truncated at 300 classifications for clarity. 1,023 registered users had more than 300 classifications which is 12.0\% of all registered volunteers. The most frequent number of classifications by registered volunteers and non-logged-in sessions is 1.}
    \label{fig:class_hist}
\end{figure}

Since the launch of \phngts\ on 2021 October 18, there have been 2,626,380 individual classifications completed across the three workflows on the NGTS Public Data. 87.6\% of the classifications were made by 8,559 registered volunteers, with the rest made in 3,319 non-logged-in sessions (tracked by anonymized IP address). Across the three workflows 138,198 subjects were classified, which comprised 85,000 individual target stars. Figure~\ref{fig:class_hist} shows the distribution of the number of classifications made by registered volunteers. This distribution is common in citizen science projects in that we find that many volunteers classify only a few subjects while a small number of users classify a large number of subjects \citep[][and references therein]{Spiers2019GiniZoo}. Registered volunteers classified a mean of 268 subjects and a median of 40 while non-logged-in sessions classified a mean of 98 subjects and a median of 31. For comparison, registered Planet Hunters volunteers classified a mean of 68 and median of 5 \kepler\ Q1 light curves \citep{Schwamb2012PHOverview} and registered Planet Hunters TESS volunteers submitted a mean of 647 and median of 33 classifications \citep{Eisner2021PHT2Overview}. We assessed the distribution of the number of classifications made by registered users using the Gini coefficient, which ranges from 0 (equal contributions from all users) to 1 (large disparity in the contributions). The Gini coefficient for the \phngts\ project is 0.85. This is similar to the mean Gini coefficient of 0.82 among other astronomy Zoonvierse projects \citep{Spiers2019GiniZoo} and the mean value for individual sectors of 0.87 in the Planet Hunters TESS project \citep{Eisner2021PHT2Overview}. To compare the time taken by the \phngts\ project to classify the data with the original NGTS vetting process, we measure time taken for 99\% of subjects in the Exoplanet Transit Search to receive their 20th classification (i.e. the time for 99\% of the dataset to be retired). The Secondary Eclipse Check and Odd/Even Transit Check workflows had data uploaded at irregular intervals therefore we do not include measurements of how long these stages took. The Exoplanet Transit Search dataset was uploaded and classified in two distinct sets in order to initially prioritise \orion\ candidates with higher SDE. The first 85,316 subjects uploaded had SDE$>=10$ and 99\% of subjects were retired by 2021 December 12, that is 56 days after the launch of the project. The second set of 20,218 subjects had $8<=$ SDE $<10$ with the first classification submitted on 2022 April 25 and 99\% of subjects being retired by 2022 September 18 (147 days later). The initial dataset greatly benefited from media coverage driving engagement in the first weeks of the project, hence the much higher rate of classification. For comparison, the original NGTS vetting process of the dataset took 1.5 years. We stress however that the aim of the \phngts\ project is not to classify the data faster but rather uncover any candidates that may have been missed in the traditional vetting process.

%% file: sections/4_idcands.tex
In this section, we describe the user weighting and subject scoring scheme used to combine multiple user classifications and select the best candidates to be advanced from the Exoplanet Transit Search to the additional workflows and then to the visual vetting stage.

\subsection{Weighting Scheme}\label{sec:weighting}
We use a user weighting and subject scoring scheme based on the scheme implemented by \citet{Schwamb2018Planet4Terrains} for the Planet Four: Terrains citizen science project, which in turn is based on the schemes developed by \citet{Lintott2008GalaxyZoo,Lintott2011GalaxyZoo1} for the Galaxy Zoo project and \citet{Schwamb2012PHOverview} for the original Planet Hunters project. This iterative weighting scheme assumes users who agree with the majority vote to be better at the tasks and up-weights them accordingly while those who disagree are down-weighted. This is not a perfect assumption, however citizen science in general relies on the assumption that the majority of users are correct. By implementing a user weighting scheme, we can determine which users are better at identifying U-shaped dips, for example, and pay more attention to their responses compared with others when attempting to identify which subjects have these features present. In this scheme, the classifications of a user are linked via their username if they are logged into Zooniverse, or by their anonymized IP address if they are not logged in. If a user excels at spotting U-shaped dips in the Exoplanet Transit Search, it does not necessarily mean that they are adept at spotting other features in this workflow or the additional workflows. Therefore, we assign and assess the following separate user weights. For the Exoplanet Transit Search, user weights are assigned for each of the five possible responses in the workflow. For the Secondary Eclipse Check, a single weight per user is calculated as users can select only one of the three possible responses in this workflow. For the Odd/Even Transit Check, two user weights are calculated for each user, one weight for the first task, where users check whether there is sufficient data coverage for both the odd and even transits, and one weight for the second task, where users check whether the depths of the odd and even transits match. Both these tasks are binary Yes/No questions and therefore only one weight per task is required. These various user weights are applied when combining the multiple volunteer assessments for each subject to calculate subject scores, which can then be used to apply various threshold cuts to select the best candidates for further vetting. By assigning weights independently, we can calculate subject scores for each response on each subject across all three workflows to identify the best ranked candidates for each of the possible responses.

Each user starts with weights equal to 1 and these weights are adjusted based on how often the user agrees with the majority assessment of all volunteers for the subjects the given user classified. User weights typically lie in a range from 0 to 1.6, with a scaling factor applied such that the mean user weight is 1. Note that user weights cannot be equal to 0. This is not to be confused with the subject scores that can be in the range $[0,1]$.
Users with only one classification for a given task retain a weight of 1 as we do not have enough information to evaluate the ability of the user to identify features in the light curves. The equations for calculating the subject scores and user weights in each of the workflows differ due to the differences in how many responses can be selected in each workflow and the number of tasks that each workflow consists of. We describe the weighting schemes used in each of the workflows below.

\subsubsection{Exoplanet Transit Search}\label{sec:weighting_ets}
We describe the process for calculating scores and weights for the `U-shaped' response in the Exoplanet Transit Search as an example. We use the `U-shaped' response as the primary criteria for spotting exoplanet-like transit features. The other possible responses in the Exoplanet Transit Search are `V-shaped,' `Data gap,' `No dip' and `Stellar variability' for which the process of calculating scores and weights is identical to the method described below. For each subject $i$ in the Exoplanet Transit Search, a subject score, $s_i(U)$, is calculated as,
\begin{align*}
     s_i(U) &= \frac{1}{U_i} \sum_k w_k(U) \numberthis \label{eq:subj_scores}\\*
     k &= \parbox[t]{0.6\columnwidth}{users who selected `U-shaped' for subject \textit{i}}
\end{align*}
% \begin{gather}
%      \label{eq:subj_scores}
%      \begin{aligned}
%           s_i(U) &= \frac{1}{U_i} \sum_k w_k(U) \\
%           k &= \parbox[t]{0.6\columnwidth}{users who selected `U-shaped' for subject \textit{i}}
%      \end{aligned}
% \end{gather}
where $w_k(U)$ is the weight for user $k$ for the `U-shaped' response and $U_i$ is the sum of the weights of all users who classified subject $i$:
\begin{align*}
     U_i &= \sum_{k=j} w_k(U) \numberthis \label{eq:weightsum}\\*
     j &= \parbox[t]{0.6\columnwidth}{all users who classified subject \textit{i}}
\end{align*}
Initially, as all user weights are equal to 1, these subject scores will effectively be a simple counting of votes, however as user weights change through the iterative scheme, these subject scores will vary from these initial values.
The subject scores can be in the range $[0,1]$, where a score of 1 means all users who classified the subject agreed that the given feature (in this case a U-shaped dip) was present, while a score of 0 indicates that all users agreed that the given feature is not visible in the subject presented.
Once the subject scores have been calculated, the next step is to assign new user weights as follows,
\begin{align*}
w_j(U) &=
\begin{cases}
\frac{A}{N_j} \big(\sum\limits_{i=p} s_i(U) + \sum\limits_{i=q} [1-s_i(U)]\big) &\text{ if } N_j>1\\*
1 &\text{ if } N_j = 1
\end{cases} \numberthis \label{eq:user_weights} \\*
p &= \parbox[t]{0.6\columnwidth}{subjects user \textit{j} classified as `U-shaped'} \\*
q &= \parbox[t]{0.6\columnwidth}{subjects user \textit{j} did not classify as `U-shaped'}
\end{align*}
where $N_j$ is the number of subjects classified in this workflow by user $j$. The scaling factor $A$ is chosen such that the mean user weight of users who classified more than one subject is 1. This equation up-weights a user when they agree with the majority of users who classified a given subject and down-weights them when their response differs from the majority assessment of the users who classified the subject. Using the new user weights calculated from Equation~\ref{eq:user_weights}, the subject scores are recalculated with Equations~\ref{eq:subj_scores} and \ref{eq:weightsum}. The user weights are then recalculated with Equation~\ref{eq:user_weights}. This process is iterated until convergence is achieved, which we define as when the median absolute difference between the old and new weights is less than or equal to $10^{-4}$ for all of the possible responses in the workflow. Once convergence is achieved and the final user weights have been calculated, these weights are applied to calculate the final subject scores that we use to implement the threshold cuts described in Section~\ref{sec:cand_selection}.

\subsubsection{Secondary Eclipse Check}\label{sec:weighting_sec}
The equations for calculating the subject scores and user weights for the Secondary Eclipse Check (Section~\ref{sec:secondary_eclipse}) differ from above as it is only possible to select one of the three possible responses (`A secondary eclipse,' `No secondary eclipse' and `A large data gap'). For this workflow, each user $j$ is assigned a single weight, $w_j(SE),$ that is applied to the calculation of the subject scores. For each subject $i$ in the Secondary Eclipse Check, a set of three subject scores, $s_i(\mathit{YS_{sec}})$, $s_i(\mathit{NS_{sec}})$ and $s_i(\mathit{DG_{sec}})$, are calculated for the three responses, `A secondary eclipse,' `No secondary eclipse' and `A large data gap,' respectively. These scores are calculated as,
\begin{align*}
    s_i(\mathit{YS_{sec}}) &= \frac{1}{\mathit{SE_{i}}} \sum_k w_k(\mathit{SE}) \numberthis \label{eq:sec_score_ys}\\*
     k &= \parbox[t]{0.6\columnwidth}{users who selected `A secondary eclipse' for subject \textit{i}}
\end{align*}
\begin{align*}
    s_i(\mathit{NS_{sec}}) &= \frac{1}{\mathit{SE_{i}}} \sum_k w_k(\mathit{SE}) \numberthis \label{eq:sec_score_ns}\\*
     k &= \parbox[t]{0.6\columnwidth}{users who selected `No secondary eclipse' for subject \textit{i}}
\end{align*}
\begin{align*}
    s_i(\mathit{DG_{sec}}) &= \frac{1}{\mathit{SE_{i}}} \sum_k w_k(\mathit{SE}) \numberthis \label{eq:sec_score_dg}\\*
     k &= \parbox[t]{0.6\columnwidth}{users who selected `A large data gap' for subject \textit{i}}
\end{align*}
where $w_k(\mathit{SE})$ is the weight for user $k$ for this workflow and $\mathit{SE_{i}}$ is the sum of the weights for all users who classified subject $i$:
\begin{align*}
     SE_i &= \sum_{k=j} w_k(SE) \numberthis \label{eq:sec_weightsum}\\*
     j &= \parbox[t]{0.6\columnwidth}{all users who classified subject \textit{i}}
\end{align*}
These subject scores can be in the range $[0,1]$, with a score of 1 indicating unanimous agreement that the given feature is present and a score of 0 meaning that all users agreed that the given feature is not visible in the subject presented.
As in the Exoplanet Transit Search, all user weights are initially set to 1 and the weights of users who classified only one subject remain as 1. Once the subject scores have been calculated, we assign new user weights as follows,
\begin{align*}
w_j(SE) &=
\begin{cases}
\frac{B}{N_j} \big(\sum\limits_{i=p} s_i(\mathit{YS_{sec}}) + \sum\limits_{i=q} s_i(\mathit{NS_{sec}}) + \sum\limits_{i=r} s_i(\mathit{DG_{sec}})\big) &\text{ if } N_j>1\\*
1 &\text{ if } N_j = 1
\end{cases} \numberthis \label{eq:sec_user_weights} \\*
p &= \parbox[t]{0.6\columnwidth}{subjects user \textit{j} classified as `A secondary eclipse'} \\*
q &= \parbox[t]{0.6\columnwidth}{subjects user \textit{j} classified as `No secondary eclipse'} \\*
r &= \parbox[t]{0.6\columnwidth}{subjects user \textit{j} classified as `A large data gap'}
\end{align*}
where $N_j$ is the number of subjects classified in this workflow by user $j$. The scaling factor $B$ is chosen such that the mean user weight of users who classified more than one subject is 1. The principles of this scheme are the same as those used for the Exoplanet Transit Search. Users are up-weighted when they agree with the majority of users who classified a given subject and down-weighted otherwise. A higher subject score for a given response indicates greater consensus among the users who classified the given subject that the given feature is present. The scheme proceeds iteratively calculating subject scores using Equations~\ref{eq:sec_score_ys},~\ref{eq:sec_score_ns},~\ref{eq:sec_score_dg}~and~\ref{eq:sec_weightsum} and weights using Equation~\ref{eq:sec_user_weights} until convergence is achieved and a final set of subject scores are calculated, as described above in Section~\ref{sec:weighting_ets}.

\subsubsection{Odd/Even Transit Check}\label{sec:weighting_odd}
In the Odd/Even Transit Check (Section~\ref{sec:odd_even}), users are presented with a subject and first asked to classify whether the data for both the odd and even transits cover the middle portion of the plot. If the user selects `No' to this first task, then they move on to reviewing the next subject, however, if they select `Yes,' the user is then asked whether the depths of the odd and even transits match. The options for responses in both tasks are simply `Yes' or `No'. Since both tasks are separate Yes/No questions, each user $j$ who classifies subjects in this workflow is assigned a weight for the first task, $w_j(OC),$ and a weight for the second task, $w_j(OD).$
The subject scores for the first task, $s_i(\mathit{YC_{odd}})$ for `Yes, the data for both the odd and even transits cover the middle portion of the plot' and $s_i(\mathit{NC_{odd}})$ for `No, the data for the odd and/or even transits do not cover the middle portion of the plot,' are calculated as follows,
\begin{align*}
    s_i(\mathit{YC_{odd}}) &= \frac{1}{\mathit{OC_{i}}} \sum_k w_k(\mathit{OC}) \numberthis \label{eq:odd_score_yc}\\*
     k &= \parbox[t]{0.6\columnwidth}{subjects user \textit{j} classified as `Yes, the data for both the odd and even transits cover the middle portion of the plot'}
\end{align*}
\begin{align*}
    s_i(\mathit{NC_{odd}}) &= \frac{1}{\mathit{OC_{i}}} \sum_k w_k(\mathit{OC}) \numberthis \label{eq:odd_score_nc}\\*
     k &= \parbox[t]{0.6\columnwidth}{subjects user \textit{j} classified as `No, the data for the odd and/or even transits do not cover the middle portion of the plot'}
\end{align*}
where $w_k(\mathit{OC})$ is the weight for user $k$ and $\mathit{OC_{i}}$ is the sum of the weights for all users who classified subject $i$:
\begin{align*}
     OC_i &= \sum_{k=j} w_k(OC) \numberthis \label{eq:oddcov_weightsum}\\*
     j &= \parbox[t]{0.6\columnwidth}{all users who classified subject \textit{i}}
\end{align*}
These subject scores can be in the range $[0,1]$, with higher scores indicating greater agreement amongst the users who classified the subject that the given feature is present. As in the other workflows, all user weights are initially set to 1 and the weights of users who classified only one subject remain as 1. The weights for this task are calculated as follows using the subject scores determined above,
\begin{align*}
w_j(OC) &=
\begin{cases}
\frac{C}{N_j} \big(\sum\limits_{i=p} s_i(\mathit{YC_{odd}}) + \sum\limits_{i=q} s_i(\mathit{NC_{odd}})\big) &\text{ if } N_j>1\\*
1 &\text{ if } N_j = 1
\end{cases} \numberthis \label{eq:odd_cov_user_weights} \\*
p &= \parbox[t]{0.6\columnwidth}{subjects user \textit{j} classified as `Yes, the data for both the odd and even transits cover the middle portion of the plot'}\\*
q &= \parbox[t]{0.6\columnwidth}{subjects user \textit{j} classified as `No, the data for the odd and/or even transits do not cover the middle portion of the plot'}
\end{align*}
where $N_j$ is the number of subjects in this task classified by user $j$. The scaling factor $C$ is chosen such that the mean user weight of users who classified more than one subject is 1. We follow the iterative scheme of recalculating the subject scores and user weights until convergence is achieved as described above in Section~\ref{sec:weighting_ets}.

Once the subject scores and user weights for the first task of the Odd/Even Transit Check have been finalised, we apply a threshold cut of $s_i(\mathit{YC_{odd}}) \geq 0.5$. That is, the subject scores for the second task, $s_i(\mathit{YD_{odd}})$ for `Yes, the odd/even transits have similar depths' and $s_i(\mathit{ND_{odd}})$ for `No, the odd/even transits do not have similar depths,' are only calculated for subjects that pass the criteria of $s_i(\mathit{YC_{odd}}) \geq 0.5$. This ensures that only subjects that have sufficient data coverage are evaluated for depth differences. This in turn means that users are neither up-weighted or down-weighted when classifying subjects in the second task where only a few volunteers were able to provide a response. The equations for calculating the subject scores for a given subject $i$ are as follows,
\begin{align*}
    s_i(\mathit{YD_{odd}}) &= \frac{1}{\mathit{OD_{i}}} \sum_k w_k(\mathit{OD}) \numberthis \label{eq:odd_score_yd}\\*
     k &= \parbox[t]{0.6\columnwidth}{subjects user \textit{j} classified as `Yes, the odd/even transits have similar depths'}
\end{align*}
\begin{align*}
    s_i(\mathit{ND_{odd}}) &= \frac{1}{\mathit{OD_{i}}} \sum_k w_k(\mathit{OD}) \numberthis \label{eq:odd_score_nd}\\*
     k &= \parbox[t]{0.6\columnwidth}{subjects user \textit{j} classified as `No, the odd/even transits do not have similar depths'}
\end{align*}
where $w_k(\mathit{OD})$ is the weight for user $k$ and $\mathit{OD_{i}}$ is the sum of the weights for all users who classified subject $i$:
\begin{align*}
     OD_i &= \sum_{k=j} w_k(OD) \numberthis \label{eq:odddep_weightsum}\\*
     j &= \parbox[t]{0.6\columnwidth}{all users who classified subject \textit{i}}
\end{align*}
These subject scores can be in the range $[0,1]$, with higher scores indicating greater agreement amongst the users who classified the subject that the given feature is present. All user weights are initially set to 1 and the weights of users who classified only one subject remain as 1. The weights for this task are calculated as follows using the subject scores determined above,
\begin{align*}
w_j(OD) &=
\begin{cases}
\frac{D}{N_j} \big(\sum\limits_{i=p} s_i(\mathit{YD_{odd}}) + \sum\limits_{i=q} s_i(\mathit{ND_{odd}})\big) &\text{ if } N_j>1\\*
1 &\text{ if } N_j = 1
\end{cases} \numberthis \label{eq:odd_dep_user_weights} \\*
p &= \parbox[t]{0.6\columnwidth}{subjects user \textit{j} classified as `Yes, the odd/even transits have similar depths'}\\*
q &= \parbox[t]{0.6\columnwidth}{subjects user \textit{j} classified as `No, the odd/even transits do not have similar depths'}
\end{align*}
where $N_j$ is the number of subjects in this task classified by user $j$. The scaling factor $D$ is chosen such that the mean user weight of users who classified more than one subject is 1. We again follow the iterative scheme of recalculating the subject scores and user weights until convergence is achieved as described above in Section~\ref{sec:weighting_ets}.

\subsubsection{Distribution of Scores and Weights}\label{sec:score_weight_dists}
\begin{figure}
    \centering
    \includegraphics[width=\columnwidth]{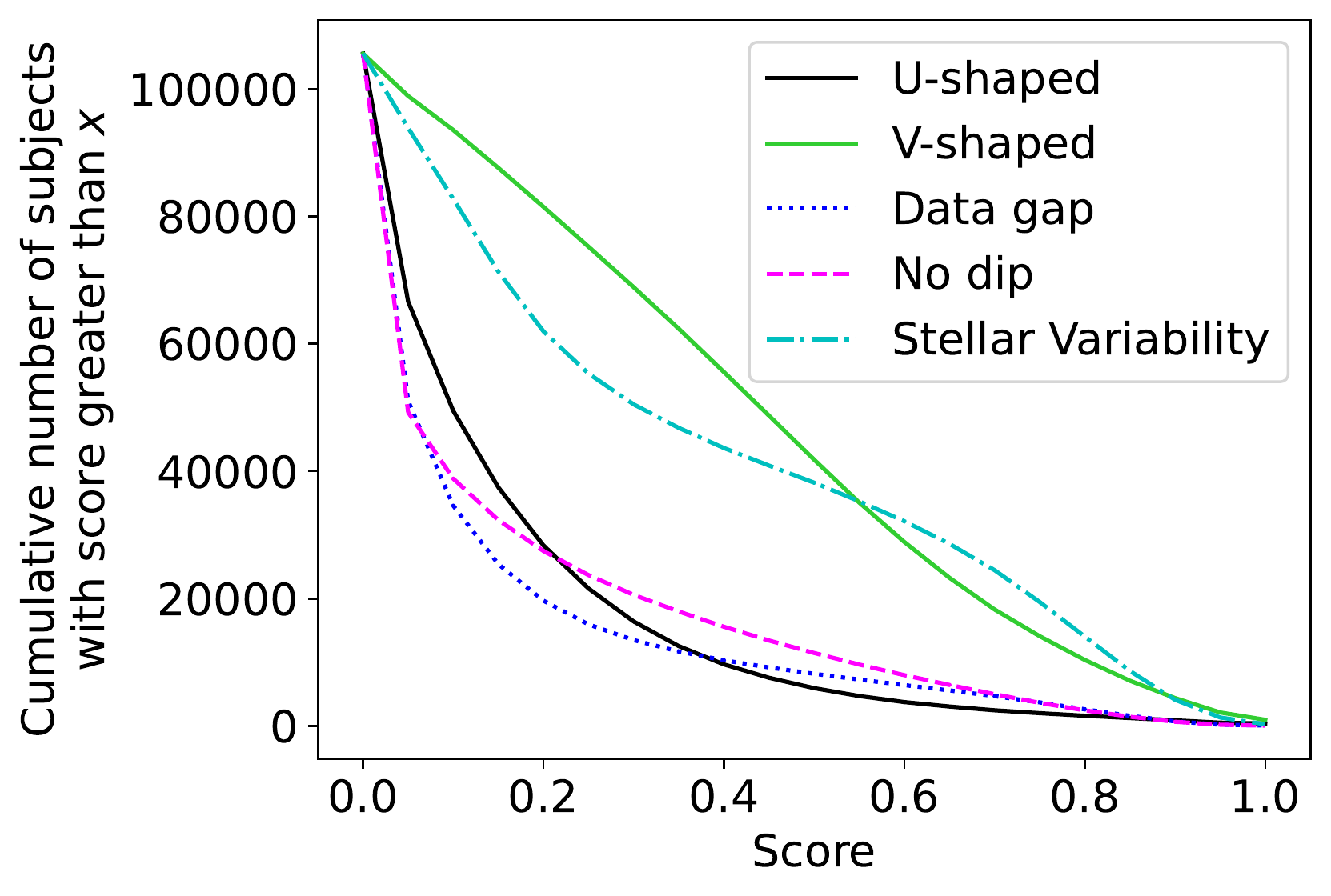}
    \caption{Cumulative distribution of subject scores greater than the given score on the x-axis. The scores for each response of the Exoplanet Transit Search  ($s_i(U)$, $s_i(V)$, $s_i(\mathit{DG})$, $s_i(\mathit{ND})$, $s_i(\mathit{SV})$) are shown.}
    \label{fig:prim_score_dist}
\end{figure}

\begin{figure*}
    \centering
    \includegraphics[width=\textwidth,height=0.9\textheight,keepaspectratio]{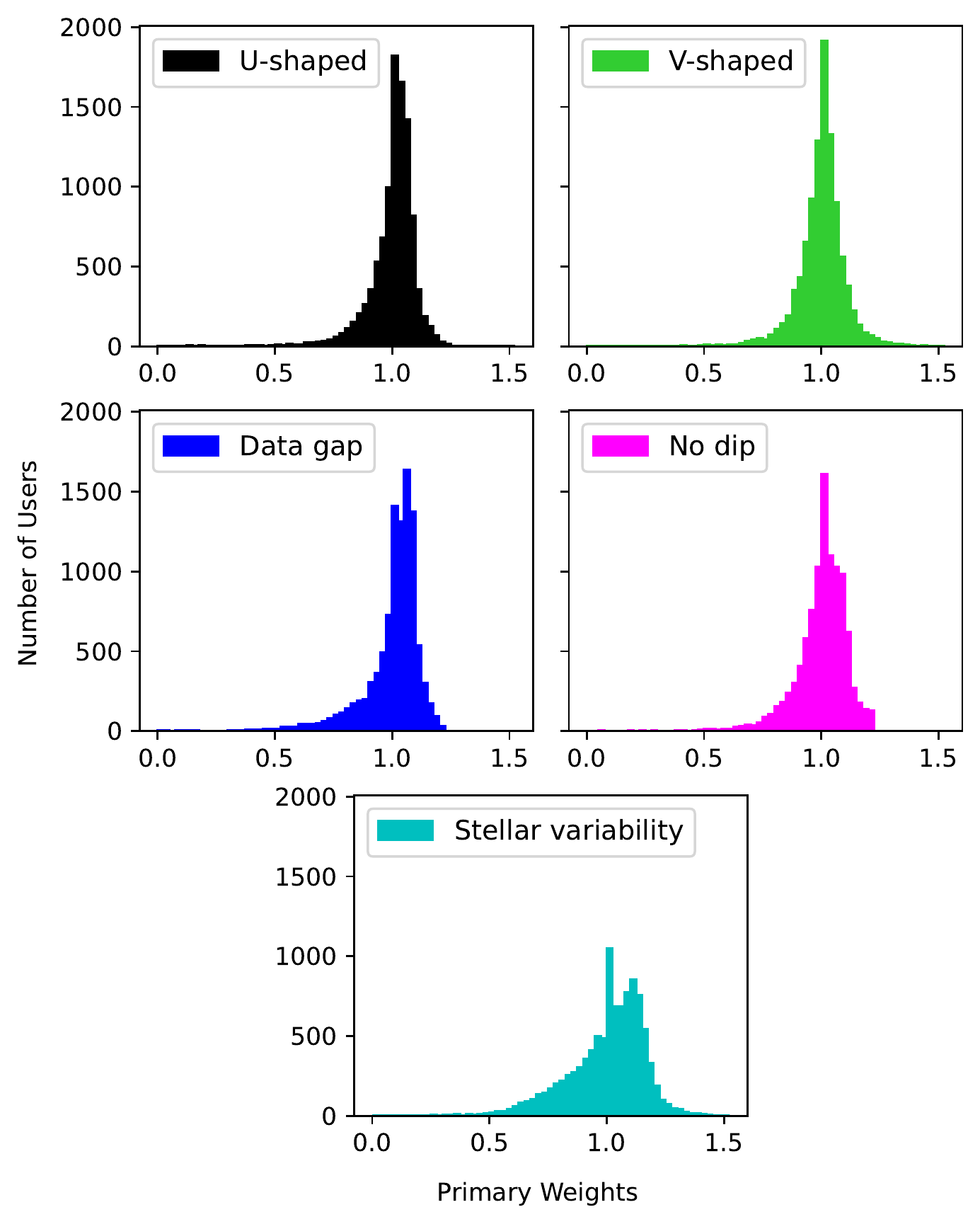}
    \caption{Histograms of user weights for each response in the Exoplanet Transit Search.}
    \label{fig:prim_weights}
\end{figure*}

Table~\ref{tab:subject_scores} provides the final scores for each candidate, with the scores for subjects in the Secondary Eclipse Check and Odd/Even Transit Check provided where available.
Figure~\ref{fig:prim_score_dist} shows the cumulative distributions of the subject scores for each response in the Exoplanet Transit Search. V-shaped dips are much more common in the sample compared with U-shaped dips, reflecting the prevalence of eclipsing binary systems that are typically found by exoplanet transit searches \citep{Brown2003FalseAlarmRates,Alonso2004STARE_FPruleout}.
The histograms for the user weights for each response in the Exoplanet Transit Search are shown in Figure~\ref{fig:prim_weights}. The apparent spike at $w_j(\mathit{SV}) = 1$ is due to the larger spread in weights for this response, resulting in the fixed weights of users who only classified one subject to be more obvious compared with the distribution of weights for the other 4 responses. For $w_j(U),$ 96\% of users have weights greater than 0.8 and 59\% have weights greater than 1. Table~\ref{tab:weights} shows the percentage of users with weights above 0.8 and 1 for all weights in the \phngts\ project. The cumulative distributions of scores and histograms of user weights for the Secondary Eclipse Check and Odd/Even Transit Check are provided in Appendix~\ref{sec:scoresweights}.

\begin{longrotatetable}
\begin{deluxetable}{c|cccccc|cccc|ccccc}
\tabletypesize{\tiny}
\tablecaption{\phngts\ subject information and scores for each workflow where available. \label{tab:subject_scores}}
\tablehead{\colhead{} & \multicolumn{6}{c}{Exoplanet Transit Search} & \multicolumn{4}{c}{Secondary Eclipse Check} & \multicolumn{5}{c}{Odd/Even Transit Check}}
\startdata
TIC ID & Subject ID & $s_i(U)$ & $s_i(V)$ & $s_i(\mathit{ND})$ & $s_i(\mathit{DG})$ & $s_i(\mathit{SV})$ & Subject ID & $s_i(\mathit{YS_{sec}})$ & $s_i(\mathit{NS_{sec}})$ & $s_i(\mathit{DG_{sec}})$ & Subject ID & $s_i(\mathit{YC_{odd}})$ & $s_i(\mathit{NC_{odd}})$ & $s_i(\mathit{YD_{odd}})$ & $s_i(\mathit{ND_{odd}})$ \\ \hline
57908727 & 74820792 & 0.183673 & 0.130785 & 0.647565 & 0.082170 & 0.058148 &          - &        - &            - &   - &          - &        - &        - &        - &        - \\
14092921 & 69494436 & 0.141035 & 0.387247 & 0.090076 & 0.680189 & 0.134000 &          - &        - &            - &   - &          - &        - &        - &        - &        - \\
333018311 & 69674773 & 0.039031 & 0.901027 & 0.000000 & 0.000000 & 0.049919 & 73015380 & 0.965594 & 0.034406 & 0.0 & 73029773 & 0.955156 & 0.044844 & 0.932479 & 0.067521 \\
165374842 & 69660130 & 0.650532 & 0.094216 & 0.000000 & 0.038786 & 0.418303 & 71368514 & 0.955075 & 0.044925 & 0.0 & 71374760 & 0.978074 & 0.021926 & 1.000000 & 0.000000 \\
60768993 & 69541419 & 0.000000 & 0.649549 & 0.000000 & 0.235270 & 0.659147 &          - &        - &            - &   - &          - &        - &        - &        - &        - \\
31141153 & 74812193 & 0.000000 & 0.637888 & 0.088087 & 0.033987 & 0.445865 &          - &        - &            - &   - &          - &        - &        - &        - &        - \\
135310361 & 69662716 & 0.228356 & 0.641675 & 0.039714 & 0.000000 & 0.189841 &          - &        - &            - &   - &          - &        - &        - &        - &        - \\
242027017 & 74817247 & 0.397940 & 0.092060 & 0.000000 & 0.601868 & 0.230841 &          - &        - &            - &   - &          - &        - &        - &        - &        - \\
441311367 & 69641162 & 0.189460 & 0.192876 & 0.000000 & 0.033981 & 1.000000 &          - &        - &            - &   - &          - &        - &        - &        - &        - \\
404020484 & 69674877 & 0.000000 & 0.845810 & 0.048033 & 0.263466 & 0.051691 & 91086040 & 1.000000 & 0.000000 & 0.0 & 91086092 & 0.946373 & 0.053627 & 0.930785 & 0.069215 \\
\vdots & \vdots & \vdots & \vdots & \vdots & \vdots & \vdots & \vdots & \vdots & \vdots & \vdots & \vdots & \vdots & \vdots & \vdots & \vdots \\
\enddata
\tablecomments{This table in its entirety can be found in the online supplemental material. We note that prior to the implementation of the user weighting and subject scoring scheme, subjects with $\geq 50\%$ of votes for `U-shaped' and $< 60\%$ of votes for `Stellar variability' were pushed to the Secondary Eclipse Check and Odd/Even Transit Check. The subject scores calculated across all workflows for the 745 subjects that met this vote criteria but did not meet the score thresholds are included in the online supplementary material.}
\end{deluxetable}
\end{longrotatetable}

\begin{deluxetable*}{cccc}
\tabletypesize{\normalsize}
\tablecaption{Percentage of users with weights above 0.8 and 1. \label{tab:weights}}
\tablehead{\colhead{Workflow} & \colhead{Weight} & \colhead{$w_j(X)>0.8$} & \colhead{$w_j(X)>1$}}
\startdata
& U-shaped & 96\% & 59\% \\
& V-shaped & 97\% & 52\% \\
Exoplanet Transit Search & Data gap & 93\% & 63\% \\
& No dip & 95\% & 55\% \\
& Stellar Variability & 88\% & 56\% \\
\hline
Secondary Eclipse Check & Secondary Eclipse & 80\% & 50\% \\
\hline
\multirow{2}{*}{Odd/Even Transit Check} & Odd/Even Data Gap & 81\% & 50\% \\
& Odd/Even Depths & 90\% & 46\% \\
\enddata
\end{deluxetable*}

\subsection{Candidate Selection}\label{sec:cand_selection}
\begin{figure*}
    \centering
    \includegraphics[width=\textwidth, height=0.93\textheight,keepaspectratio]{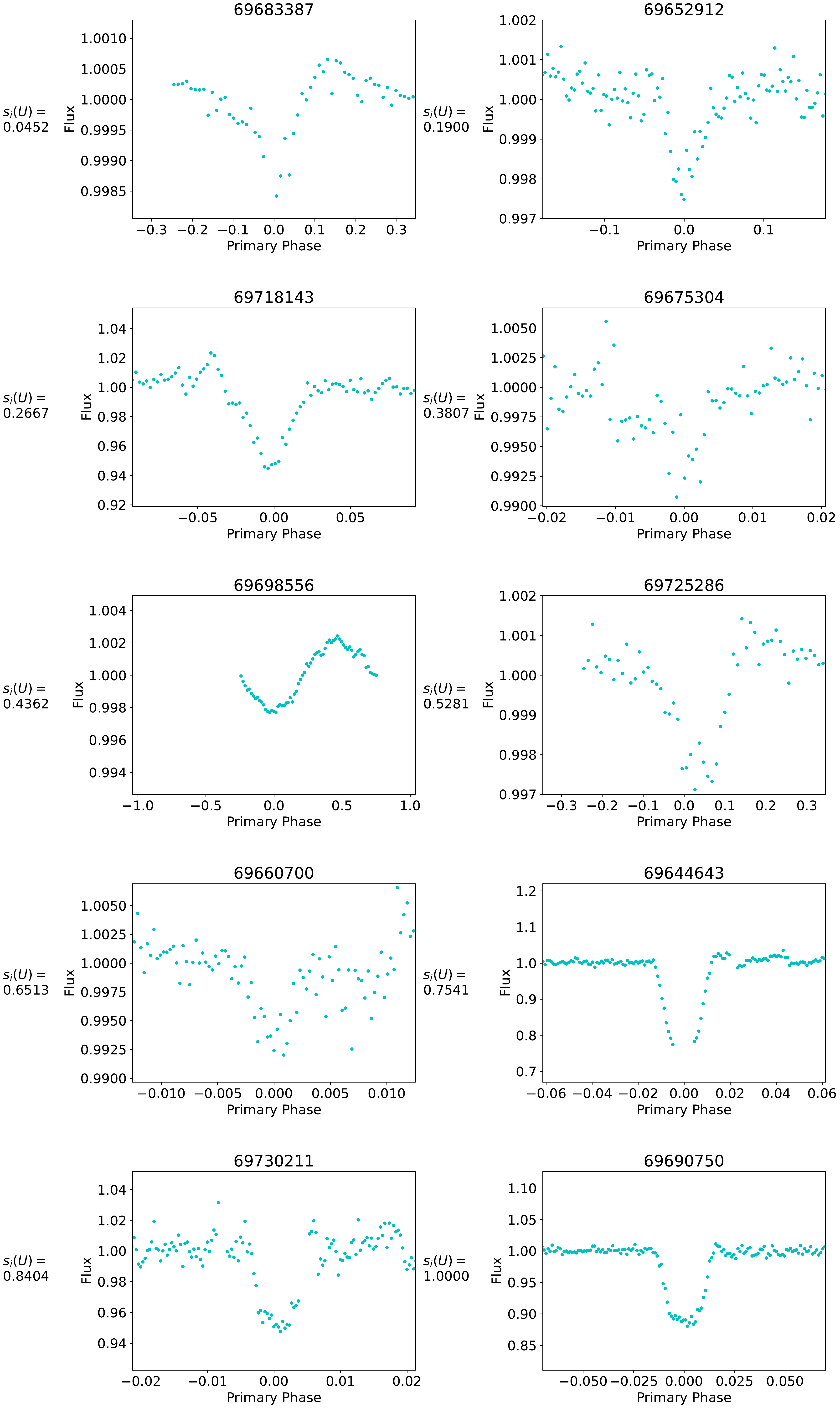}
    \caption{Random sample of Exoplanet Transit Search subjects per U-shaped score $s_i(U)$ binned with bin size of 0.1. The x-axes show Phase and the y-axes show Normalized Flux. Titles display Subject ID. The U-shaped subject score, $s_i(U),$ is shown to the left of each image.}\label{fig:example_scores}
\end{figure*}

We use the subject scores calculated from the weighting scheme to impose a series of threshold cuts in order to select a list of candidates that advance to the additional workflows and then to subsequent visual vetting by the \phngts\ science team. The aim of these cuts is to generate a list of potential exoplanet transit candidates without also including an overwhelming number of false positives.
% In Section~\ref{sec:planet_cands} we provide details for a select number of the most interesting planet candidates identified by the \phngts\ project.
We used the information from visual vetting of a random subset of preliminary candidates from the Exoplanet Transit Search to set these threshold cuts for candidate selection.
Figure~\ref{fig:example_scores} shows example light curves for ten subjects that were classified in the Exoplanet Transit Search for $s_i(U)$ randomly selected from bins of size 0.1. We see that subjects with a score that is closer to 1 tend to have more obvious U-shaped transits present and therefore we select candidates with $s_i(U) \geq 0.5$ and $s_i(\mathit{SV}) < 0.6$ for further vetting in the additional workflows.
The imposed limit on stellar variability ($s_i(\mathit{SV}) < 0.6$) reduces the number of false positives from light curves that show more variability around the primary transit, which is typically indicative of systematics or eclipsing binary systems that exhibit ellipsoidal modulation \citep{Morris1985EllipsoidalVariable,Mazeh2008CloseBinaries}.
% The relatively low number of subjects with high U-shaped scores (Figure~\ref{fig:prim_score_dist}), indicating an increased likelihood of the presence of U-shaped dips, provides us with a straightforward method for applying threshold cuts to significantly reduce the number of subjects that require vetting in the additional workflows.
% The imposed cut of $s_i(U) \geq 0.5$ indicates that the weighted majority of users have determined there to be a U-shaped transit, which is more indicative of an exoplanet transit compared with an eclipsing binary system which typically produce V-shaped transits \citep{MandelAgol2002AnalyticLCs}.

We also search for candidates that may exhibit a grazing or near-grazing exoplanet transit, which typically display a V-shaped transit. We achieve this by using the scores for the V-shaped response ($s_i(V)$) to select candidates with $s_i(V) \geq 0.8$ and $s_i(\mathit{SV}) < 0.6$ to be advanced to the additional workflows. We do not include subjects that have already been advanced to the additional workflows with the $s_i(U) \geq 0.5$ cut again. We impose the same limit as before on stellar variability ($s_i(\mathit{SV}) < 0.6$) to reduce the number of false positives. These cuts were selected after visual vetting of a preliminary selection of candidates from the Exoplanet Transit Search. Although this cut will also identify eclipsing binaries, the majority of these will be filtered by the Secondary Eclipse Check and Odd/Even Transit Check. In the scenario where only the primary eclipse of an eclipsing binary system is visible, further vetting tests by the \phngts\ science team (such as estimates of the radius of the secondary body) can be used to rule out these false positives.

The candidates selected by the above cuts are classified in the Secondary Eclipse Check and Odd/Even Transit Check workflows that, as described in Sections~\ref{sec:secondary_eclipse}~and~\ref{sec:odd_even}, are designed to check for secondary eclipses and differences in the depths of the odd and even transits. Once candidates are retired from these workflows, we apply the user weighting schemes to each of the workflows as described in Sections~\ref{sec:weighting_sec}~and~\ref{sec:weighting_odd}. We apply another set of threshold cuts using the scores calculated for each candidate in these workflows in order to generate a list of potential planet candidates to be reviewed by the \phngts\ science team, while aiming to rule out as many false positives as possible. As described in Section~\ref{sec:weighting_odd}, we only calculate scores for the second task of the Odd/Even Transit Check if a subject has sufficient data coverage, i.e. $s_i(\mathit{YC_{odd}}) \geq 0.5$. In addition to this cut, candidates with both $s_i(\mathit{NS_{sec}}) \geq 0.5$ and $s_i(\mathit{YD_{odd}}) \geq 0.5$ are selected for review. These cuts were chosen following visual vetting of a random subset of the preliminary candidate lists from these workflows. These cuts are designed to select candidates that have been classified as showing no secondary eclipse and appear to have matching depths for the odd and even transits, both of which are designed to rule out obvious eclipsing binary systems.

The final stage of the vetting process involves the \phngts\ science team reviewing the remaining potential planet candidates. Prior to this vetting stage, we remove 1144 known NGTS Objects of Interest (NOIs) that constitute known planets and planet candidates as well as eclipsing binaries and objects that were initially identified as possible candidates but have since been rejected in the original NGTS candidate identification process. We review the remaining 4726 potential planet candidates using the internal NGTS vetting portal \citep{Wheatley2018NGTS}.
This allows the \phngts\ science team to do the following: review the estimated stellar and planet candidate parameters; review the phase-folded light curves for other significant periods detected by \orion; determine whether the detected period is common within the field (indicative of systematics); identify nearby objects that may be the true source of the observed signal; review the light curves for individual nights of observation; and review the centroid vetting \citep{Guenther2017Centroid}. A full description of the vetting process is described in \citet{Bayliss2018NGTS1b}.
Table~\ref{tab:stages_subjects} outlines the number of subjects that remain at each step of this candidate selection process.

\begin{deluxetable*}{l|c|c}
\tabletypesize{\normalsize}
\tablecaption{Number of \phngts\ subjects that remain at each stage of the candidate selection process. $s_i(U)$, $s_i(V)$ and $s_i(\mathit{SV})$ are the subject scores for the U-shaped, V-shaped and Stellar Variability responses in the Exoplanet Transit Search, respectively. $s_i(\mathit{NS_{sec}})$, $s_i(\mathit{YC_{odd}})$ and $s_i(\mathit{YD_{odd}})$ are the subject scores for `No Secondary Eclipse,' `Yes, the data for both the odd and even transits cover the middle portion of the plot' and `Yes, the odd/even transits have similar depths,' respectively. The 5 remaining planet candidates are selected after visual vetting by the \phngts\ science team. \label{tab:stages_subjects}}
\tablehead{\colhead{} & \multicolumn{1}{p{3cm}}{\centering Number of subjects remaining} & \multicolumn{1}{p{3cm}}{\centering Percentage remaining \\ from previous step}}
\startdata
All                                                                              & 105534  &                      \\
Exoplanet Transit Search scores: $s_i(U) \geq 0.5$ \& $s_i(\mathit{SV}) < 0.6$ \dag   & 6303    & 5.97                    \\
\multicolumn{1}{r|}{$s_i(V) \geq 0.8$ \& $s_i(\mathit{SV}) < 0.6$}               & \multicolumn{1}{c|}{10029} & \multicolumn{1}{c}{9.50}  \\
Secondary Eclipse and Odd/Even Transit Check scores: & \\
\multicolumn{1}{r|}{$s_i(\mathit{NS_{sec}}) \geq 0.5$ \& $s_i(\mathit{YC_{odd}}) \geq 0.5$ \& $s_i(\mathit{YD_{odd}}) \geq 0.5$} & 5870  & 36.94                      \\
Removing known NGTS objects of interest (NOIs)                            & 4726 & 80.51\\
Planet candidates                                                         & 5 & 0.11 
\enddata
\tablecomments{\dag\ Includes 745 subjects that were moved to the additional workflows using the initial vote counting method.}
\end{deluxetable*}

%% file: sections/5_planetcands.tex
The \phngts\ science team review stage identified 5 planet candidates for further analysis from the 4726 subjects that passed all threshold cuts. The 4721 subjects not carried forward consisted primarily of spurious signals due to telescope systematics and eclipsing binary candidates as well as a number of possible planet candidates that were deemed infeasible for further observations. Those deemed infeasible typically had estimated secondary radii above 2\,\rjup\ and therefore a higher likelihood of being brown dwarfs or very low-mass stars, thus follow-up observations would be an inefficient use of limited observing resources. Additionally some candidates showed only a single convincing transit event and therefore we could not constrain the period or confidently determine possible planetary parameters.
Here we discuss the details of the most interesting planet candidates that were identified via the science team review. At the time of discovery, none of these candidates were NGTS Objects of Interest (NOIs) or \tess\ Objects of Interest (TOIs\footnote{https://tess.mit.edu/toi-releases/} -- \citealt{Guerrero2021TOIRelease}). We note that, in addition to the planet candidates described here that were discovered through \phngts, the full online version of Table~\ref{tab:subject_scores} includes additional candidate/solved systems that are to be included in forthcoming publications, e.g. TIC-388917846 (Battley et al., in preparation).
Table~\ref{tab:planet_cands_orion} provides the initial transit parameters derived by \orion\ for each of the 5 most interesting planet candidates and an estimate of their planetary radii, calculated by combining the transit depth and the stellar radius from the \tess\ Input Catalog \citep[TIC v8.2;][]{Stassun2019RevisedTIC,Paegert2021TICv8.2}.
\begin{deluxetable}{cccccc}
\tabletypesize{\scriptsize}
\tablecaption{\orion\ parameters for \phngts\ planet candidates. \label{tab:planet_cands_orion}}
\tablehead{\colhead{TIC ID} & \colhead{Epoch} & \colhead{Period} & \colhead{Transit Depth} & \colhead{SDE} & \colhead{$R_p$}}
\startdata
 & BJD - 2457000 & days & \% &  & \rjup \\ \hline
165227846 & 1098.557674 & 2.096882 & 5.18 & 50.66 & 0.71 \\ %NOI-106847
135251751 & 1104.019630 & 4.048836 & 0.27 & 25.91 & 1.21 \\ %NOI-106855
125925505 &  789.818866 & 1.743484 & 2.19 & 27.55 & 1.11 \\ %NOI-106844
125759305 &  795.630509 & 9.992979 & 2.35 & 32.48 & 1.15 \\ %NOI-106868
180997904 & 1100.341400 & 4.936048 & 0.96 & 33.24 & 1.27 \\ %NOI-106869
% 392695186 &  758.868322 & 3.951575 & 2.46 & 25.87 & 1.28 \\ %NOI-106922
\enddata
\end{deluxetable}

\subsection{Additional follow-up data}\label{ref:followup_data}
Here we describe the facilities used to obtain additional follow-up data of our noteworthy planet candidates. Table~\ref{tab:planet_cands_follow} outlines the follow-up obtained for each planet candidate to date. We use available \tess\ data and obtain additional ground-based photometric observations using the 1\,m telescope at the South African Astronomical Observatory (SAAO). High-resolution speckle imaging was obtained using the Zorro instrument on Gemini South \citep{Scott2021Zorro,Howell2022Zorro}. Spectroscopic follow-up was carried out using CORALIE \citep{Queloz2001CORALIE} and the Fibre-fed Extended Range Optical Spectrograph \citep[FEROS;][]{Kaufer1998FEROS}.

\begin{deluxetable*}{cccc}
\tabletypesize{\scriptsize}
\tablecaption{An overview of the observing facilities used for \phngts\ candidate follow-up. More details on the observations can be found in the main text and subsequent tables. \label{tab:planet_cands_follow}}
\tablehead{\colhead{TIC ID} & \colhead{Photometry} & \colhead{Spectroscopy} & \colhead{Speckle}}
\startdata
165227846 & \tess\ (10, 37, 63, 64); SAAO (2022-04-28, $z^\prime$; 2023-02-28, $g^{\prime}$; 2023-04-11, $i^{\prime}$; 2023-05-23, $V$) & & Zorro (2022-05-22) \\ %NOI-106847  
135251751 & \tess\ (10, 37, 63) & CORALIE; FEROS & Zorro (2023-05-27) \\ %NOI-106855
125925505 & SAAO (2022-04-30, $z^\prime$; 2022-07-16, $r^\prime$; 2023-07-24, $V$) & FEROS & Zorro (2022-05-22) \\ %NOI-106844
125759305 & \tess\ (11,38) & & \\ %NOI-106868  
180997904 & \tess\ (10, 36, 37, 63); SAAO (2023-04-10, $i^\prime$) & & \\ %NOI-106869  
% 392695186 & \tess\ (11,38) &  & \\ %NOI-106922 
\enddata
\tablecomments{Numbers in brackets for each candidate with \tess\ photometry indicate which \tess\ sectors this target was observed in.}
\end{deluxetable*}

\subsubsection{\tess}\label{sec:tess}
\tess\ \citep{Ricker2015TESS} is a space-based NASA survey telescope. \tess\ monitors $96\,\times\,24$\,deg sectors of the sky for $\mysim27.4$\,days at a time with near-continuous coverage during this observing window. None of our targets were observed with 2\,minute or 20\,second cadence. We utilise the \tess\ Full Frame Images (FFIs), which can be used to extract light curves for any target in the \tess\ field of view (see Table~\ref{tab:planet_cands_follow}). We use the light curves generated from the FFIs by the Quick-Look Pipeline \citep[QLP;][]{Huang2020QLPPrime,Kunimoto2021QLPFirstExt,Kunimoto2022QLPSecondExt} where available and use the \texttt{eleanor} \citep{Feinstein2019eleanor}, TESSCut \citep{Brasseur2019TESSCut} and \texttt{lightkurve} \citep{Lightkurve2018lightkurve} packages to access and analyse the light curves for targets without QLP light curves.
%For data collected in Sectors 63 onward, pre-made \texttt{eleanor} and TESSCut light curves are yet to be generated therefore we manually use TESSCut to download a cutout of the \tess\ FFI around our target. We use \texttt{lightkurve} to create an aperture mask and employ its \texttt{RegressionCorrector} class to remove the dominant scattered light background signal, allowing us to produce light curves suitable for further analysis.

% The survey strategy of \tess\ and cadence used to observe targets has evolved through the prime mission (Sectors 1 to 26), first extended mission (Sectors 27 to 55) and second extended mission (Sectors 56 to 83). The satellite measures the brightness of a catalog of pre-selected stars with 2\,minute cadence in the prime mission and extended missions, plus 20\,second cadence for select targets in the extended missions. In addition, Full Frame Images (FFIs) are made available with 30\,minute cadence for the prime mission, 10\,minute cadence for the first extended mission and 200\,second cadence for the second extended mission which can be used to extract light curves for any target in the \tess\ field of view.

\subsubsection{SAAO}\label{sec:saao}
We obtained follow-up photometry for a selection of our targets (see Table~\ref{tab:planet_cands_follow}) between 2022 April 28 and 2023 July 24 using the 1\,m telescope at the South African Astronomical Observatory (SAAO).
The 1\,m telescope was equipped with different versions of the Sutherland High-speed Optical Camera (SHOC) at different times. Observations carried out in 2022 were obtained using the `shocnawe' camera while observations from 2023 were obtained using `shocndisbelief'. These instruments are nearly identical with 2.85\,arcmin by 2.85\,arcmin fields of view and pixel scales of 0.167\,arcsec/pixel \citep{Coppejans2013SHOC}. 
Observations were carried out using a range of photometric filters (see Table~\ref{tab:planet_cands_follow}) to check for differences in the transit depth across multiple passbands. A difference in the measured transit depth in differing filters would indicate that the system is an eclipsing binary consisting of two stars with differing colours \citep{Drake2003ColorPhotom_FPs,Tingley2004ColorPhotom_FPs,Parviainen2019ColorPhotom_FPs}. We were able to simultaneously observe at least two comparison stars of similar brightness for each target. Calibration frames for the data reduction were taken each night at sunset and/or sunrise. Each light curve was bias and flat-field corrected using the local Python-based SAAO SHOC pipeline, which uses IRAF \citep{Tody1986IRAF} photometry tasks \citep[PyRAF][]{STScI2012PyRAF} and facilitates the extraction of raw and differential light curves. We used the Starlink package AUTOPHOTOM \citep{Currie2014StarlinkSoftware} to perform aperture photometry on both our target and comparison stars, and chose apertures that gave the maximum signal-to-noise ratio. Background annuli were chosen around the target and comparison stars that avoided any contaminating faint sources in the field of view. Finally, the measured fluxes of the comparison stars were used in order to conduct differential photometry of the target.

\subsubsection{Gemini/Zorro}\label{sec:zorro}
We performed high-resolution speckle imaging of TIC-125925505, TIC-165227846 and TIC-135251751 using the Zorro instrument \citep{Scott2021Zorro,Howell2022Zorro} on the 8.1\,m Gemini South telescope on Cerro Pach\'{o}n, Chile. Speckle imaging observations provide extremely high-resolution images by taking multiple sets of 1000 rapid (60\,ms) exposures in quick succession. This effectively `freezes out' the atmosphere such that the diffraction limit of the telescope can be reached and we can search for close-in companion stars at up to 8\,mag contrast levels. We can quantify the contribution of any nearby stars to the photometric measurements made with NGTS, \tess\ and SAAO instruments, allowing for more accurate estimates of the planet candidate radii. Simultaneous observations were obtained with the 562 and 832\,nm filters. These data were reduced as described in \citet{Howell2011SpeckleKepler} to produce contrast curves for both filters.

\subsubsection{CORALIE}\label{sec:coralie}
We monitored TIC-135251751 to obtain radial velocity measurements using the CORALIE high-resolution ($R\,\mysim\,60,000$) \'{e}chelle spectrograph mounted on the Swiss 1.2\,m Euler telescope at La Silla Observatory \citep{Queloz2001CORALIE}.
% CORALIE is fed by two fibers: one fiber is used to observe the target while a second fiber can connect to either a Fabry-P\'{e}rot etalon for simultaneous wavelength calibration or on-sky for background subtraction.
We obtained 8 spectra between 2023 January 3 and 2023 May 15.
% with exposure times ranging from 1500 to 1800\,s, resulting in average signal-to-noise ratios in the range 10 to 15.
% These spectra were obtained as part of \sob{brief follow-up program details}.
The spectra were reduced using the standard CORALIE Data Reduction Software (DRS) before being cross-correlated with an A0 stellar mask close to the spectral type of the host star to obtain a cross-correlation function \citep[CCF; e.g.][]{Baranne1996ELODIE,Pepe2002HARPS}. The pipeline calculates the radial velocity and its associated error as well as parameters such as full width half maximum (FWHM), contrast and bisector inverse slope (BIS) which are derived from the CCF. These additional parameters are used as diagnostics of stellar activity that may be affecting the RV measurements, and have previously been used to detrend the impact of activity on the radial velocities of stars when searching for orbiting planets \citep[e.g.][]{Queloz2001BisectorInverseSlope,Melo2007StellarActivityReduction,Diaz2018StellarActivity}. We opt for an A0 template to measure the RVs of an early F-type star as this is the closest match of the templates available (A0, G2, K5, M2). Early F-types undergo less magnetic braking compared with Solar-type stars. Furthermore, the use of the `incorrect' mask has been shown to have little effect on the overall trends of parameters, with only small changes in the absolute values \citep[e.g.][]{Costes2021StellarActivityMaskChanges}.

\subsubsection{FEROS}\label{sec:feros}
The Fibre-fed Extended Range Optical Spectrograph \citep[FEROS;][]{Kaufer1998FEROS} is a high-resolution ($R\,\mysim\,48,000$) \'{e}chelle spectrograph mounted on the MPG/ESO 2.2\,m telescope at La Silla Observatory.
% FEROS is fed by two fibers to observe the target and either a Th-Ar lamp or the sky to allow for simultaneous wavelength calibration or background subtraction.
We obtained radial velocity measurements of TIC-125925505 and TIC-135251751. A total of 7 FEROS spectra were collected between 2022 April 10 and 2023 March 31.
% with exposure times \sob{X}\,seconds leading to S/N ranging from \sob{X to Y}.
% These spectra were obtained as part of \sob{brief follow-up program details}.
FEROS data were processed with the \texttt{CERES} pipeline \citep{Brahm2017CERESPipeline}, which uses the cross-correlation technique to calculate radial velocities, its associated error and the BIS (sometimes referred to as Bisector Span, however the methods of calculating BIS for CORALIE and FEROS are identical and as described in \citet{Queloz2001BisectorInverseSlope}).

% \subsubsection{HARPS}\label{sec:harps}
% % The Near Infra-Red Planet Searcher \citep[NIRPS;][]{Bouchy2017NIRPS,Wildi2017NIRPS} and High Accuracy Radial velocity Planet Searcher \citep[HARPS;][]{Mayor2003HARPS} are high-resolution \'{e}chelle spectrographs which are both mounted on the ESO 3.6\,m telescope at La Silla Observatory, allowing simultaneous spectra in the NIR (971 to 1854\,nm) and VIS (378 to 691\,nm) channels. For TIC-165227846, we obtained 18 spectra (across 9 individual nights) with NIRPS between 2023 April 01 and 2023 April 27, with simultaneous HARPS spectra also taken, under program number 111.254E.001 (PI: O'Brien). We obtained two consecutive exposures using NIRPS with exposure times ranging from 600 to 900\,s to \sob{techincal reason why we did this.} The simultaneous HARPS spectra were obtained with exposure times ranging from 1200 to 1700\,s, with only one spectra taken per night. We did not obtain a simultaneous HARPS spectra on 2023 April 08 due to \sob{why?}.
% The High Accuracy Radial velocity Planet Searcher \citep[HARPS;][]{Mayor2003HARPS} is a high-resolution \'{e}chelle spectrographs which is mounted on the ESO 3.6\,m telescope at La Silla Observatory.
% \sob{Other HARPS observations summary}.
% \sob{Data reduction description}.

% \subsubsection{\spupnic}\label{sec:spupnic}
% \sob{TIC-165227846 got \spupnic\ follow-up but it did not reveal anything. TIC-135251751 possibly got \spupnic\ follow-up? Need to check with Matt}

\subsection{Modelling}\label{sec:modelling}
We performed Spectral Energy Distribution (SED) fitting for each candidate to obtain more accurate stellar parameters using the \ariadne\ package \citep{Vines2022ARIADNE}. \ariadne\ fits broadband photometry measurements from catalogs (see \citet{Vines2022ARIADNE} for photometric bandpasses used) to different stellar atmosphere models such as Phoenix V2 \citep{Husser2013PhoenixV2}, BT-Stell, BT-Cond, BT-NextGen \citep{Hauschildt1999BTNextGen,Allard2012BTModels}, \citet{Kurucz1993SED} and \citet{Castelli2004SED}. The results of these analyses, along with additional parameters from the TIC v8.2 catalog where noted \citep{Stassun2019RevisedTIC,Paegert2021TICv8.2}, are provided in Table~\ref{tab:planet_cands_stellar}.

\begin{deluxetable*}{ccccccccc}
\tabletypesize{\scriptsize}
\tablecaption{Stellar parameters for \phngts\ planet candidates. \label{tab:planet_cands_stellar}}
\tablehead{\colhead{TIC ID} & \colhead{Component} & \colhead{Stellar Mass} & \colhead{Stellar Radius} & \colhead{Stellar $T_{\text{eff}}$} & \colhead{Stellar log$(g)$} & Stellar Luminosity & \colhead{Stellar ${V_{\text{mag}}}^1$} & \colhead{Stellar ${J_{\text{mag}}}^1$}}
\startdata
 & & \msun\ & \rsun\ & K & & L$_{\odot}$ & mag & mag \\ \hline
165227846 & - & $0.3257^{+0.0193}_{-0.0134}$ & $0.3299^{+0.0291}_{-0.0174}$ & $3244.4892^{+80.4376}_{-117.2090}$ & $3.7335^{+0.4008}_{-0.1179}$ & $0.0109^{+0.0022}_{-0.0018}$ & $16.365\pm1.133$  & $11.743\pm0.021$ \\ %NOI-106847
135251751 & (TIC) & $1.44\pm0.23855$ & $2.39031\pm0.124894$ & $6742\pm133$ & $3.83952\pm0.0910193$  & $10.63554\pm0.8621945$ & $11.125\pm0.01$   & $10.222\pm0.022$ \\ %NOI-106855-CAT
135251751 & A & $1.50$ & $1.679$ & $7020$ & - & 5.75 & -   & - \\ %NOI-106855-A
135251751 & B & $1.33$ & $1.473$ & $6550$ & - & 3.64 & -   & - \\ %NOI-106855-B
125925505 & - & $0.7555^{+0.0375}_{-0.0276}$ & $0.5643^{+0.0576}_{-0.0442}$ & $4741.7769^{+126.9399}_{-126.9399}$ & $4.2991^{+0.4636}_{-0.3883}$ & $0.1441^{+0.0352}_{-0.0258}$   & $15.482\pm0.103$  & $13.443\pm0.026$ \\ %NOI-106844
125759305 & - & $0.8250^{+0.0380}_{-0.0350}$ & $0.7832^{+0.0154}_{-0.0161}$ & $4931.5295^{+73.4975}_{-62.1902}$ & $4.2963^{+0.5860}_{-0.4133}$ & $0.3270^{+0.0238}_{-0.0206}$ & $14.73\pm0.206$   & $12.914\pm0.023$ \\ %NOI-106868
180997904 & - & $0.9946^{+0.0543}_{-0.0376}$ & $1.3033^{+0.0400}_{-0.0307}$ & $5390.6185^{+91.7171}_{-74.4120}$ & $4.6502^{+0.2154}_{-0.6755}$ & $1.3012^{+0.1204}_{-0.0985}$ & $14.376\pm0.069$  & $12.979\pm0.023$ \\ %NOI-106869
\enddata
\tablecomments{Refs: 1 - TIC v8.2 \citep{Paegert2021TICv8.2}. TIC parameters for \cosworthid\ which assume a single star are provided for comparison. The derivation of the parameters of the individual components of \cosworthid\ are described in Section~\ref{sec:cosworthsec}.}
\end{deluxetable*}

We derive planetary parameters for each system using the \allesfitter\ package \citep{Guenther2019allesfittercode,Guenther2021allesfitterpaper}. \texttt{Allesfitter} combines \texttt{ellc} \citep[light curve and RV models;][]{Maxted2016ellc}, \texttt{emcee} \citep{ForemanMackey2013emcee}, \texttt{dynesty} \citep[Nested Sampling (NS);][]{Speagle2020dynesty} and \texttt{celerite} \citep{ForemanMackey2017celerite} to simultaneously fit photometric and spectroscopic data, with models available for a variety of signals including multiple planet systems and stellar variability.
Prior to fit with \allesfitter\, we use the \bruce\footnote{https://github.com/samgill844/bruce} package to compute preliminary planetary and orbital parameters from the NGTS data only. \bruce\ is an open-source package for the modelling of binary stars and exoplanets, which employs \texttt{emcee} \citep[Markov Chain Monte Carlo (MCMC) sampling;][]{ForemanMackey2013emcee} and \texttt{celerite} \citep[Gaussian Process (GP) models;][]{ForemanMackey2017celerite}. We also use the stellar parameters obtained from the SED fits with \texttt{ARIADNE} as priors for the \allesfitter\ modelling.
Given the length of time for the initial \bruce\ MCMC fits to converge, a nested sampling approach with \allesfitter\ was taken to perform the global analysis.
For all candidates, we adopted a quadratic limb darkening law for the photometric data as parameterized in \citet{Kipping2013LimbDarkening}. We use the \texttt{PyLDTk} package \citep{Parviainen2015LDTk}, which uses Phoenix V2 models \citep{Husser2013PhoenixV2} and transmission curves from the Spanish Virtual Observatory (SVO) Filter Service \citep{Rodrigo2012SVO,Rodrigo2020SVO}, to estimate priors for the limb darkening coefficients.
We implement the Mat\'{e}rn 3/2 GP kernel to model any out-of-transit variability as this kernel is versatile in its ability to model both long and short-term trends \citep{Guenther2021allesfitterpaper}.
Due to the large pixel scale of \tess\ (21\,arcsec\,per\,pixel, Section~\ref{sec:tess}), the \tess\ data are fit with a dilution factor due to the possibility of contribution from nearby stars.
% We fit each of our systems with a circular model (eccentricity fixed to be zero) as theoretical predictions and the observed population of hot Jupiters both show orbital eccentricities are often consistent with zero \cite[e.g.][]{Anderson2012WASP44.45.46Eccentricity}.

We also utilise the \triceratops\ \citep{Giacalone2020TriceratopsSoftware,Giacalone2021TriceratopsSoftware} validation package to calculate False Positive Probabilities (FPPs) for each of our planet candidates where \tess\ data are available (Section~\ref{sec:tess}). \triceratops\ uses Bayesian analysis to calculate the probabilities that a transiting signal was produced by a transiting planet or a number of false positive scenarios such as nearby or background eclipsing binaries. \triceratops\ can also include high-resolution speckle imaging data to provide stronger constraints on false positive scenarios such as unresolved companions. Therefore, we include Gemini/Zorro data where available in our calculations (Section~\ref{sec:zorro}). We perform the FPP calculation 1000 times for each planet candidate and compute the mean and standard deviation of the results. The parameters derived using \allesfitter\ and the fractional FPP values computed using \triceratops\ are shown in Table~\ref{tab:planet_cands_fits}. We note that the FPP values presented are higher than any generally accepted validation threshold \citep[FPP$\lesssim 0.01$;][]{Montet2015vespa,Giacalone2020TriceratopsSoftware,Giacalone2021TriceratopsSoftware}, however \triceratops\ is known to penalise giant planets due to the degeneracy between their radii and those of brown dwarfs and very low-mass stars \citep{Bryant2023OccurrenceGiantsMDwarfs}. We are presenting these systems as candidates only and plan where possible to gather further, more robust follow-up observations to confirm or rule out these systems as planetary.

\begin{deluxetable*}{cccccccc}
\tabletypesize{\scriptsize}
\tablecaption{\phngts\ planet candidate fitting results. \label{tab:planet_cands_fits}}
\tablehead{\colhead{TIC ID} & \colhead{Component} & \colhead{Epoch} & \colhead{Period} & \colhead{$R_p/R_*$} & \colhead{$R_p$} & \colhead{\triceratops\ FPP}}
\startdata
 & & BJD - 2457000 & days &  & \rjup & \\ \hline
165227846 & - & $2094.49836_{-0.00005}^{+0.00005}$ & $2.0966799_{-1E-7}^{+1E-7}$ & $0.50050_{-0.02385}^{+0.01664}$ & $1.61406_{-0.12850}^{+0.13200}$ & $0.7267^{+0.4316}_{-0.2733}$ \\ %NOI-106847
135251751 & A & $2091.84231_{-0.00065}^{+0.00066}$ & $4.04841_{-0.000004}^{+0.000004}$ & $0.068013_{-0.00191}^{+0.00221}$ & $1.111_{-0.031}^{+0.036}$ & - \\ %NOI-106855-A
135251751 & B & $\cdots$ & $\cdots$ & $0.08552_{-0.00241}^{+0.00278}$ & $1.226_{-0.035}^{+0.040}$ & - \\ %NOI-106855-B
125925505 & - & $1970.07766_{-0.00040}^{+0.00040}$  & $1.7433513_{-8E-7}^{+8E-7}$ & $0.15059_{-0.00383}^{+0.00392}$ & $0.83261_{-0.07627}^{+0.07892}$ & - \\ %NOI-106844
125759305 & - & $1585.26417_{-0.00070}^{+0.00071}$  & $9.995842_{-8E-6}^{+5E-6}$ & $0.13589_{-0.00092}^{+0.00122}$ & $1.03640_{-0.02241}^{+0.02250}$ & $0.1617\pm0.0298$ \\ %NOI-106868
180997904 & - & $2072.85973_{-0.00153}^{+0.00145}$ & $4.93668_{-0.00001}^{+0.00001}$ & $0.08773_{-0.00234}^{+0.00200}$ & $1.11353_{-0.04033}^{+0.03922}$ & $0.6714\pm0.0114$ \\ %NOI-106869
\enddata
\tablecomments{We list the possible planetary parameters for the candidate in the \cosworthid\ system, where it may be orbiting either the primary or secondary star of this binary system. The derivation of these parameters is described in Section~\ref{sec:cosworthsec}. We do not include \triceratops\ FPP values for \cosworthid\ due to the uncertain nature of the system. We do not know which star the transiting candidate orbits nor do we have secure estimates of the stellar radii of the possible hosts.}
\end{deluxetable*}

\subsection{Interesting Systems}
%transition
The five candidates identified by the \phngts\ project are all consistent with being hot giant planets, orbiting their host stars with periods of less than 10\,days and with estimated radii similar to those of Jupiter or Saturn. We report the discovery of three hot giant planet candidates orbiting late-G/early-K-type host stars (\fernandoid, \felipeid\ and \sebastianid) as well as the detection of a transiting companion likely in an S-type orbit around one component of a binary star system (\cosworthid). In addition, we highlight the discovery of \nigelid, a hot Jupiter candidate orbiting an M-dwarf that would be the lowest mass star to host a giant planet if confirmed.
% \fernandoid, \felipeid\ and \sebastianid\ have late-G/early-K-type host stars while \nigelid\ is an M-dwarf host which would be the lowest mass star to host a giant planet if confirmed. \cosworthid\ orbits either an \sob{F1V or F5V} star in an S-type orbit.

\subsubsection{TIC-165227846}
\begin{figure*}
    \centering
    \includegraphics[width=\textwidth]{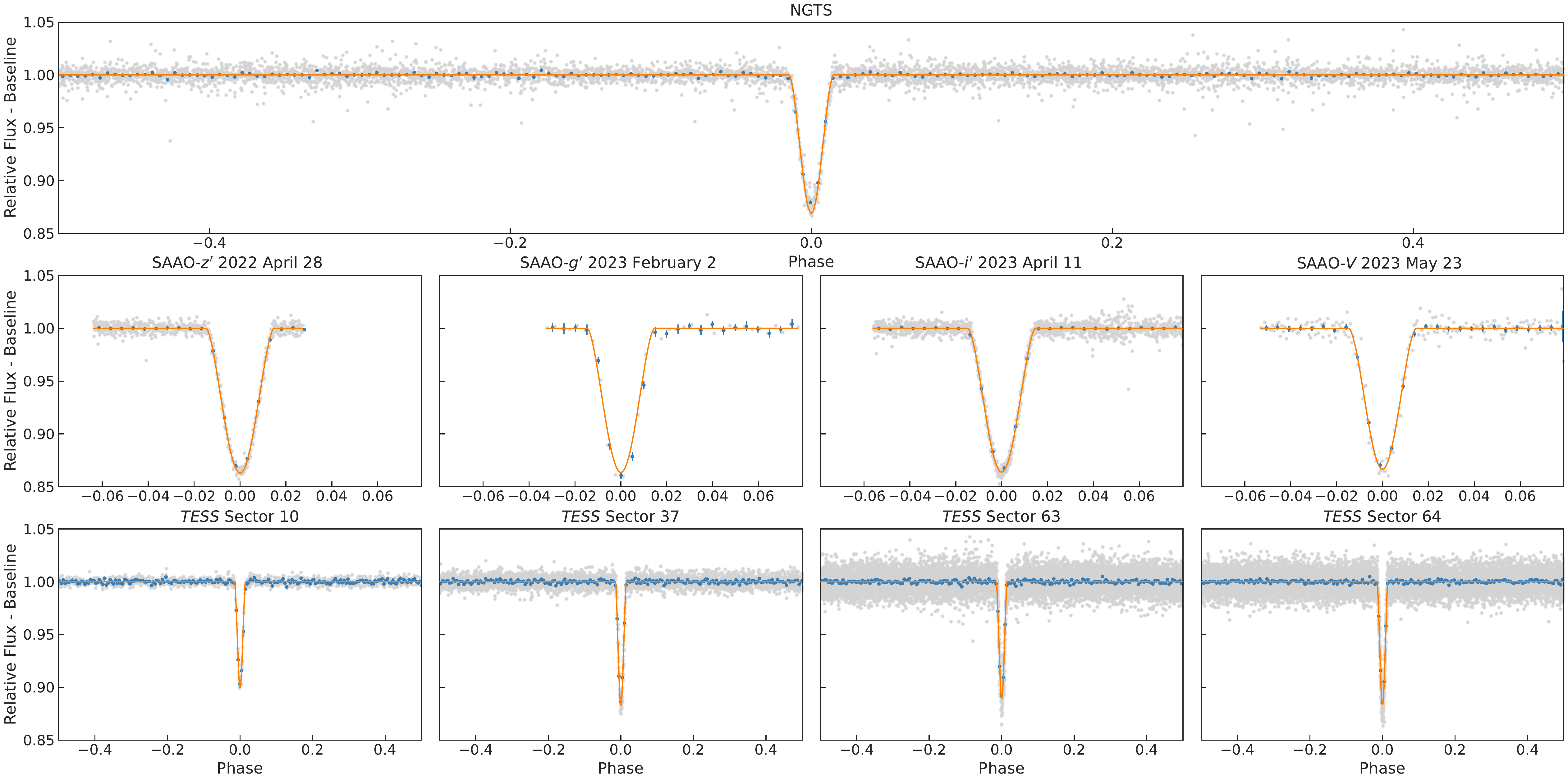}
    \caption{Discovery and follow-up data obtained for \nigelid. The phase folded light curves from NGTS, SAAO and each \tess\ sector are shown with the median best fit circular model from \allesfitter\ in orange.}
    \label{fig:nigel_photom}
\end{figure*}

\begin{figure}
    \centering
    \includegraphics[width=\columnwidth]{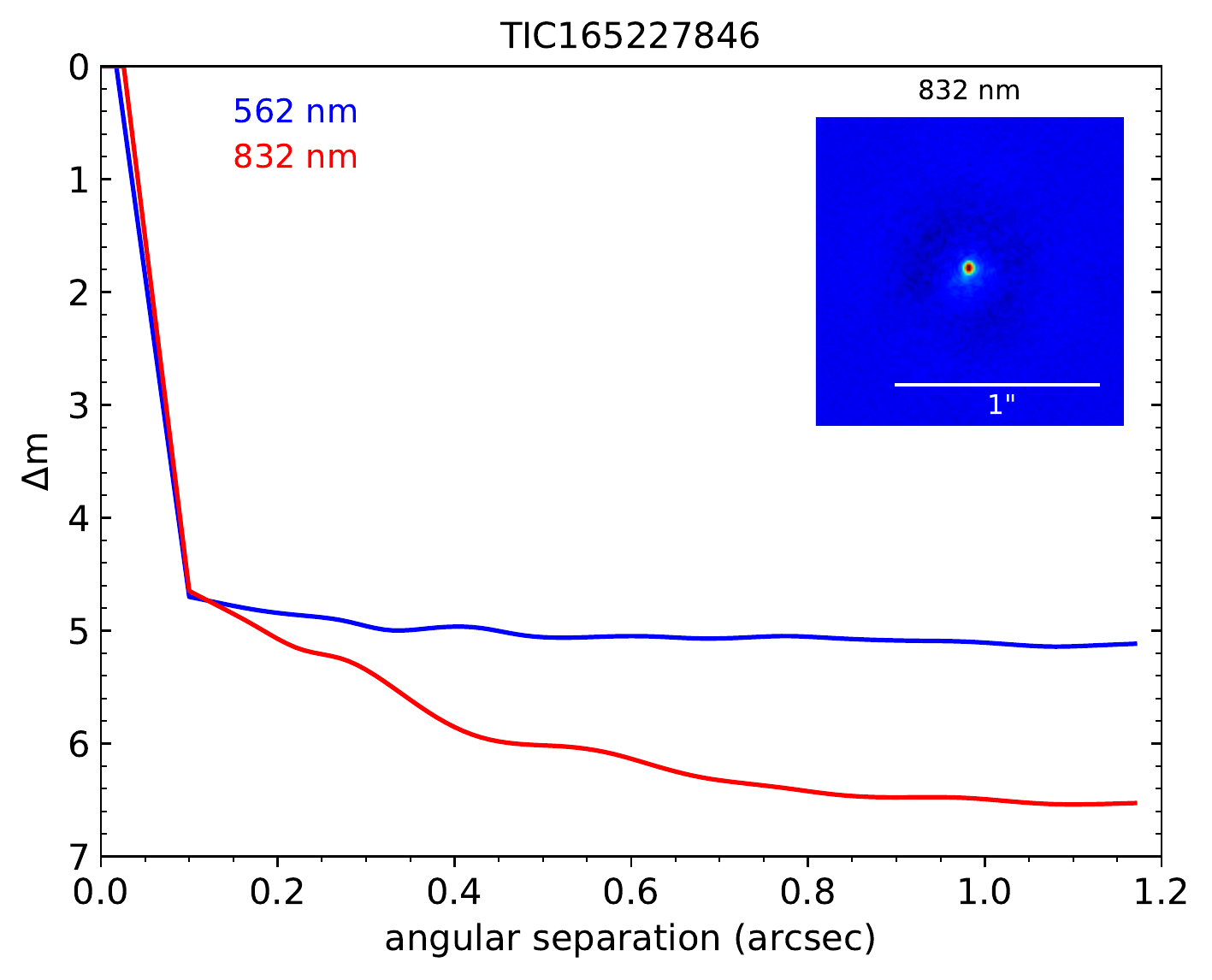}
    \caption{The Zorro constrast curve for \nigelid. The blue and red lines show the contrast curves for the 562\,nm and 832\,nm filters, respectively. The inset shows the reconstructed image of the star from Zorro.}
    \label{fig:nigel_zorro}
\end{figure}

\orion\ detected a transit signal in the NGTS data with a period of 2.097\,days, depth of 5.18\% and SDE of 50.66 from a total of 27 full or partial transits. We note that \orion\ underestimated the depth of the transit. The stellar parameters of \nigelid\ indicate that it is a mid M-dwarf, therefore the large transit depth remains consistent with a planetary radius for the transiting companion.
Using 4 \tess\ sectors and observations of 4 individual transits in different filters using SHOC at SAAO, we measure a transit depth of 13.1\% that, when combined with the stellar radius of 0.32\,\rsun, gives a companion radius of \nigelplrad\,\rjup.
The transit depth measured in Sector 10 is shallower compared with the other datasets ($\mysim10\%$) however the data has a cadence of 30\,minutes resulting in poor sampling of the full transit. Figure~\ref{fig:nigel_photom} shows the photometric data obtained for \nigelid.
Zorro imaging reveals no companions within 1.17\,arcsec of the target at the 4–5\,mag limit in the 562\,nm filter and at the 4-7\,mag limit in the 832\,nm filter. These data are shown in Figure~\ref{fig:nigel_zorro}.
With a mass of 0.33\,\msun, \nigelid\ would be the lowest mass star to host a close-in giant planet if the companion is confirmed to be planetary \citep{Bryant2023OccurrenceGiantsMDwarfs}. We note that this candidate was independently detected in \citet{Bryant2023OccurrenceGiantsMDwarfs}. We are actively seeking radial velocity measurements and the further analysis of this system will be the subject of future work.

% %NOI-106847
% NGTS, TESS, SAAO, Zorro, (NIRPS/HARPS)

% One light curve was obtained for TIC-165227846 using the SHOC instrument on the 1\,m telescope at SAAO. On 2022 April 28 observations were taken with the $z^\prime$ filter and an exposure time of 30\,seconds. 550 images were taken.

%Zorro 2022 May 22. 18 sets (I think. 23 minutes total on Science target according to OT)

\subsubsection{TIC-135251751}\label{sec:cosworthsec} %NOI-106855, Subject 69658020 discovery subject.
\begin{figure}
    \centering
    \includegraphics[width=\columnwidth]{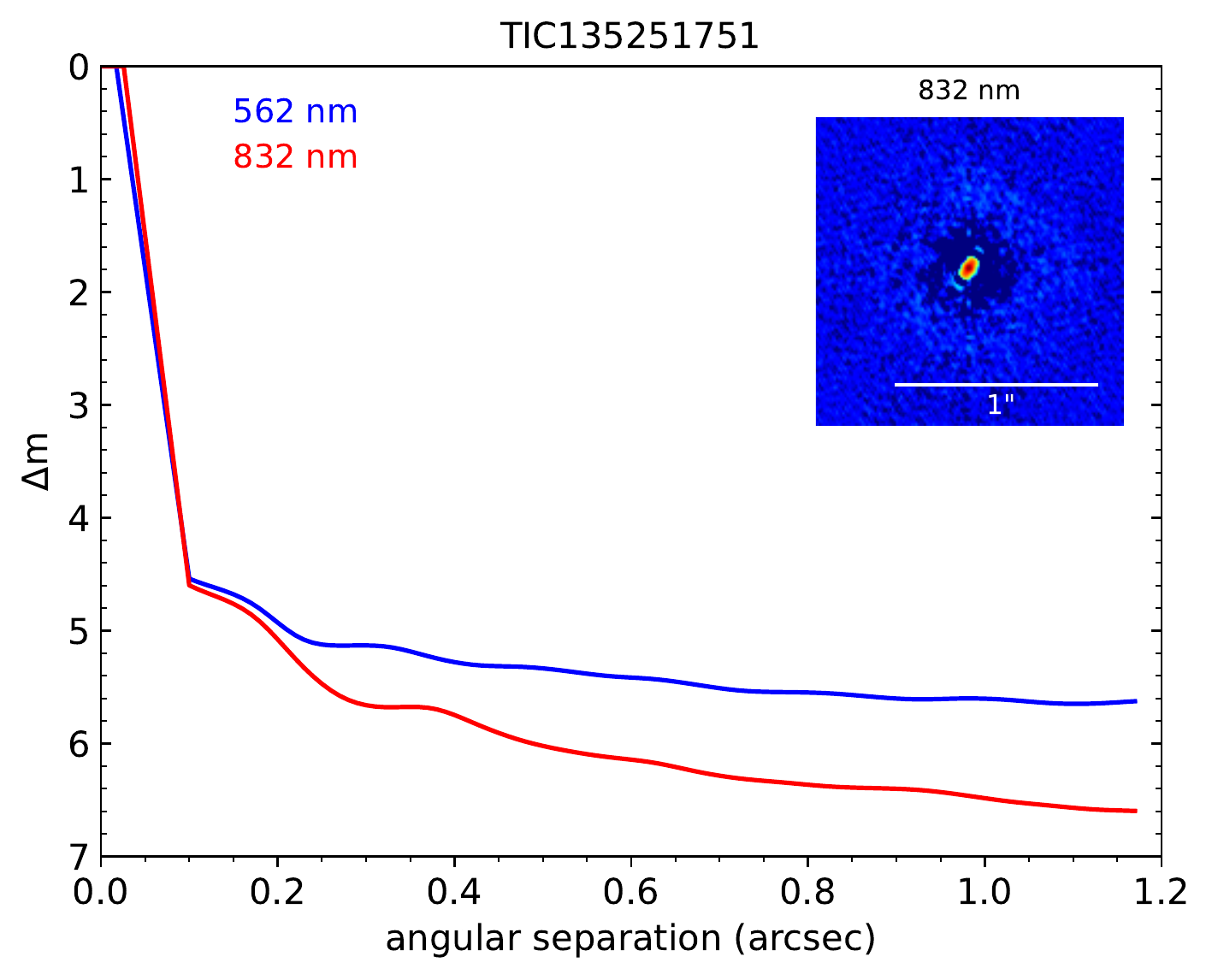}
    \caption{The Zorro contrast curve for \cosworthid. The blue and red lines show the contrast curves for the 562\,nm and 832\,nm filters, respectively. The inset shows the reconstructed image of the system from Zorro. The elongated shape of the reconstructed image is due to this being a binary star system.}
    \label{fig:cosworth_zorro}
\end{figure}

% \cosworthid\ was observed by NGTS between 2017 December 16 and 2018 July 28 with 243,447 images collected during this observing campaign.
\orion\ detected a transit signal in the NGTS data with a period of 4.049\,days, depth of 0.27\% and SDE of 25.91 from a total of 15 full or partial transits.
Catalog values for \cosworthid\ quote a stellar radius of $2.39\pm0.125$\,\rsun, $T_{\text{eff}} = 6742$\,K, log$(g) = 3.84\pm0.091$ and luminosity of $10.64\pm0.862$\,L$_{\odot}$ indicating that it is likely a subgiant.
Combining this stellar radius and the measured transit depth gives an initial estimate for the radius of the orbiting body to be $1.21$\,\rjup, suggesting the possibility of this candidate being a hot Jupiter orbiting a subgiant.
However, high-resolution speckle imaging of \cosworthid\ using Zorro, shown in Figure~\ref{fig:cosworth_zorro}, reveals that this is a binary star system with a projected separation of 0.033\,arcsec and flux ratios of 0.662 and 0.603 in the 562\,nm and 832\,nm filters, respectively. These flux ratios correspond to magnitude differences between the two stars of 0.45 and 0.55 for the 562\,nm and 832\,nm filters, respectively, indicating that the two stars are likely of similar spectral type.

\begin{figure*}
    \gridline{\fig{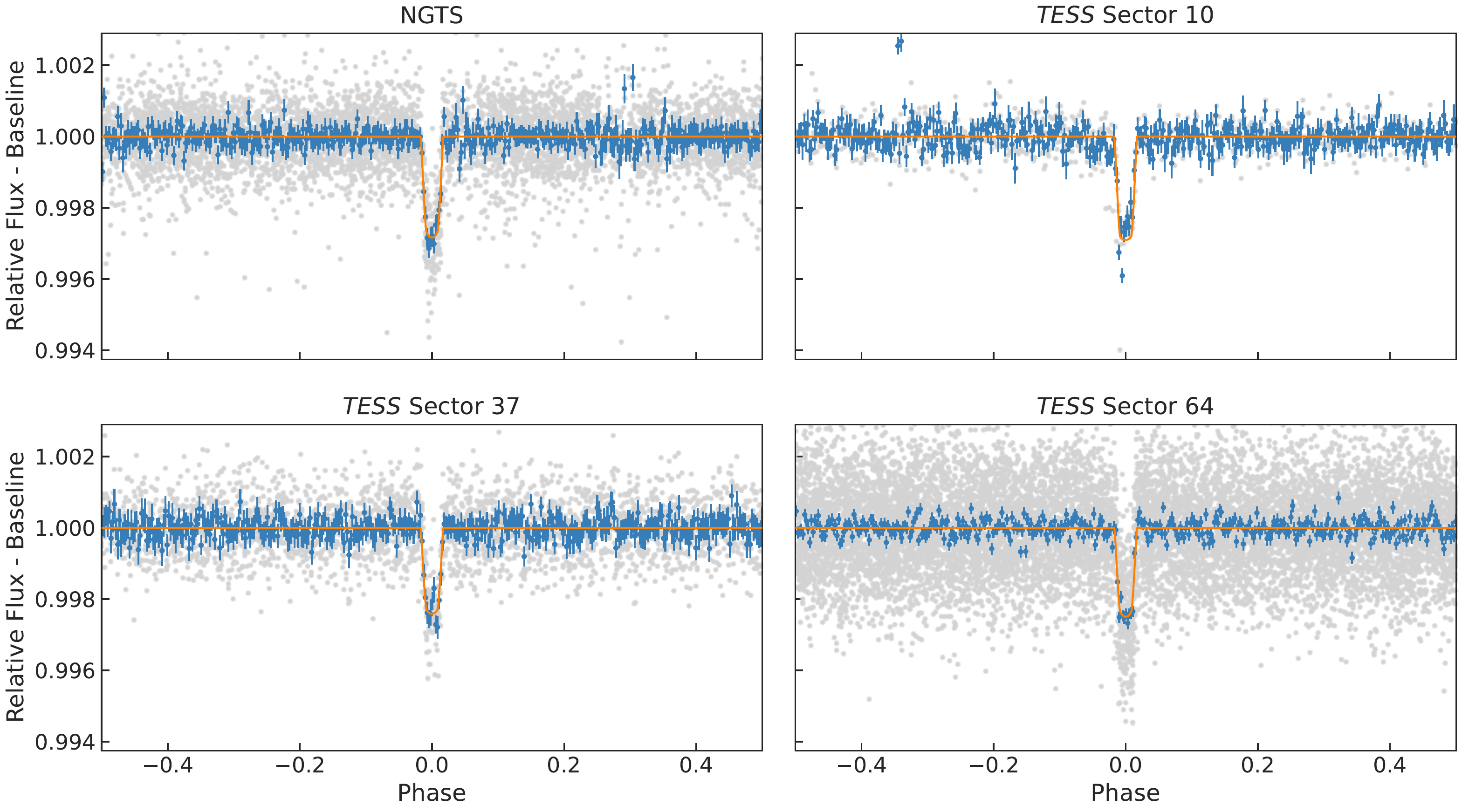}{\textwidth}{}}
    \vspace{-0.8cm}
    \gridline{
            \fig{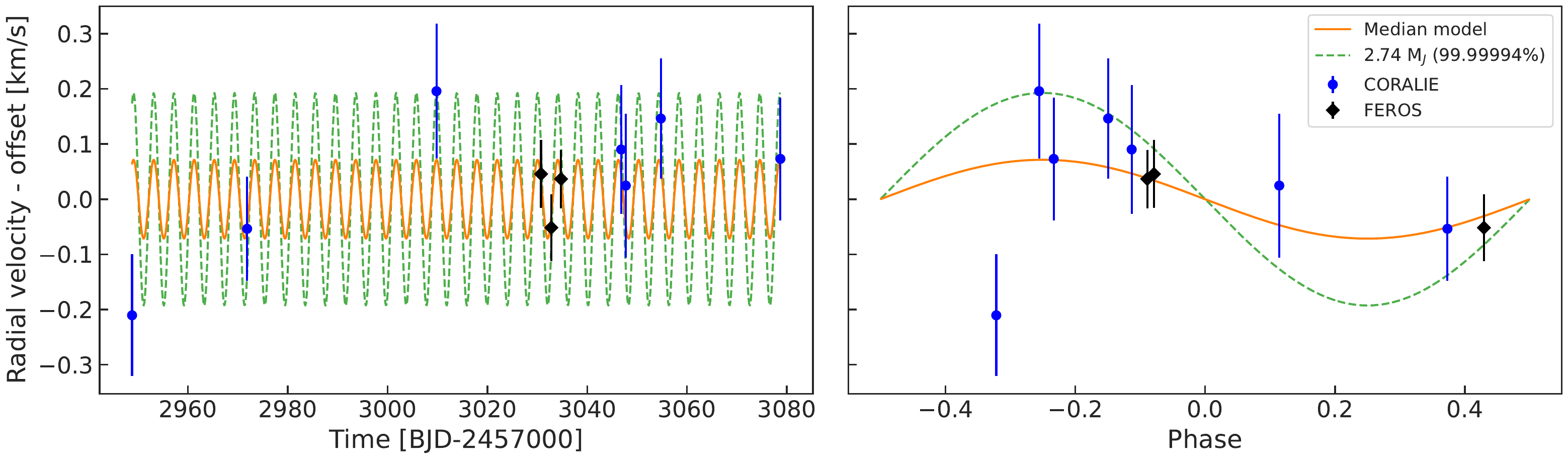}{0.99\textwidth}{}}
    \vspace{-0.8cm}
    \caption{Discovery and follow-up data obtained for TIC-135251751. The phase folded light curves from NGTS and each \tess\ sector are shown in the top row. The radial velocity data is shown in time and phase folded in the bottom row. The model shown in orange is the median best fit circular model from \allesfitter. The green dashed line shows the RV model for the 99.99994\% upper mass limit, corresponding to the \cosworthplmass\,\mjup\ companion mass limit. The radial velocity measurements and \allesfitter\ models assume a single star with catalog values.} \label{fig:cosworth_data}
\end{figure*}

In addition to the NGTS data, the target was observed in \tess\ Sectors 10, 37 and 64. We identify a total of 19 transits across the three \tess\ sectors.
We obtained 8 CORALIE spectra (Section~\ref{sec:coralie}) and 3 FEROS spectra (Section~\ref{sec:feros}) between 2023 January 4 and 2023 May 16. We excluded the CORALIE spectra obtained on 2023 May 16 from further analysis due to high instrumental drift.
Given the binarity of the host system, the radial velocities will not be an accurate measure of the radial velocity induced by the transiting companion however we report the values and analysis here to highlight the importance of obtaining high-resolution speckle imaging when validating the planetary nature of candidate systems. The CORALIE RV data are shown in Table~\ref{tab:cosworth_coralie_rvs} and the FEROS data are shown in Table~\ref{tab:cosworth_feros_rvs}.
The photometric light curves and radial velocity data for \cosworthid\ are shown in Figure~\ref{fig:cosworth_data} with models from \allesfitter\ that assume a single host star with catalog values from the TIC (Table~\ref{tab:planet_cands_stellar}).

\begin{deluxetable*}{cccccc}
\tablecaption{CORALIE spectroscopic data for \cosworthid\ assuming a single star}\label{tab:cosworth_coralie_rvs}
\tablehead{\colhead{Time} & \colhead{RV} & \colhead{RV error} & \colhead{FWHM} & \colhead{BIS} & \colhead{Contrast}}
\startdata
BJD - 2457000 & \kms & \kms & \kms & \kms & \\
\hline
2948.80439868 & -13.097311973 & 0.110387058 & 40.547364878 & -0.929035261 & 17.588155 \\
2971.85923573 & -12.94049194 & 0.094379824 & 41.324681030 & -0.535237690 & 17.359035 \\
3009.79791702 & -12.691021747 & 0.121918536 & 40.535339566 & -0.873255231 & 18.005368 \\
3046.81076808 & -12.79670901 & 0.11689358 & 40.668496072 & -1.026139583 & 17.696245 \\
3047.73086323 & -12.862174746 & 0.130432863 & 40.872850012 & -1.305164977 & 17.729222 \\
3054.76065788 & -12.740658082 & 0.108603896 & 40.744196360 & 0.500147341 & 17.916060 \\
3078.71216262 & -12.81379369 & 0.111130462 & 40.222439766 & -0.225302864 & 17.983820 \\
% 3080.68213999 & -13.603837972 & 0.128812966 & 41.431278712 & -0.212453892 & 17.662660 \\
\enddata
\end{deluxetable*}
\begin{deluxetable}{cccc}
\tablecaption{FEROS spectroscopic data for \cosworthid\ assuming a single star. No measurements of Bisector Inverse Slope (BIS) were obtained for two of the observations.}\label{tab:cosworth_feros_rvs}
\tablehead{\colhead{Time} & \colhead{RV} & \colhead{RV error} & \colhead{BIS}}
\startdata
BJD - 2457000 & \kms & \kms & \kms \\
\hline
3030.755730267 & -12.99280 & 0.0615 & 0.55370 \\
3032.812970775 & -13.0902 & 0.0607 & - \\
3034.762647563 & -13.0018 & 0.0529 & - \\
\enddata
\end{deluxetable}

Assuming a single star, the modelling of the photometric and radial velocity measurements is consistent with a planetary companion with a radius of \cosworthplrad\,\rjup\ and a 99.99994\% upper mass limit on the companion body of \cosworthplmass\,\mjup. The spectra show no obvious signs of the system being a double-lined binary, likely due to the similar spectral types of the two binary components resulting in largely similar spectral lines for both that cannot be disentangled easily. In addition, the CCFs are broad with FWHM $\approx40$\,\kms, indicating that the stars are fast rotators which results in broad, overlapping features that are blended together and further hinder the ability to separate the two component stars.
We see correlation between the RVs and CCF contrast which is indicative that we are not measuring the radial velocity signal expected of a hot Jupiter around a single star. The high-resolution speckle imaging of this candidate reveals part of the true nature of this system, highlighting the importance of obtaining these data from instruments such as `Alopeke and Zorro \citep{Scott2021Zorro,Howell2022Zorro} when validating planetary candidates.

In order to obtain estimates of the possible parameters of the transiting companion, we first estimate the stellar parameters of the two possible host stars. We use the total luminosity of $9.394021\pm0.404434$\,\lsun\ provided by \gaia\ DR2 \citep{Gaia2016GaiaMission,Gaia2018GaiaDR2} and the mean flux ratio, $\frac{F_B}{F_A} = 0.6325$, from Zorro to calculate individual luminosities of the two stellar components. We find luminosities of $L_A = 5.75\,\lsun$ and $L_B = 3.64\,\lsun$ for the primary and secondary star, respectively. We compare these values to standard values provided by \citet{Pecaut2012ApJ...746..154P,PecautMamajek2013ColorsTemps} and estimate the stars to have spectral types of F1V and F5V. We perform basic spectral analysis of the CORALIE data and find the spectra are consistent with these estimated spectral types. We use these spectral types to estimate the stellar parameters that are given in Table~\ref{tab:planet_cands_stellar} for each of the stars. Due to both stars being in the photometric apertures for all instruments, we must account for the dilution of the transit depth due to the non-host star when calculating the possible parameters of the transiting companion. The diluted transit depth is measured as $\delta_{\text{dil}} = 0.00283_{-0.00016}^{+0.00018}$. The undiluted transit depth is given as $\delta_{\text{undil.}} = \delta_{\text{dil}} \times (1+ \frac{F_{\text{cont}}}{F_{\text{host}}})$ where $F_{\text{cont}}$ is the flux of the contaminant (non-host) star and $F_{\text{host}}$ is the flux of the host star. We estimate the radius of the transiting companion to be $1.111_{-0.031}^{+0.036}$\,\rjup\ if it orbits the primary (F1V) star and $1.226_{-0.035}^{+0.040}$\,\rjup\ if it orbits the secondary (F5V) star. Therefore, the transiting companion in this system is consistent with being planetary in size and may be a hot Jupiter in an S-type orbit around one of the stars of this binary star system. The discovery of exoplanets in close binary systems pose interesting questions for planet formation theories \citep{Thebault2015PlanetFormationBinaries}. Additional follow-up and analysis of the \cosworthid\ system will be the subject of future work.

\subsubsection{TIC-125925505}
\begin{figure}
    \centering
    \includegraphics[width=\columnwidth]{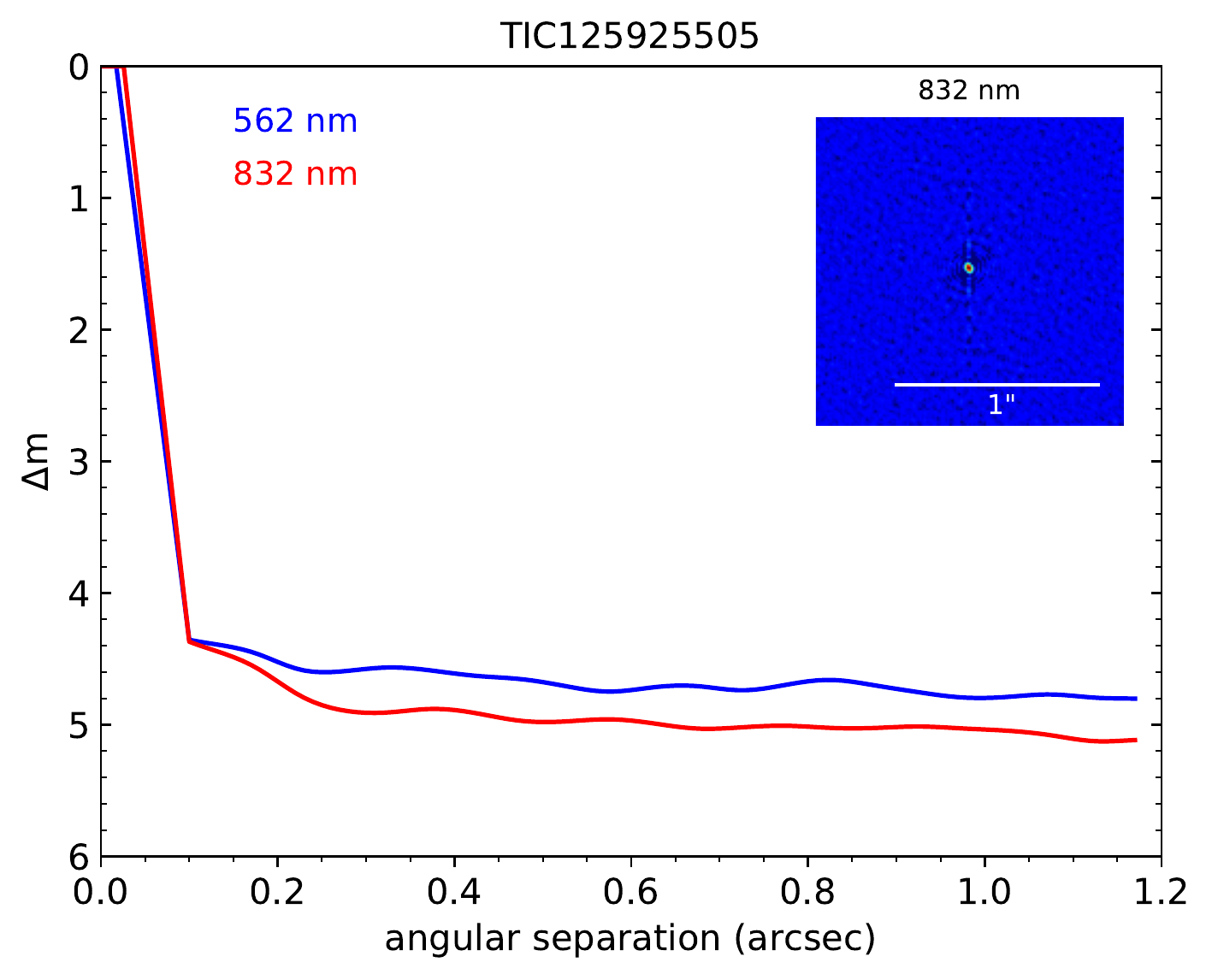}
    \caption{The Zorro contrast curve for \fernandoid. The blue and red lines show the contrast curves for the 562\,nm and 832\,nm filters, respectively. The inset shows the reconstructed image of the star from Zorro.}
    \label{fig:fernando_zorro}
\end{figure}

\begin{figure*}
    \gridline{\fig{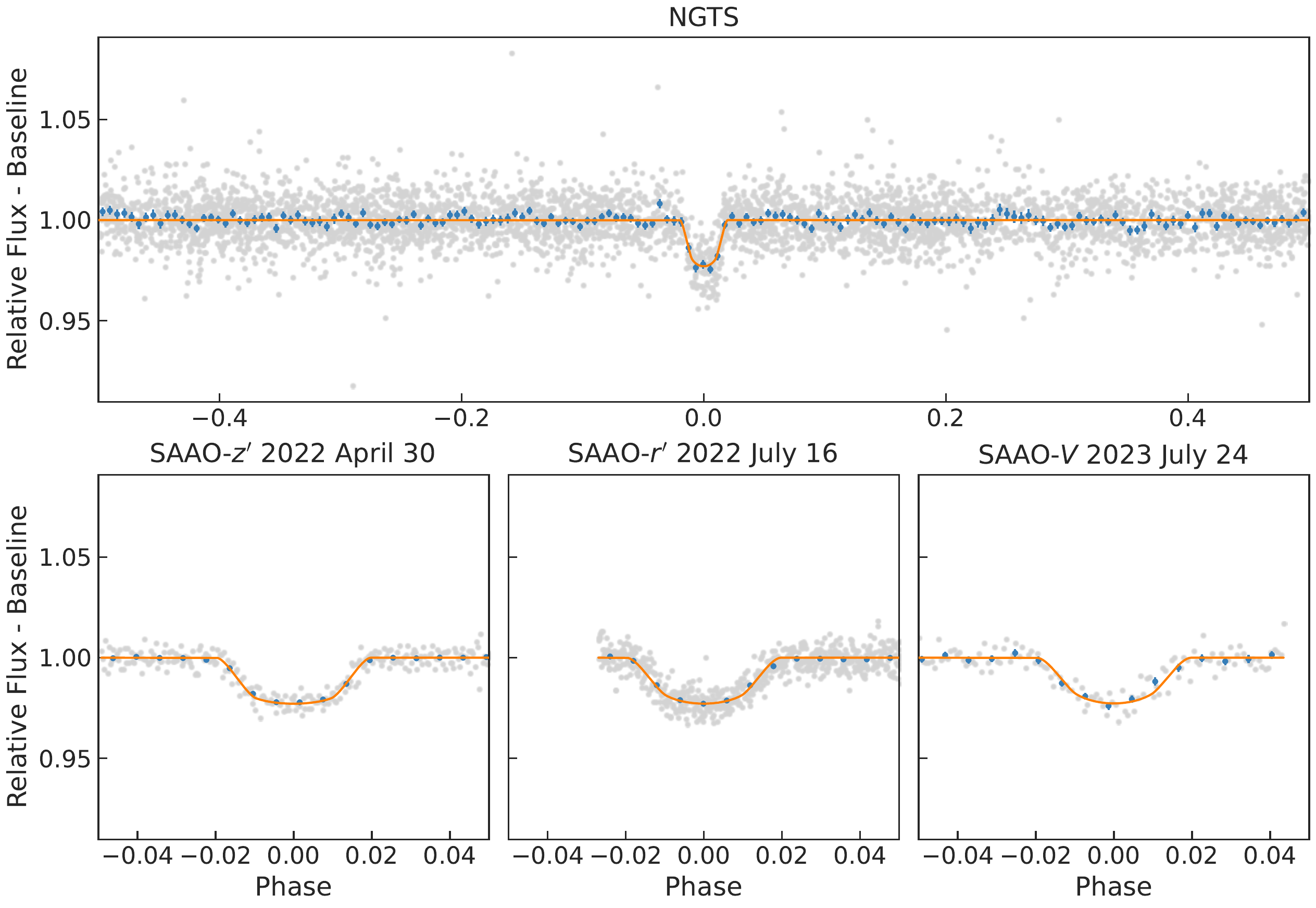}{\textwidth}{}}
    \vspace{-0.8cm}
    \gridline{
            \fig{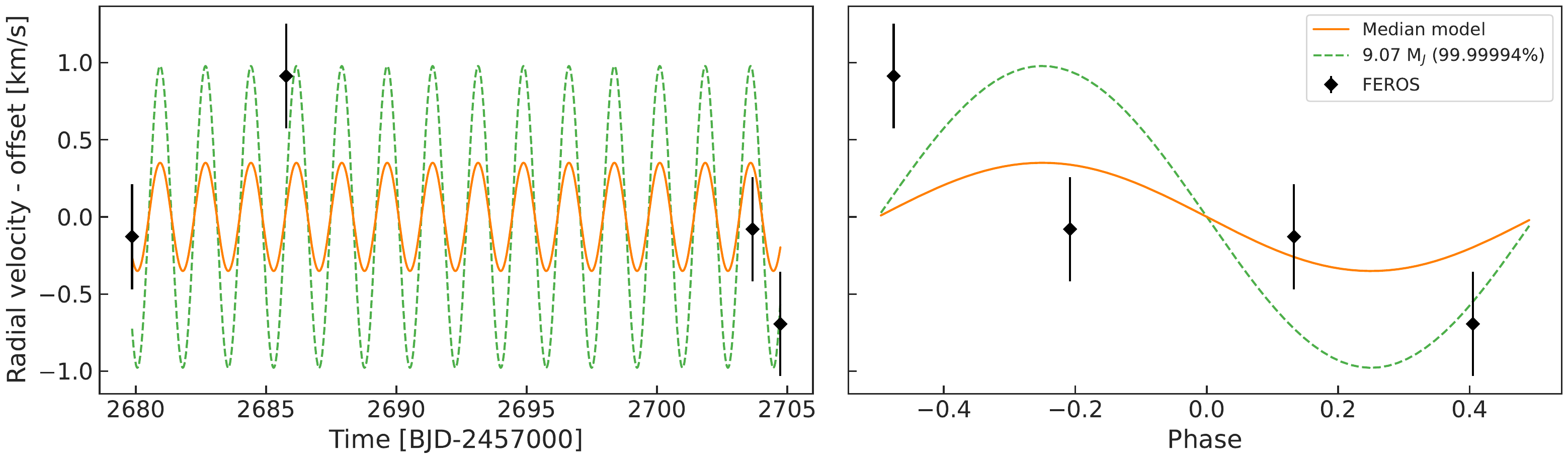}{0.99\textwidth}{}}
    \vspace{-0.8cm}
    \caption{Discovery and follow-up data obtained for \fernandoid. The phase folded light curves from NGTS is shown in the top row and each of the SAAO transit observations are shown in the middle row with the median best fit circualr model shown in orange. The radial velocity data is shown in time and phase folded in the bottom row. The model shown in orange is the median best fit circular model from \allesfitter. The green dashed line shows the RV model for the 99.99994\% upper mass limit, corresponding to the \fernandoplmass\,\mjup\ companion mass limit.} \label{fig:fernando_data}
\end{figure*}

\orion\ detected a transit signal in the NGTS data with a period of 1.743\,days, depth of 2.19\% and SDE of 27.55 from a total of 19 full or partial transits.
The host star stellar radius is 0.56\,\rsun, with SED fitting indicating that the host is a late K/early M-type star.
We observed three transits of \fernandoid\ using the 1\,m telescope at SAAO using different photometric filters ($z^\prime$, $r^\prime$, $V$).
\fernandoid\ falls in the overscan region of the CCD in \tess\ Sectors 11 and 65, and therefore we have no \tess\ data for this candidate.
The transit depth is consistent across all datasets, ruling out the possibility of this system being an eclipsing binary consisting of two stars with differing colours.
Figure~\ref{fig:fernando_zorro} shows the Zorro observations for \fernandoid. We detect no companions within 1.17\,arcsec of the target at the 4–5\,mag limit in both the 562\,nm and 832\,nm filters.
We obtain a companion radius of \fernandoplrad\,\rjup, indicating that this candidate, if planetary, is a hot giant planet.
We obtained 4 FEROS spectra of \fernandoid\ (see Table~\ref{tab:fernando_feros_rvs}). We see a maximum radial velocity variation of 1.6\,\kms\ and place a 99.99994\% upper mass limit of \fernandoplmass\,\mjup\ on the companion using the global \allesfitter\ modelling. However, we do not believe the available RV data are sufficient to consider this candidate a validated planet. This candidate requires additional RV monitoring to confirm the companion as planetary. Figure~\ref{fig:fernando_data} shows the discovery and follow-up data obtained for \fernandoid\ to date.
\begin{deluxetable}{cccc}
\tablecaption{FEROS spectroscopic data for \fernandoid.}\label{tab:fernando_feros_rvs}
\tablehead{\colhead{Time} & \colhead{RV} & \colhead{RV error} & \colhead{BIS}}
\startdata
BJD - 2457000 & \kms & \kms & \kms \\
\hline
2679.853131131  & 7.63910 & 0.0544 & -0.1776 \\
2685.764694888 & 8.67995 & 0.0275 & -0.2057 \\
2703.665987695 & 7.68750 & 0.0333 & -0.3509 \\
2704.73451741 & 7.073100 & 0.0352 & 0.1549
\enddata
\end{deluxetable}

%NOI-106844
% NGTS, SAAO, Zorro, FEROS (TESS Sector 65 hopefully)

% Two light curves were obtained for TIC-125925505 using the SHOC instrument on the 1\,m telescope at SAAO. On 2022 April 30 observations were taken with the $z^\prime$ filter and an exposure time of 60\,seconds. The telescope suffered a brief guiding loss however this was rectified before the transit ingress. 270 images were taken. On 2022 July 16 observations were taken with the $r^\prime$ filter and an exposure time of 20\,seconds. 775 images were taken.

%Zorro 2022 May 22. 16 sets (I think. 21:55 total on Science target according to OT)

% TIC-125925505 falls in the overscan region of the CCD in Sector 11 therefore we have no \tess\ data for this candidate. It is expected to be observed in Sector 65 (2023 May 04 to 2023 June 02) \sob{should be available come submission time so likely remove this sentence}.

\subsubsection{TIC-125759305}
\begin{figure*}
    \centering
    \includegraphics[width=\textwidth]{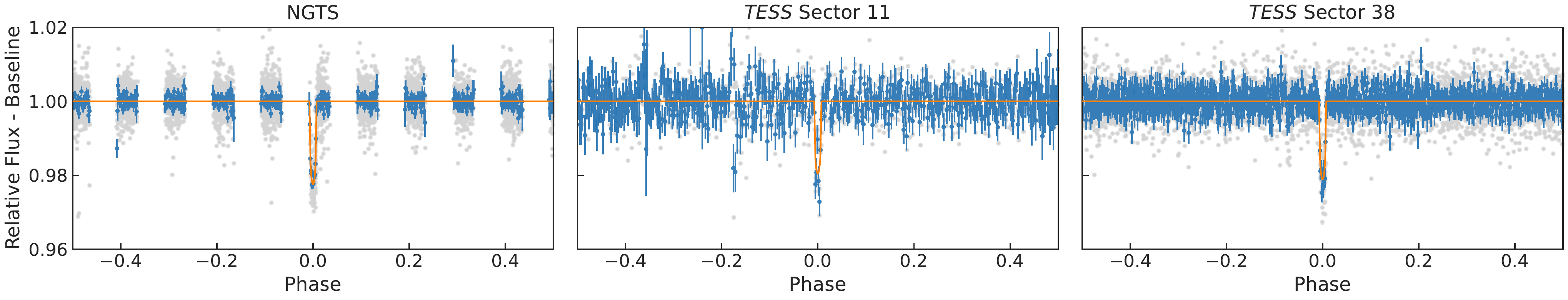}
    \caption{Discovery and follow-up data obtained for \felipeid. The phase folded light curves from NGTS and each \tess\ sector are shown with the median best fit circular model from \allesfitter\ in orange.}
    \label{fig:felipe_photom}
\end{figure*}

%NOI-106868
% NGTS, TESS
%SAAO obs junked due to bad weather
\orion\ detected a transit signal in the NGTS data with a period of 9.993\,days, depth of 2.35\% and SDE of 32.48 from a total of 10 full or partial transits.
The host star stellar radius of 0.78\,\rsun\ with SED fitting indicating that the host is a K-dwarf.
In addition to the NGTS data, the target was observed in \tess\ Sectors 11 and 38.
We identify a total of 5 transits across the two \tess\ sectors with a consistent transit depth across all datasets, ruling out the possibility of this system being an eclipsing binary consisting of two stars with differing colours. Figure~\ref{fig:felipe_photom} shows the NGTS and \tess\ photometry obtained for \felipeid.
We obtain a companion radius of \felipeplrad\,\rjup, indicating that this candidate, if planetary, is a hot Jupiter.

\subsubsection{TIC-180997904}
\begin{figure*}
    \centering
    \includegraphics[width=\textwidth]{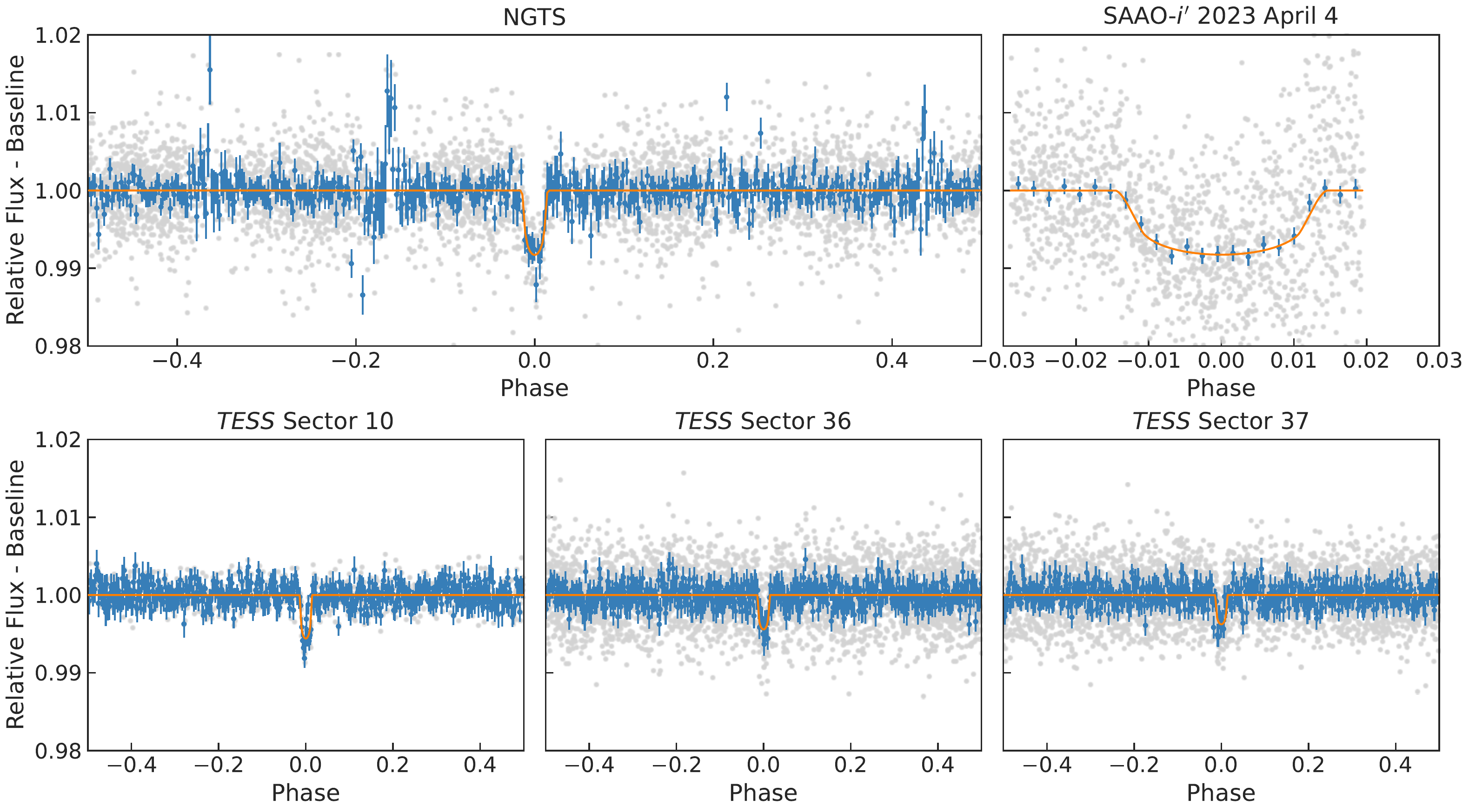}
    \caption{Discovery and follow-up data obtained for \sebastianid. The phase folded light curves from NGTS, SAAO and each \tess\ sector are shown with the median best fit circular model from \allesfitter\ in orange.}
    \label{fig:sebastian_photom}
\end{figure*}

%NOI-106869
% NGTS, TESS, SAAO
\orion\ detected a transit signal in the NGTS data with a period of 4.936\,days, depth of 0.96\% and SDE of 33.24 from a total of 7 full or partial transits.
The host star stellar radius is 1.30\,\rsun, with SED fitting indicating that the host is a G-type star.
In addition to the NGTS data, the target was observed in \tess\ Sectors 10, 36 and 37 and using the $i^{\prime}$ filter on SHOC to observe a full transit.
We identify a total of 12 transits across the three \tess\ sectors with a consistent transit depth across all datasets, ruling out the possibility of this system being an eclipsing binary consisting of two stars with differing colours. Figure~\ref{fig:sebastian_photom} shows the NGTS, SAAO and \tess\ photometry obtained for \sebastianid. The cluster of points above the normalised flux level at phase $\approx-0.18$ is due to systematics on a single night dominating this region of the light curve.
We obtain a companion radius of \sebastianplrad\,\rjup, indicating that this candidate, if planetary, is a hot Jupiter.

%% file: sections/6_deteff.tex
\begin{figure*}
    \centering
    \includegraphics[width=\textwidth]{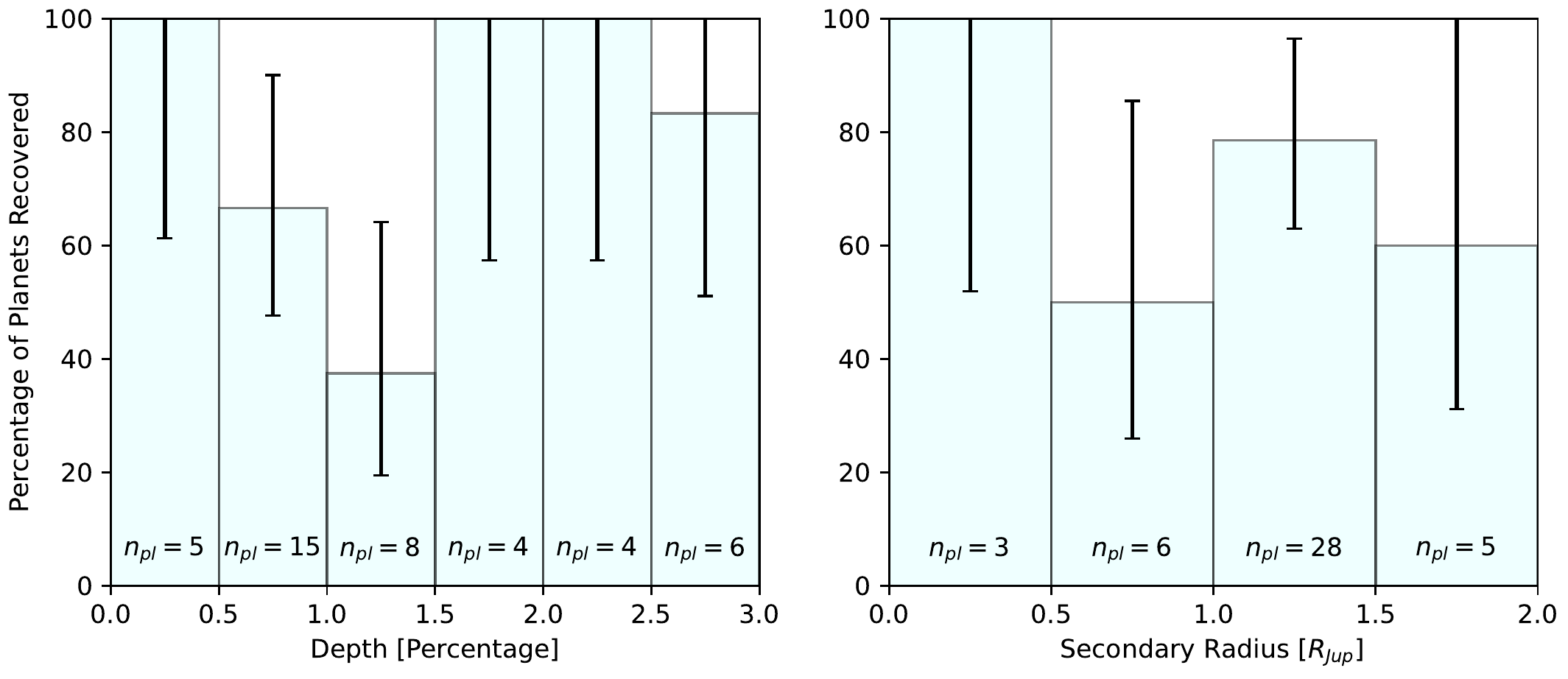}
    \caption{\textit{Left:} Recovery efficiency as a function of transit depth for the confirmed planets in the \phngts\ sample. \textit{Right:} Recovery efficiency as a function of planetary radius for the confirmed planets in the \phngts\ sample. The number of planets, $n_{pl}$, in each bin is noted on the plot. Error bars are the Poissonian 68\% uncertainty, described in \citet{Kraft1991PoissonErrors}.}
    \label{fig:confplanet_de}
\end{figure*}
\begin{figure*}
    \centering
    \includegraphics[width=\textwidth]{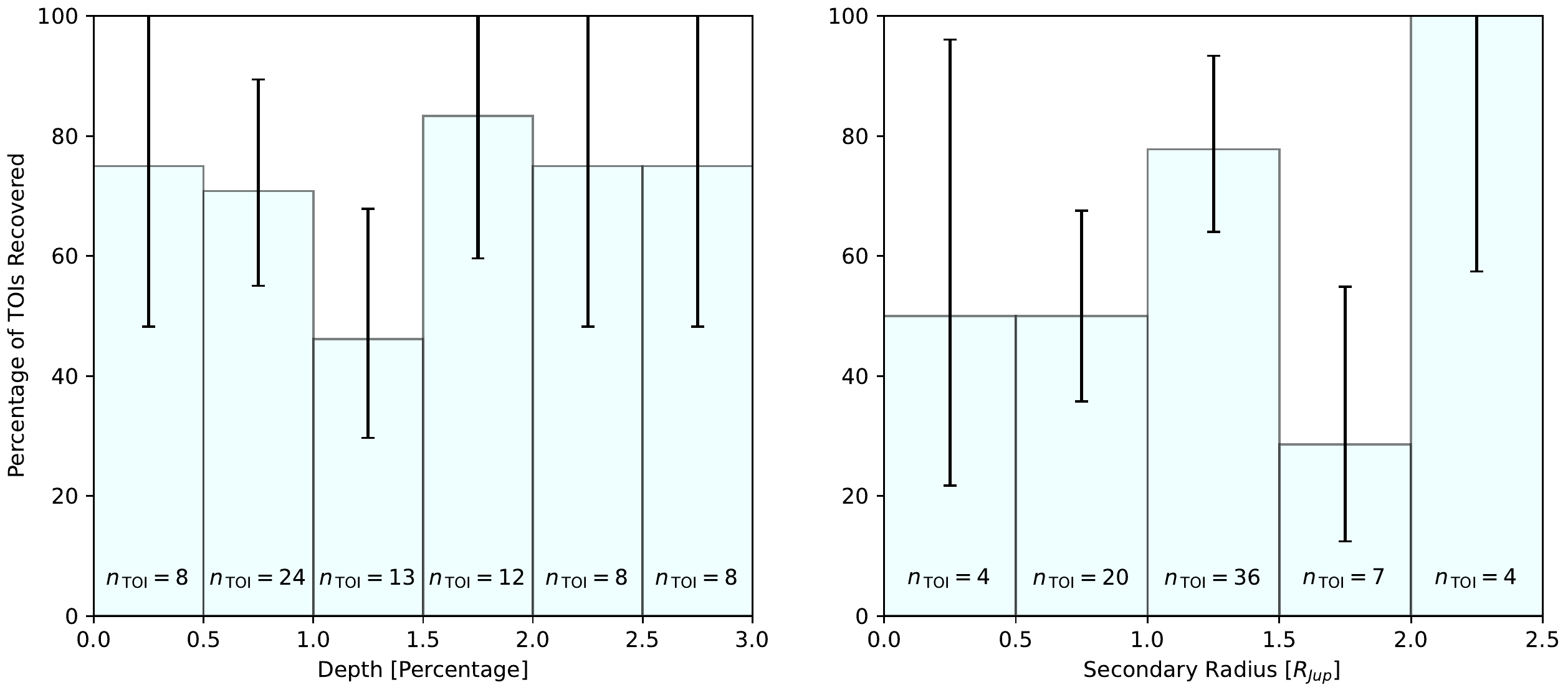}
    \caption{\textit{Left:} Recovery efficiency as a function of transit depth for TOIs in the \phngts\ sample, up to a depth of 3\%. \textit{Right:} Recovery efficiency as a function of secondary radius for TOIs in the \phngts\ sample, up to a radius of 2\,\rjup. The number of TOIs, $n_{\text{TOI}}$, in each bin is noted on the plot. Error bars are the Poissonian 68\% uncertainty, described in \citet{Kraft1991PoissonErrors}.}
    \label{fig:toi_de}
\end{figure*}

We assess the detection efficiency of the \phngts\ project by calculating the number of confirmed planets successfully recovered in our sample. We determine a planet to be successfully recovered if it passes to the additional workflows and satisfies the criteria $s_i(\mathit{NS_{sec}}) \geq 0.5$ \& $s_i(\mathit{YD_{odd}}) \geq 0.5$ \& $s_i(\mathit{YC_{odd}}) \geq 0.5$, i.e. the planet passes to the final list of candidates that are reviewed by the NGTS science team.
In the \phngts\ sample, there exists 42 \orion\ candidates that correspond to confirmed/validated planets from the NASA Exoplanet Archive\footnote{\url{https://exoplanetarchive.ipac.caltech.edu/}} \citep[Accessed 2023-08-02;][]{Akeson2013NASAExoArchive} where \orion\ identifies the published orbital period to within 5\%. \phngts\ successfully recovers 31 of these planets. 
We further assess the detection efficiency by dividing the confirmed planet sample into transit depth and planetary radius bins. We elect to use bin sizes of 0.5\% for transit depth and 0.5\,\rjup\ for planetary radius in order to preserve a sufficient number of planets in each bin to measure the detection efficiencies. Planetary radii are calculated by combining the transit depth measured by \orion\ with the stellar radius provided by the TIC. We elect to use the transit depth measured by \orion\ so that the radius value is consistent with the depth of the transit in the image presented to the volunteers. We note that NGTS-10\,b/TIC-37348844 \citep{McCormac2020NGTS10b_USP_HJ} does not have a stellar radius listed in the TIC therefore we use the value provided by the NASA Exoplanet Archive. The percentage of planets recovered in each bin are shown in Figure~\ref{fig:confplanet_de} and outlined in Tables~\ref{tab:confpl_de_tab}~and~\ref{tab:confpl_de_rad_tab}. While the recovery rate is apparently higher for shallower transits compared with, for example, the $1.0 <  \delta \leq 1.5\%$ bin, we find that the percentage of planets recovered is consistent across all bins within the Poissonian 68\% uncertainties. These uncertainties are calculated as described in \citet{Kraft1991PoissonErrors}. Our sample of confirmed planets have orbital periods between 0.77 and 7.53\,days therefore do not span a sufficient range of parameter space to analyze the recovery fraction as a function of period.
We use the \tess\ Object of Interest (TOI) catalog\footnote{https://tess.mit.edu/toi-releases/} \citep[Accessed 2023-07-27;][]{Guerrero2021TOIRelease} to further assess the detection efficiency. There are 112 TOIs in our sample that have dispositions of Known Planets (KP), Confirmed Planet (CP) or Planet Candidate (PC) and have the same detected period in NGTS and \tess\ data to within 5\%. We successfully recover 70 of these TOIs. Figure~\ref{fig:toi_de} shows the detection efficiency of TOIs as a function of transit depth and secondary radius with the values given in Tables~\ref{tab:toi_de_tab}~and~\ref{tab:toi_de_rad_tab}. As in the confirmed planet sample, we find that the recovery rate is apparently lower for the $1.0 <  \delta \leq 1.5\%$ bin compared with shallower transits however the percentage of TOIs recovered is consistent across all bins within the Poissonian 68\% uncertainties. Similarly, the percentage of TOIs recovered with $1.5 <  R_2 \leq 2.0$ is considerably lower however this measurement is limited by the small number of TOIs in this bin ($n_{\text{TOI}}=7$). Our sample of TOIs have orbital periods between 0.70 and 10.85\,days therefore do not span a sufficient range of parameter space to analyze the recovery fraction as a function of period.

\begin{deluxetable*}{cccc}
\tabletypesize{\normalsize}
\tablecaption{Number of confirmed/validated planets successfully recovered by \phngts\ per transit depth bin. Error bars are the Poissonian 68\% uncertainty, described in \citet{Kraft1991PoissonErrors}.\label{tab:confpl_de_tab}}
\tablehead{\colhead{Transit Depth (\%)} & \colhead{Number of planets} & \colhead{Planets recovered} & \colhead{Percentage of planets recovered}}
\startdata
$0 \leq \delta \leq 0.5$ & 5 & 5 & $100_{-38.67}^{+0}$ \\
$0.5 <  \delta \leq 1.0$ & 15 & 10 & $66.67_{-18.97}^{+23.46}$ \\
$1.0 <  \delta \leq 1.5$ & 8 & 3 & $37.5_{-18.01}^{+36.68}$ \\
$1.5 <  \delta \leq 2.0$ & 4 & 4 & $100_{-42.56}^{+0}$ \\
$2.0 <  \delta \leq 2.5$ & 4 & 4 & $100_{-42.56}^{+0}$ \\
$2.5 <  \delta \leq 3.0$ & 6 & 5 & $83.33_{-32.23}^{+16.67}$
\enddata
\end{deluxetable*}

\begin{deluxetable*}{cccc}
\tabletypesize{\normalsize}
\tablecaption{Number of confirmed/validated planets successfully recovered by \phngts\ per secondary radius bin. Error bars are the Poissonian 68\% uncertainty, described in \citet{Kraft1991PoissonErrors}.\label{tab:confpl_de_rad_tab}}
\tablehead{\colhead{Secondary Radius [\rjup]} & \colhead{Number of planets} & \colhead{Planets recovered} & \colhead{Percentage of planets recovered}}
\startdata
$0 \leq R_2 \leq 0.5$ & 3 & 3 & $100_{-48.02}^{+0}$ \\
$0.5 <  R_2 \leq 1.0$ & 6 & 3 & $50_{-24.01}^{+35.57}$ \\
$1.0 <  R_2 \leq 1.5$ & 28 & 22 & $78.57_{-15.56}^{+17.93}$ \\
$1.5 <  R_2 \leq 2.0$ & 5 & 3 & $60_{-28.81}^{+40}$
\enddata
\end{deluxetable*}

\begin{deluxetable*}{cccc}
\tabletypesize{\normalsize}
\tablecaption{Number of TOIs successfully recovered by \phngts\ per depth bin, up to a depth of 3\%. Error bars are the Poissonian 68\% uncertainty, described in \citet{Kraft1991PoissonErrors}.\label{tab:toi_de_tab}}
\tablehead{\colhead{Transit Depth} & \colhead{Number of TOIs} & \colhead{TOIs recovered} & \colhead{Percentage of TOIs recovered}}
\startdata
$0 \leq \delta \leq 0.5$ & 8 & 6 & $75_{-26.79}^{+25}$ \\
$0.5 <  \delta \leq 1.0$ & 24 & 17 & $70.83_{-15.81}^{+18.58}$ \\
$1.0 <  \delta \leq 1.5$ & 13 & 7 & $46.15_{-16.49}^{+21.69}$ \\
$1.5 <  \delta \leq 2.0$ & 12 & 10 & $83.33_{-23.71}^{+16.67}$ \\
$2.0 <  \delta \leq 2.5$ & 8 & 6 & $75_{-26.79}^{+25}$ \\
$2.5 <  \delta \leq 3.0$ & 8 & 6 & $75_{-26.79}^{+25}$
\enddata
\end{deluxetable*}

\begin{deluxetable*}{cccc}
\tabletypesize{\normalsize}
\tablecaption{Number of TOIs successfully recovered by \phngts\ per secondary radius bin, up to a radius of 2\,\rjup. Error bars are the Poissonian 68\% uncertainty, described in \citet{Kraft1991PoissonErrors}.\label{tab:toi_de_rad_tab}}
\tablehead{\colhead{Secondary Radius [\rjup]} & \colhead{Number of TOIs} & \colhead{TOIs recovered} & \colhead{Percentage of TOIs recovered}}
\startdata
$0 \leq R_2 \leq 0.5$ & 4 & 2 & $50_{-28.29}^{+46.05}$ \\
$0.5 <  R_2 \leq 1.0$ & 20 & 10 & $50_{-14.23}^{+17.57}$ \\
$1.0 <  R_2 \leq 1.5$ & 36 & 28 & $77.78_{-13.75}^{+15.60}$ \\
$1.5 <  R_2 \leq 2.0$ & 7 & 2 & $28.57_{-16.16}^{+26.31}$ \\
$2.0 <  R_2 \leq 2.5$ & 4 & 4 & $100_{-42.56}^{+0}$
\enddata
\end{deluxetable*}

%% file: sections/7_conclusions.tex
We present the results from the analysis of the NGTS Public Data through the Planet Hunters NGTS citizen science project. The \phngts\ project engaged 8,559 registered citizen scientists, with responses provided by an additional 3,319 non-logged-in sessions. A total of 2,626,380 individual classifications for 138,198 subjects were submitted to the project. We combined the classifications of multiple volunteers using an iterative weighting scheme to search for new planet candidates in these data.
This search has yielded 5 new planet candidates that are all consistent with being hot giant planets. Each of these candidates are undergoing active follow-up observations in an effort to characterise the systems and confirm whether the transiting companions are planetary. In particular, \nigelid\ would be the lowest mass star to host a transiting giant planet if confirmed, while \cosworthid\ may host a giant planet in an S-type orbit around one of the components of a binary star system.
We assessed the detection efficiency of our project by determining how many of the confirmed planets and TOIs present in the dataset were successfully recovered. We successfully recover 31 out of 42 confirmed planet and 70 out of 112 TOIs in the \phngts\ sample.
We provide the scores for each response for each subject classified in the \phngts\ project. These data, available in the online supplementary material, can be harnessed for a wide range of applications, for example eclipsing binary cataloging or the search for additional planet candidates not reported in this work.
Overall, we have shown that the citizen science approach can be complementary to the traditional eyeballing process of the NGTS data in finding new planet candidates not previously detected in the NGTS Public Data. Furthermore, the \phngts\ project can classify these large data sets much faster than the traditional eyeballing approach. In addition, the fresh perspective of citizen scientists compared with professional astronomers allows the detection of more unusual planet candidates such as \nigelid\ that may have been previously overlooked due to its unusually large transit depth (13.1\%). The further analysis of some of these planet candidates will be the subject of future work and the \phngts\ project will continue to search for new planet candidates as NGTS continues to observe the night sky in search of transiting planets.

%% file: sections/ack.tex
%STFC (ST/W507751/1), RAS, NASA/ADS, MAST, Zooniverse (Project builder), TESS, Exo Archive

The data presented in this paper are the result of the efforts of the \phngts\ volunteers, without whom this work would not have been possible. Their contributions are individually acknowledged at: \url{https://www.zooniverse.org/projects/mschwamb/planet-hunters-ngts/about/results}. We also thank our \phngts\ Talk moderators Arttu Sainio and See Min Lim for their time and efforts helping the \phngts\ volunteer community.

%Data access
\textbf{Data access:} All \phngts\ data supporting this study is provided as supplementary information accompanying this paper. Some of the data presented in this paper were obtained from the Mikulski Archive for Space Telescopes (MAST) at the Space Telescope Science Institute. The specific observations analyzed can be accessed via the following references: QLP: \citet{Huang2020QLPdoi}; TIC: \citet{STScI2018TIC}. STScI is operated by the Association of Universities for Research in Astronomy, Inc., under NASA contract NAS5–26555. Support to MAST for these data is provided by the NASA Office of Space Science via grant NAG5–7584 and by other grants and contracts. Data from the NASA Exoplanet Archive Planetary Systems Table can be accessed via: \dataset[https://doi.org/10.26133/NEA12]{https://doi.org/10.26133/NEA12} \citep{nasaexoarchiveDOI}.

S.M.O. is supported by a UK Science and Technology Facilities Council (STFC) Studentship (ST/W507751/1). 
The contributions at the University of Warwick by S.G., D.B., P.J.W., R.G.W. and D.A. have been supported by STFC through consolidated grants ST/P000495/1, ST/T000406/1 and ST/X001121/1.
C.A.W. would like to acknowledge support from STFC (grant number ST/X00094X/1).
J.S.J. gratefully acknowledges support by FONDECYT grant 1201371 and from the ANID BASAL project FB210003.
This work has been carried out within the framework of the National Centre of Competence in Research PlanetS supported by the Swiss National Science Foundation under grants 51NF40\_182901 and 51NF40\_205606. M.P.B. acknowledges the financial support of the SNSF.
The contributions at the Mullard Space Science Laboratory by E.M.B. have been supported by STFC through the consolidated grant ST/W001136/1.
E.G. gratefully acknowledges support from the UK Science and Technology Facilities Council (STFC; project reference ST/W001047/1).
A.O. is supported by an STFC studentship.

We thank Jean~C.~Costes for a useful discussion on RV template mask choice.

%NGTS
Based on data collected under the NGTS project at the ESO La Silla Paranal Observatory. The NGTS facility is operated by the consortium institutes with support from the UK Science and Technology Facilities Council (STFC) under projects ST/M001962/1, ST/S002642/1 and ST/W003163/1.

%Zooniverse
This publication uses data generated via the \href{https://www.zooniverse.org/}{Zooniverse.org} platform, development of which is funded by generous support, including from the National Science Foundation, NASA, the Institute of Museum and Library Services, UKRI, a Global Impact Award from Google, and the Alfred P. Sloan Foundation. The code base for the Zooniverse Project Builder Platform is available under an open-source license at \url{https://github.com/zooniverse/Panoptes} and \url{https://github.com/zooniverse/Panoptes-Front-End}.

%TESS
This paper includes data collected by the TESS mission, which are publicly available from the Mikulski Archive for Space Telescopes (MAST). Funding for the TESS mission is provided by the NASA's Science Mission Directorate.

%SAAO
This paper uses observations made at the South African Astronomical Observatory (SAAO).

%Gemini
Some of the observations in the paper made use of the High-Resolution Imaging instrument Zorro. Zorro was funded by the NASA Exoplanet Exploration Program and built at the NASA Ames Research Center by Steve B. Howell, Nic Scott, Elliott P. Horch, and Emmett Quigley. Zorro was mounted on the Gemini South telescope of the international Gemini Observatory, a program of NSF’s NOIRLab, which is managed by the Association of Universities for Research in Astronomy (AURA) under a cooperative agreement with the National Science Foundation on behalf of the Gemini partnership: the National Science Foundation (United States), National Research Council (Canada), Agencia Nacional de Investigación y Desarrollo (Chile), Ministerio de Ciencia, Tecnología e Innovación (Argentina), Ministério da Ciência, Tecnologia, Inovações e Comunicações (Brazil), and Korea Astronomy and Space Science Institute (Republic of Korea).

% %CORALIE
% \changes{This publication makes use of The Data \& Analysis Center for Exoplanets (DACE), which is a facility based at the University of Geneva (CH) dedicated to extrasolar planets data visualization, exchange, and analysis. DACE is a platform of the Swiss National Centre of Competence in Research (NCCR) PlanetS, federating the Swiss expertise in Exoplanet research. The DACE platform is available at \url{https://dace.unige.ch}.}

%Gaia
This work has made use of data from the European Space Agency (ESA) mission
{\it Gaia} (\url{https://www.cosmos.esa.int/gaia}), processed by the {\it Gaia}
Data Processing and Analysis Consortium (DPAC,
\url{https://www.cosmos.esa.int/web/gaia/dpac/consortium}). Funding for the DPAC
has been provided by national institutions, in particular the institutions
participating in the {\it Gaia} Multilateral Agreement.

%Lightkurve
This research made use of Lightkurve, a Python package for Kepler and TESS data analysis \citep{Lightkurve2018lightkurve}.

%NASA/ADS
This research has made use of NASA’s Astrophysics Data System.

%SVO
This research has made use of the Spanish Virtual Observatory (https://svo.cab.inta-csic.es) project funded by MCIN/AEI/10.13039/501100011033/ through grant PID2020-112949GB-I00.

%Publication notification throughout the nation (and beyond)
%NGTS: NASA/ADS Library (email Pete)
%Zooniverse: https://help.zooniverse.org/getting-started/lab-policies/
%SAAO: https://www.saao.ac.za/astronomers/observing/
%Gemini: https://www.gemini.edu/observing/phase-iii/acknowledging-gemini#Acknowledging%20Gemini
%ESO: https://www.eso.org/sci/observing/policies/publications.html

\vspace{5mm}
\facilities{NGTS, \tess, SAAO:1.0m (SHOC), Gemini:South (Zorro), Euler1.2m (CORALIE), Max Planck:2.2m (FEROS)}
%https://journals.aas.org/facility-keywords/


\software{
astropy \citep{Astropy2013,Astropy2018v2,Astropy2022v5},
matplotlib \citep{Hunter2007matplotlib},
Jupyter Notebook \citep{Kluyver2016Jupyter},
NumPy \citep{Walt2011Numpy,Harris2020Numpy},
pandas \citep{pandas2023},
triceratops \citep{Giacalone2020TriceratopsSoftware,Giacalone2021TriceratopsSoftware},
\href{https://github.com/zooniverse/Panoptes}{panoptes},
\href{https://github.com/zooniverse/panoptes-python-client}{panoptes-python-client},
lightkurve \citep{Lightkurve2018lightkurve},
eleanor \citep{Feinstein2019eleanor},
tess-point \citep{Burke2020tesspoint},
TESSCut \citep{Brasseur2019TESSCut},
NASA GSFC-eleanor lite \citep{Powell2022NASAGSFCeleanor},
allesfitter \citep{Guenther2019allesfittercode,Guenther2021allesfitterpaper},
ellc \citep{Maxted2016ellc},
emcee \citep{ForemanMackey2013emcee},
dynesty \citep{Speagle2020dynesty},
celerite \citep{ForemanMackey2017celerite}
}

%% file: sections/8_scoresweights.tex
\begin{figure}
    \centering
    \includegraphics[width=\columnwidth]{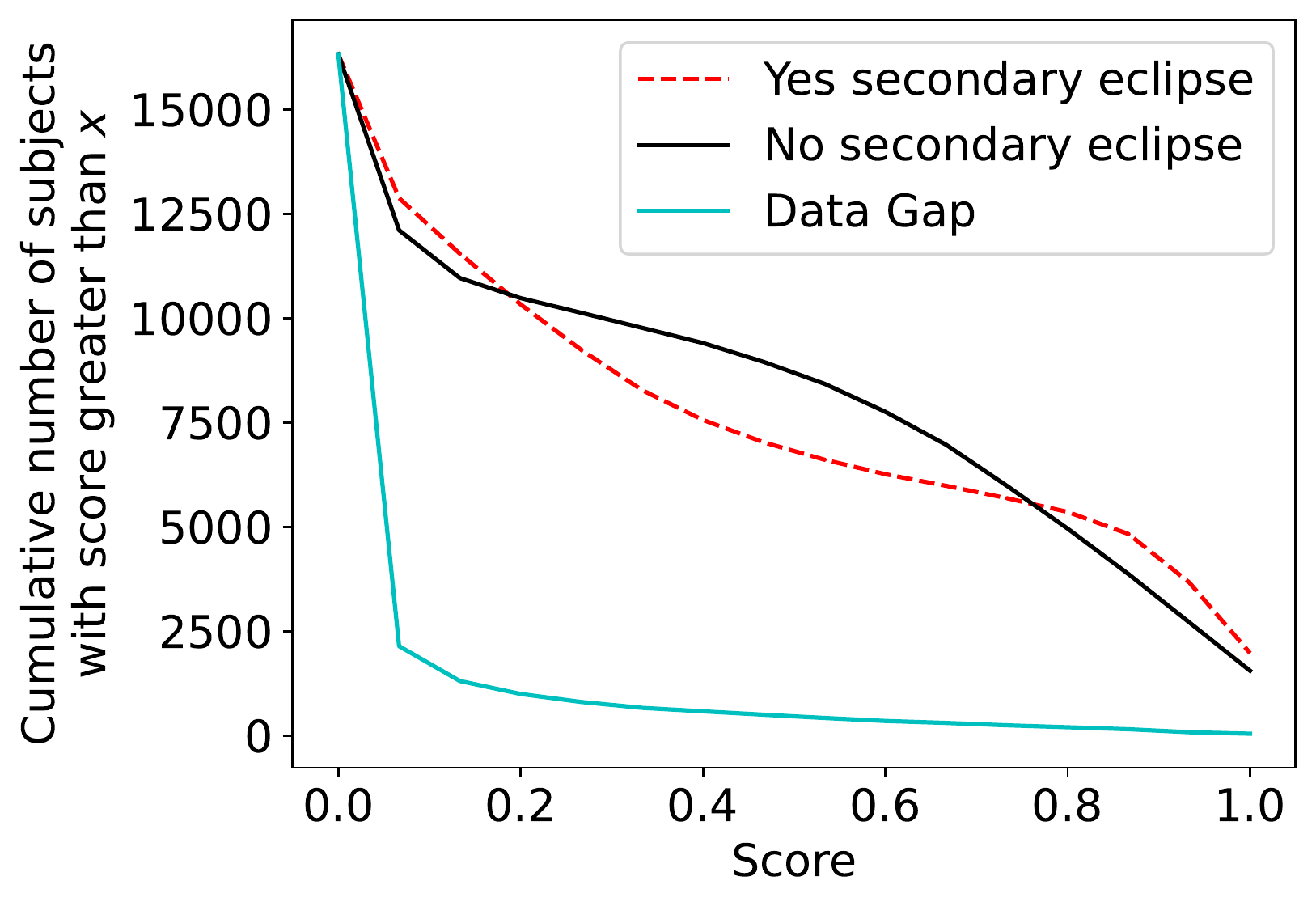}
    \caption{Cumulative distribution of subject scores greater than the given score on the x-axis for the Secondary Eclipse Check.}
    \label{fig:sec_score_dist}
\end{figure}
\begin{figure}
    \centering
    \includegraphics[width=\columnwidth]{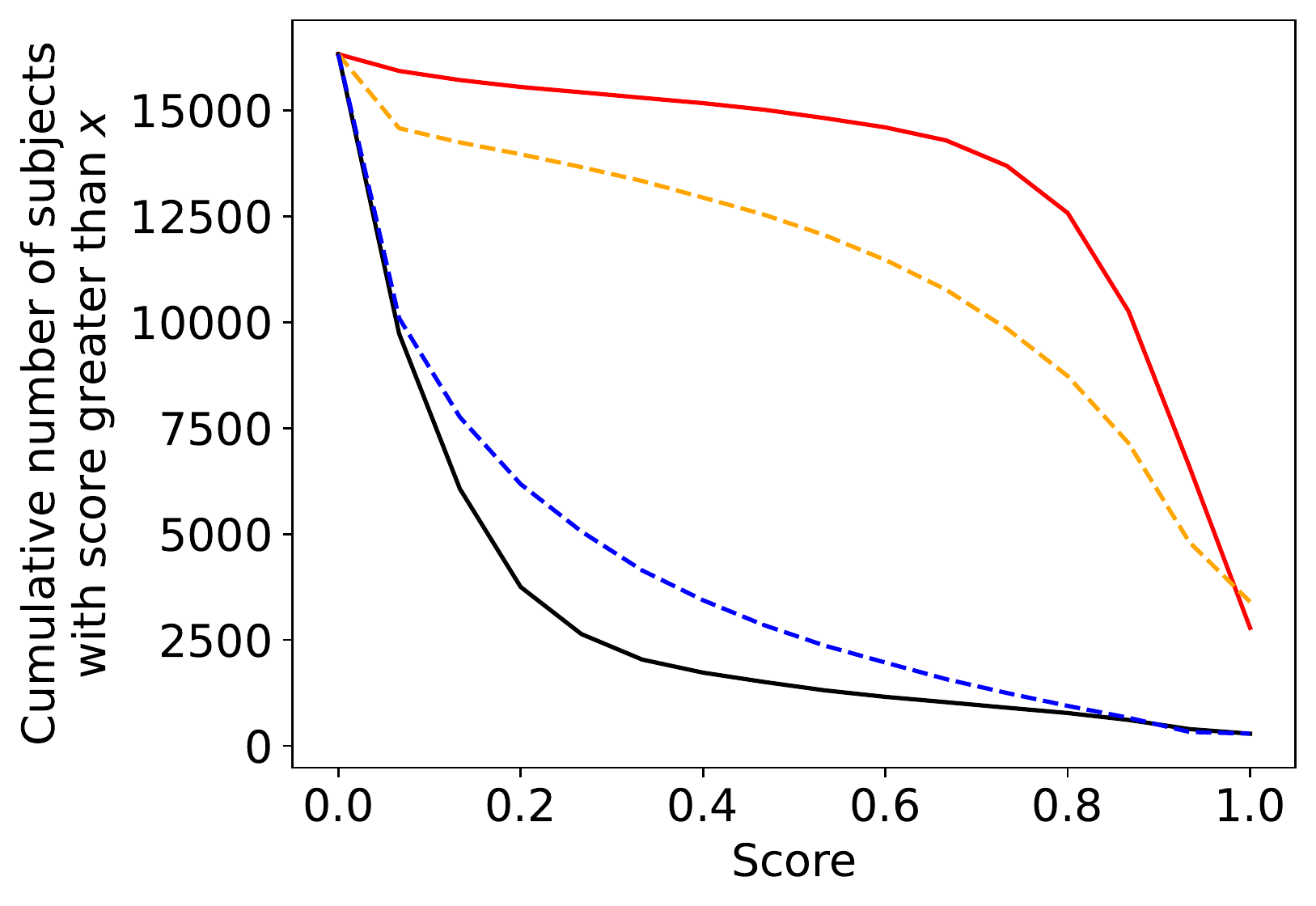}
    \caption{Cumulative distribution of subject scores greater than the given score on the x-axis for the Odd/Even Transit Check. The solid red line shows the distribution for `Yes, the data for both the odd and even transits cover the middle portion of the plot.' The solid black line shows the distribution for `No, the data for the odd and/or even transits do not cover the middle portion of the plot.' The orange dashed line shows the distribution for `Yes, the odd/even transits have similar depths.' The blue dashed line shows the distribution for `No, the odd/even transits do not have similar depths.'}
    \label{fig:odd_score_dist}
\end{figure}
\begin{figure}
    \centering
    \includegraphics[width=\columnwidth]{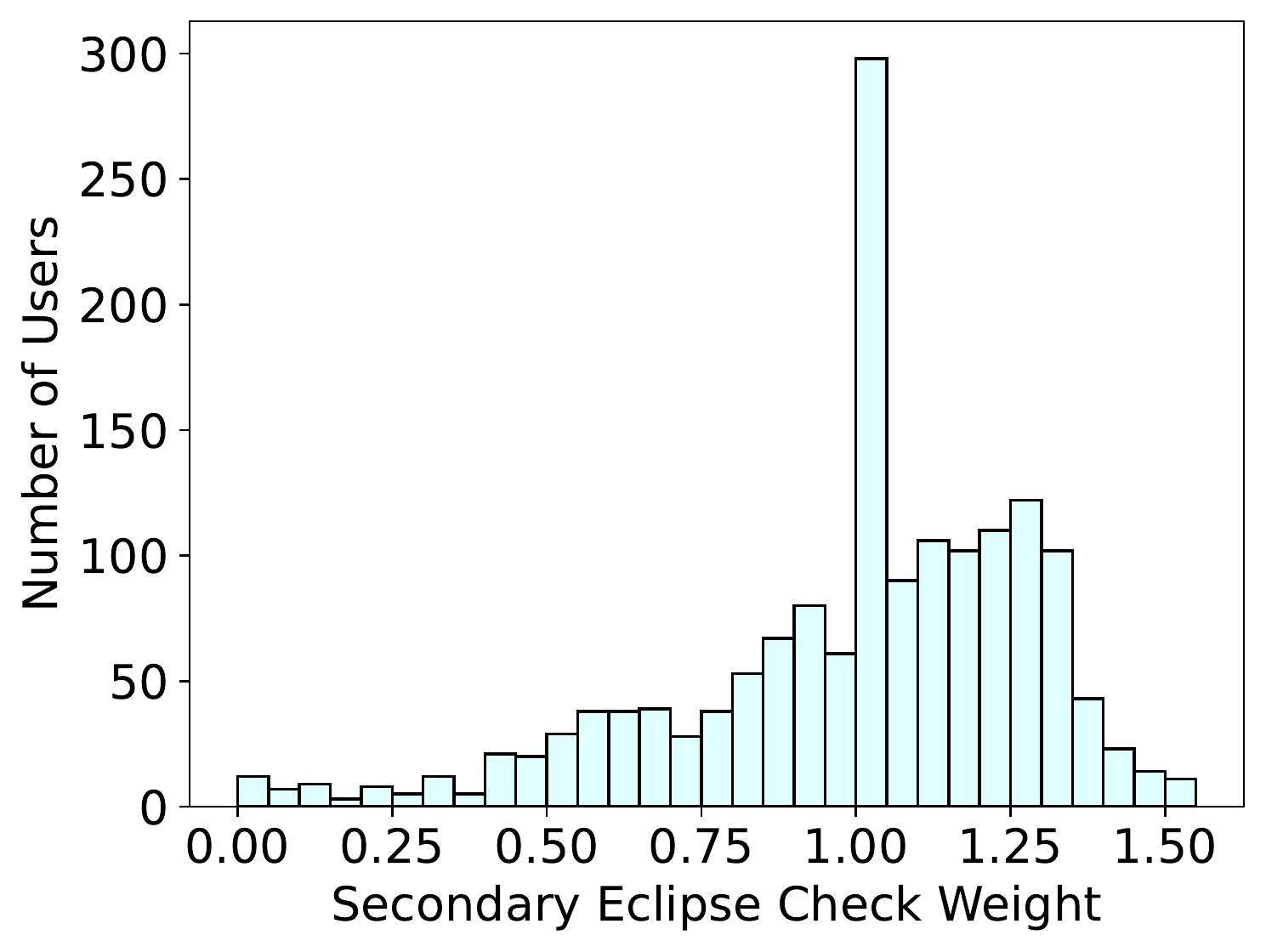}
    \caption{Histogram of user weights in the Secondary Eclipse Check. The apparent spike at $w_j(\mathit{SE}) = 1$ is due to the large spread in weights for this response, resulting in the fixed weights of users who only classified one subject to be more obvious compared with the full distribution of weights.}
    \label{fig:sec_weights}
\end{figure}
\begin{figure}
    \centering
    \includegraphics[width=\columnwidth]{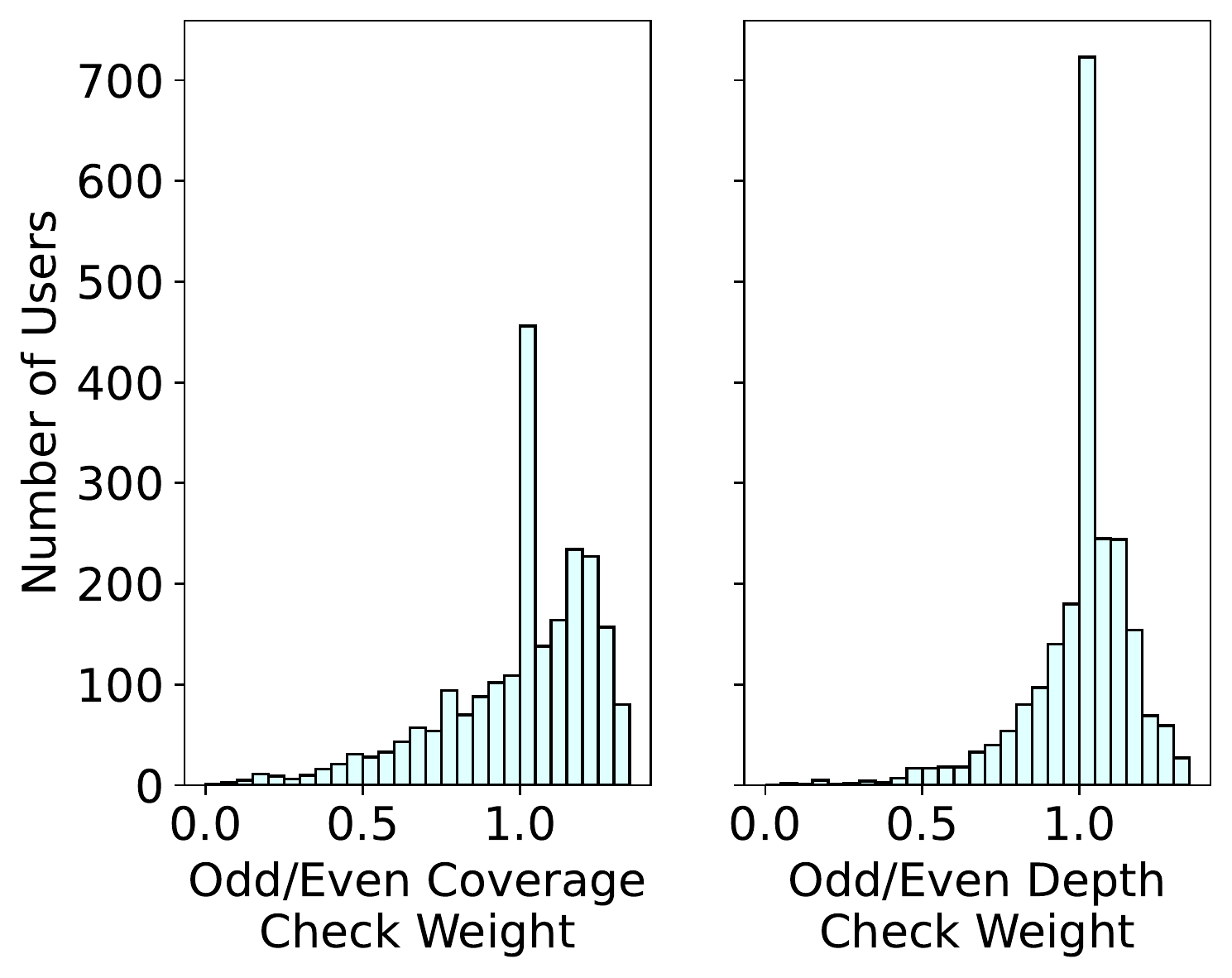}
    \caption{Histograms of user weights for both the data coverage check and transit depth check in the Odd/Even Transit Check. The apparent spike at $w_j(\mathit{OC}) = 1$ and $w_j(\mathit{OD}) = 1$ is due to the large spread in weights for these responses, resulting in the fixed weights of users who only classified one subject to be more obvious compared with the full distribution of weights.}
    \label{fig:odd_weights}
\end{figure}

Cumulative distributions of subject scores, as in Figure~\ref{fig:prim_score_dist}, for the Secondary Eclipse Check (Figure~\ref{fig:sec_score_dist}) and Odd/Even Transit Check (Figure~\ref{fig:odd_score_dist}).
Histograms of user weights, as in Figure~\ref{fig:prim_weights}, for the Secondary Eclipse Check (Figure~\ref{fig:sec_weights}) and Odd/Even Transit Check (Figure~\ref{fig:odd_weights}).

%% file: sections/9_classtables.tex
Tables~\ref{tab:ets_annotations},~\ref{tab:sec_annotations}~and~\ref{tab:oec_annotations} show the raw \phngts\ classifications submitted for the Exoplanet Transit Search, Secondary Eclipse Check and Odd/Even Transit Check, respectively. 
\begin{deluxetable}{ccccc}
\tabletypesize{\tiny}
\tablecaption{\phngts\ classifications submitted for the Exoplanet Transit Search \label{tab:ets_annotations}}
\tablehead{\colhead{User Name} & \colhead{Classification Timestamp} & \colhead{Subject ID} & \colhead{Responses} & \colhead{Classification ID}}
\startdata
beyondcommunication & 2021-10-16 17:25:37 UTC & 69473214 & [A U-shaped or box-shaped dip in the middle,A V-shaped dip in the middle,Stellar variability] & 1 \\
beyondcommunication & 2021-10-16 17:27:32 UTC & 69475120 & [A U-shaped or box-shaped dip in the middle,Stellar variability] & 2 \\
beyondcommunication & 2021-10-16 17:28:46 UTC & 69475064 & [A U-shaped or box-shaped dip in the middle,A V-shaped dip in the middle,Stellar variability] & 3 \\
beyondcommunication & 2021-10-16 17:29:19 UTC & 69474371 & [A U-shaped or box-shaped dip in the middle,Stellar variability] & 4 \\
beyondcommunication & 2021-10-16 17:29:45 UTC & 69474972 & [A V-shaped dip in the middle] & 5 \\
\vdots & \vdots & \vdots & \vdots & \vdots
\enddata
\tablecomments{This table in its entirety can be found in the online supplemental material.}
\end{deluxetable}

\begin{deluxetable}{ccccc}
\tabletypesize{\tiny}
\tablecaption{\phngts\ classifications submitted for the Secondary Eclipse Check \label{tab:sec_annotations}}
\tablehead{\colhead{User Name} & \colhead{Classification Timestamp} & \colhead{Subject ID} & \colhead{Response} & \colhead{Classification ID}}
\startdata
astro-sobrien & 2021-10-26 15:45:44 UTC & 69757462 & No secondary eclipse & 1 \\
astro-sobrien & 2021-10-26 15:48:22 UTC & 69757703 & A secondary eclipse & 2 \\
astro-sobrien & 2021-10-26 16:03:36 UTC & 69757434 & No secondary eclipse & 3 \\
Vidar87 & 2021-10-26 23:15:51 UTC & 69757265 & A secondary eclipse & 4 \\
Vidar87 & 2021-10-26 23:15:56 UTC & 69757551 & No secondary eclipse & 5 \\
\vdots & \vdots & \vdots & \vdots & \vdots
\enddata
\tablecomments{This table in its entirety can be found in the online supplemental material.}
\end{deluxetable}

\begin{deluxetable}{cccccc}
\tabletypesize{\tiny}
\tablecaption{\phngts\ classifications submitted for the Odd/Even Transit Check \label{tab:oec_annotations}}
\tablehead{\colhead{User Name} & \colhead{Classification Timestamp} & \colhead{Subject ID} & \colhead{Data Coverage Response} & \colhead{Depth Match Response} & \colhead{Classification ID}}
\startdata
mschwamb & 2021-10-26 15:39:01 UTC & 69760298 & Yes & Yes & 1 \\
not-logged-in-66f08cfbb0dc0936544d & 2021-10-26 23:23:53 UTC & 69760264 & Yes & Yes & 2 \\
not-logged-in-66f08cfbb0dc0936544d & 2021-10-26 23:23:59 UTC & 69759935 & Yes & Yes & 3 \\
not-logged-in-66f08cfbb0dc0936544d & 2021-10-26 23:24:12 UTC & 69760084 & Yes & No & 4 \\
not-logged-in-66f08cfbb0dc0936544d & 2021-10-26 23:24:50 UTC & 69760288 & Yes & Yes & 5 \\
\vdots & \vdots & \vdots & \vdots & \vdots & \vdots
\enddata
\tablecomments{This table in its entirety can be found in the online supplemental material.}
\end{deluxetable}